\newif\ifsiam
\siamfalse

\ifsiam
\documentclass[final]{siamltex}
\else
\documentclass[a4paper]{article}
\fi

\usepackage{amsmath}
\usepackage{amsfonts}
\usepackage{amssymb}         
\usepackage{amsopn}
\usepackage{latexsym}
\usepackage{graphicx}        
\usepackage{mathrsfs}		
\usepackage{psfrag}
\usepackage{algorithmic}
\usepackage{algorithm}
\usepackage{color}
\usepackage{verbatim}
\usepackage{url}

\ifsiam
\else
\fi

\newtheorem{result}{\ }[section]
\newtheorem{thm}[result]{Theorem}

\newtheorem{lem}[result]{Lemma}
\newtheorem{cor}[result]{Corollary}
\newtheorem{prop}[result]{Proposition}

\ifsiam
\else
\newenvironment{proof}
 {{\sl Proof.}\hspace*{1 ex}}%
 {{\nopagebreak\hspace*{\fill}$\Box$\par\vspace{12pt}}}
\fi

\newcommand{\transpose}[1]{{#1}^{\top}}
\newcommand{\kDMDGP}{${}^{\mbox{\sf\tiny K}}$DMDGP}
\newcommand{\CM}[1]{\mbox{\sf CM}(#1)}

\title{Euclidean distance geometry and applications}

\author{Leo
  Liberti\thanks{LIX, \'Ecole Polytechnique, 91128 Palaiseau,
    France. E-mail: {\tt liberti@lix.polytechnique.fr}.} 
  \and Carlile Lavor\thanks{Dept.~of Applied Math.~(IMECC-UNICAMP),
    State University of Campinas, 13081-970, Campinas - SP,
    Brazil. E-mail: {\tt clavor@ime.unicamp.br}.}\and Nelson
  Maculan\thanks{Federal University of Rio de Janeiro (COPPE--UFRJ),
    C.P. 68511, 21945-970, Rio de Janeiro - RJ, Brazil. E-mail: {\tt
      maculan@cos.ufrj.br}.}\and Antonio Mucherino\thanks{IRISA, 
    Univ.~of Rennes I, France. E-mail: {\tt antonio.mucherino@irisa.fr}.}}

\begin{document}

\maketitle

\begin{abstract} 
Euclidean distance geometry is the study of Euclidean geometry based
on the concept of distance. This is useful in several applications
where the input data consists of an incomplete set of distances, and
the output is a set of points in Euclidean space that realizes the
given distances. We survey some of the theory of Euclidean distance
geometry and some of the most important applications: molecular
conformation, localization of sensor networks and statics.
\end{abstract}

\pagestyle{myheadings}
\thispagestyle{plain}

\markboth{LIBERTI, LAVOR, MACULAN, MUCHERINO}{DISTANCE GEOMETRY PROBLEMS}


\ifsiam
\else
\tableofcontents
\fi

\section{Introduction}
\label{s:intro}
In 1928, Menger gave a characterization of several geometric
concepts (e.g.~congruence, set convexity) in terms of distances
\cite{menger28}. The results found by Menger, and eventually completed
and presented by Blumenthal \cite{blumenthal}, originated a body of
  knowledge which goes under the name of {\it Distance Geometry} (DG).
  This survey paper is concerned with what we believe to be the
  fundamental problem in DG:
\begin{quote}
  {\sc Distance Geometry Problem} (DGP). Given an integer $K>0$ and a
  simple undirected graph $G=(V,E)$ whose edges are weighted by a
  nonnegative function $d:E\to\mathbb{R}_+$, determine whether there
  is a function $x:V\to\mathbb{R}^K$ such that:
  \begin{equation}
    \forall \{u,v\}\in E\quad \|x(u)-x(v)\|=d(\{u,v\}). \label{eq:mdgp}
  \end{equation}
\end{quote}
Throughout this survey, we shall write $x(v)$ as $x_v$ and
$d(\{u,v\})$ as $d_{uv}$ or $d(u,v)$; moreover, norms $\|\cdot\|$ will
be Euclidean unless marked otherwise (see \cite{deza} for an account
of existing distances).

Given the vast extent of this field, we make no claim nor attempt to
exhaustiveness. This survey is intended to give the reader an idea of
what we believe to be the most important concepts of DG, keeping in
mind our own particular application-oriented slant (i.e.~molecular
conformation). 

The function $x$ satisfying \eqref{eq:mdgp} is also called a {\it
  realization} of $G$ in $\mathbb{R}^K$. If $H$ is a subgraph of $G$
and $\bar{x}$ is a realization of $H$, then $\bar{x}$ is a {\it
partial} realization of $G$. If $G$ is a given graph, then we
sometimes indicate its vertex set by $V(G)$ and its edge set by $E(G)$.

We remark that, for Blumenthal, the fundamental problem of DG was what
he called the ``subset problem'' \cite[Ch.~IV \S 36,
p.91]{blumenthal}, i.e.~finding necessary and sufficient conditions to
decide whether a given matrix is a distance matrix (see
Sect.~\ref{s:prelim:matrix}). Specifically, for Euclidean distances,
necessary conditions were (implicitly) found by
Cayley \cite{cayley1841}, who proved that five points in
$\mathbb{R}^3$, four points on a plane and three points on a line will
have zero Cayley-Menger determinant (see Sect.~\ref{s:mathapps}). Some
sufficient conditions were found by Menger \cite{menger31}, who proved
that it suffices to verify that all $(K+3)\times(K+3)$ square
submatrices of the given matrix are distance matrices
(see \cite[Thm.~38.1]{blumenthal}; other necessary and sufficient
conditions are given in Thm.~\ref{thm:edm}). The most prominent
difference is that a distance matrix essentially represents a {\it
complete weighted graph}, whereas the DGP does not impose any
structure on $G$. The first explicit mention we found of the DGP as
defined above dates 1978:
\begin{quote}
  {\it The positioning problem arises when it is necessary to locate a
    set of geographically distributed objects using measurements of
    the distances between some object pairs.} (Yemini,
  \cite{yemini78})
\end{quote}
The explicit mention that only some object pairs have known distance
makes the crucial transition from classical DG lore to the DGP. In the
year following his 1978 paper, Yemini wrote another paper on the
computational complexity of some problems in graph rigidity
\cite{yemini}, which introduced the {\it position-location problem} as
the problem of determining the coordinates of a set of objects in
space from a sparse set of distances. This was in contrast with
typical structural rigidity results of the time, whose main focus was
the determination of the rigidity of given frameworks (see
\cite{whiteleyI} and references therein). Meanwhile, Saxe had
published a paper in the same year \cite{saxe79} where the DGP was
introduced as the {\it $K$-embeddability problem} and shown to be
strongly {\bf NP}-complete when $K=1$ and strongly {\bf NP}-hard for
general $K>1$.

The interest of the DGP resides in the wealth of its applications
(molecular conformation, wireless sensor networks, statics, data
visualization and robotics among others), as well as in the beauty of
the related mathematical theory. Our exposition will take the
standpoint of a specific application which we have studied for a
number of years, namely the determination of protein structure using
Nuclear Magnetic Resonance (NMR) data. Two of the pioneers in this
application of DG are Crippen and Havel \cite{CH88}. A discussion
about the relationship between DG and real-world problems in
computational chemistry is presented in \cite{crippenbook}.

NMR data \label{p:nmr} is usually presented in current DG literature
as consisting of a graph whose edges are weighted with intervals,
which represent distance measurements with errors. This, however, is
already the result of data manipulation carried out by the NMR
specialists. The actual situation is more complex: the NMR machinery
outputs some frequency readings for distance values related to pairs
of atom types. Formally, one could imagine the NMR machinery as a
black box whose input is a set of distinct atom type pairs $\{a,b\}$
(e.g.~$\{\mbox{H},\mbox{H}\}$, $\{\mbox{C},\mbox{H}\}$ and so on), and
whose output is a set of triplets $(\{a,b\},d,q)$. Their meaning is
that $q$ pairs of atoms of type $a,b$ were observed to have (interval)
distance $d$ within the molecule being analysed. The chemical
knowledge about a protein also includes other information, such as
covalent bond and angles, certain torsion angles, and so on
(see \cite{schlick} for definitions of these chemical terms). Armed
with this knowledge, NMR specialists are able to output an interval
weighted graph which represents the molecule with a subset of its
uncertain distances (this process, however, often yields errors, so
that a certain percentage of interval distances might be outright
wrong \cite{berger}). The problem of finding a protein structure given
all practically available information about the protein is not
formally defined, but we name it anyway, as the {\sc Protein Structure
from Raw Data} (PSRD) for future reference. Several DGP variants
discussed in this survey are abstract models for the PSRD.

The rest of this paper is organized as follows. Sect.~\ref{s:prelim}
introduces the mathematical notation and basic definitions.
Sect.~\ref{s:map}-\ref{s:inclusion} present a taxonomy of problems in
DG, which we hope will be useful in order for the reader not to get
lost in the scores of acronyms we use.  Sect.~\ref{s:mathapps}
presents the main fundamental mathematical results in
DG. Sect.~\ref{s:molecular} discusses applications to molecular
conformation, with a special focus to proteins. Sect.~\ref{s:engapps}
surveys engineering applications of DG: mainly wireless sensor
networks and statics, with some notes on data visualization and
robotics.

\subsection{Notation and definitions}
\label{s:prelim}
In this section, we give a list of the basic mathematical definitions
employed in this paper. We focus on graphs, orders, matrices,
realizations and rigidity.

\subsubsection{Graphs}
\label{s:prelim:graphs}
The main objects being studied in this survey are weighted
graphs. Most of the definitions below can be found on any standard
textbook on graph theory \cite{diestelbook}. We remark that we only
employ graph theoretical notions to define paths (most definitions of
paths involve an order on the vertices).
\begin{enumerate}
\item A {\it simple undirected graph} $G$ is a couple $(V,E)$ where
  $V$ is the set of {\it vertices} and $E$ is a set of unordered pairs
  $\{u,v\}$ of vertices, called {\it edges}. For $U\subseteq V$, we
  let $E[U]=\{\{u,v\}\in E\;|\;u,v\in U\}$ be the set of edges {\it
  induced} by $U$.
\item $H=(U,F)$ is a {\it subgraph} of $G$ if $U\subseteq V$ and
  $F\subseteq E[U]$. The subgraph $H$ of $G$ is {\it induced} by $U$
  (denoted $H=G[U])$ if $F=E[U]$.
\item A graph $G=(V,E)$ is {\it complete} (or a {\it clique} on $V$) if
  $E=\{\{u,v\}\;|\;u,v\in V\land u\not=v\}$.
\item Given a graph $G=(V,E)$ and a vertex $v\in V$, we let $N_G(v)=\{u\in
  V\;|\;\{u,v\}\in E\}$ be the {\it neighbourhood} of $v$ and
  $\delta_G(v)=\{\{u,w\}\in E\;|\;u=v\}$ be the {\it star} of $v$ in
  $G$. If no ambiguity arises, we simply write $N(v)$ and $\delta(v)$.
\item We extend $N_G$ and $\delta_G$ to subsets of vertices: given a
  graph $G=(V,E)$ and $U\subseteq V$, we let $N_G(U)=\bigcup_{v\in U}
  N_G(v)$ be the neighbourhood of $U$ and $\delta_G(U)=\bigcup_{v\in
  U}\delta_G(v)$ be the {\it cutset} induced by $U$ in $G$. A cutset
  $\delta(U)$ is {\it proper} if $U\not=\varnothing$ and $U\not=V$.
  If no ambiguity arises, we write $N(U)$ and $\delta(U)$. 
\item A graph $G=(V,E)$ is {\it connected} if no proper
  cutset is empty.
\item Given a graph $G=(V,E)$ and $s,t\in V$, a {\it simple path} $H$
  with {\it endpoints} $s,t$ is a connected subgraph $H=(V',E')$ of $G$
  such that $s,t\in V'$, $|N_H(s)|=|N_H(t)|=1$, and $|N_H(v)|=2$ for all
  $v\in V'\smallsetminus\{s,t\}$. 
\item A graph $G=(V,E)$ is a {\it simple cycle} if it is connected and
  for all $v\in V$ we have $|N(v)|=2$.
\item Given a simple cycle $C=(V',E')$ in a graph $G=(V,E)$, a {\it
  chord} of $C$ in $G$ is a pair $\{u,v\}$ such that $u,v\in U$ and
  $\{u,v\}\in E\smallsetminus E'$.
\item A graph $G=(V,E)$ is {\it chordal} if every simple cycle
  $C=(V',E')$ with $|E'|>3$ has a chord.
\item Given a graph $G=(V,E)$, $\{u,v\}\in E$ and $z\not\in V$, the
  graph $G'=(V',E')$ such that $V'=(V\cup\{z\})\smallsetminus\{u,v\}$
  and $E'=(E\cup\{\{w,z\}\;|\;w\in N_G(u)\cup
  N_G(v)\})\smallsetminus\{\{u,v\}\}$ is the {\it edge contraction} of
  $G$ w.r.t.~$\{u,v\}$. 
\item Given a graph $G=(V,E)$, a {\it minor} of $G$ is any graph
  obtained from $G$ by repeated edge contraction, edge deletion and
  vertex deletion operations.
\item Unless otherwise specified, we let $n=|V|$ and $m=|E|$.
\end{enumerate}

\subsubsection{Orders}
\label{s:prelim:orders}
Algorithms for realizing graphs in Euclidean spaces are often
iterative on the graph vertices, and therefore require (or define) a
vertex order. The names of the orders listed below refer to acronyms
that indicate the problems they originate from; the acronyms
themselves will be explained in Sect.~\ref{s:map}. Orders are defined
with respect to a graph and sometimes an integer (which will turn out
to be the dimension of the embedding space). 
\begin{enumerate}
\item For any positive integer $p\in\mathbb{N}$, we let
  $[p]=\{1,\ldots,p\}$.
\item For a set $V$, a total order $<$ on $V$, and $v\in V$, we let
  $\gamma(v)=\{u\in V\;|\;u<v\}$ be the set of {\it predecessors} of
  $v$ w.r.t.~$<$, and let $\rho(v)=|\gamma(v)|+1$ be the {\it rank} of
  $v$ in $<$. We also define $\eta(v)=\{u\in V\;|\;v<u\}$ to be the
  set of {\it successors} of $v$ w.r.t.~$<$.
\item The notation $N(v)\cap\gamma(v)$ indicates the set of adjacent
  predecessors of a vertex $v$; $N(v)\cap\eta(v)$ indicates the set of
  adjacent successors of $v$.
\item It is easy to show that if $G=(V,E)$ is a
  simple path then there is an order $<$ on $V$ such that for all
  $\{u,v\}\in E$ we have $\rho(u)=\rho(v)-1$, and that the vertices of
  minimum and maximum rank in $<$ are the endpoints of the path.
\item A {\it perfect elimination order} (PEO) on $G=(V,E)$ is an order
  on $V$ such that, for each $v\in V$, $G[N(v)\cap\eta(v)]$ is a
  clique in $G$. 
\item A {\it DVOP order} on $G=(V,E)$ w.r.t.~an integer $K\in[n]$ is an
  order on $V$ where (a) the first $K$ vertices induce a clique in $G$
  and (b) each $v\in V$ of rank $\rho(v)>K$ has
  $|N(v)\cap\gamma(v)|\ge K$. 
\item A {\it Henneberg type I order} is a DVOP order where each $v$
  with $\rho(v)>K$ has $|N(v)\cap\gamma(v)|=K$.
\item A {\it $K$-trilateration} (or {\it $K$-trilaterative}) order 
  is a DVOP order where (a) the first $K+1$ vertices induce a clique
  in $G$ and (b) each $v$ with $\rho(v)>K+1$ has
  $|N(v)\cap\gamma(v)|\ge K+1$. 
\item A {\it DDGP order} is a DVOP order where for each $v$ with
  $\rho(v)>K$ there exists $U_v\subseteq N(v)\cap\gamma(v)$ with
  $|U_v|=K$ and $G[U_v]$ a clique in $G$.
\item A {\it \kDMDGP\; order} is a DVOP order where, for each $v$ with
  $\rho(v)>K$, there exists $U_v\subseteq N(v)\cap\gamma(v)$ with (a)
  $|U_v|=K$, (b) $G[U_v]$ a clique in $G$, (c) $\forall u\in
  U_v\;(\rho(v)-K-1\le\rho(u)\le\rho(v)-1)$.
\end{enumerate}
Directly from the definitions, it is clear that:
\begin{itemize}
\setlength{\parskip}{-0.1cm}
\item \kDMDGP\; orders are also DDGP orders;
\item DDGP, $K$-trilateration and Henneberg type I orders are also
DVOP orders;
\item \kDMDGP\; orders on graphs with a minimal number of edges are
inverse PEOs where each clique of adjacent successors has size $K$;
\item $K$-trilateration orders on graphs with a minimal number of
edges are inverse PEOs where each clique of adjacent successors has
size $K+1$.
\end{itemize}
Furthermore, it is easy to show that DDGP, $K$-trilateration and
Henneberg type I orders have a non-empty symmetric difference, and
that there are PEO instances not corresponding to any
inverse \kDMDGP\; or $K$-trilateration orders.

\subsubsection{Matrices}
\label{s:prelim:matrix}
The incidence and adjacency structures of graphs can be well
represented using matrices. For this reason, DG problems on graphs can
also be seen as problems on matrices.
\begin{enumerate}
\item A {\it distance space} is a pair $(X,d)$ where
  $X\subseteq\mathbb{R}^K$ and $d:X\times X\to\mathbb{R}_+$ is a
  distance function (i.e., a metric on $X$).
\item A {\it distance matrix} for a finite distance space
  $(X=\{x_1,\ldots,x_n\},d)$ is the $n\times n$ square matrix
  $D=(d_{uv})$ where for all $u,v\le|X|$ we have $d_{uv}=d(x_u,x_v)$.
\item A {\it partial matrix} on a field $\mathbb{F}$ is a pair $(A,S)$
  where $A=(a_{ij})$ is an $m\times n$ matrix on $\mathbb{F}$ and $S$
  is a set of pairs $(i,j)$ with $i\le m$ and $j\le n$; the {\it
  completion} of a partial matrix is a pair $(\alpha,B)$, where
  $\alpha:S\to\mathbb{F}$ and $B=(b_{ij})$ is an $m\times n$ matrix on
  $\mathbb{F}$, such that $\forall (i,j)\in S\;(b_{ij}=\alpha_{ij})$
  and $\forall (i,j)\not\in S\;(b_{ij}=a_{ij})$.
\item An $n\times n$ matrix $D=(d_{ij})$ is a {\it Euclidean
  distance matrix} if there exists an integer $K>0$ and a set
  $X=\{x_1,\ldots,x_n\}\subseteq\mathbb{R}^K$ such that for all
  $i,j\le n$ we have $d_{ij}=\|x_i-x_j\|$.
\item An $n\times n$ symmetric matrix $A=(a_{ij})$ is {\it positive
  semidefinite} if all its eigenvalues are nonnegative.
\item Given two $n\times n$ matrices $A=(a_{ij})$, $B=(b_{ij})$, the
{\it Hadamard product} $C=A\circ B$ is the $n\times n$ matrix
$C=(c_{ij})$ where $c_{ij}=a_{ij}b_{ij}$ for all $i,j\le n$. 
\item Given two $n\times n$ matrices $A=(a_{ij})$, $B=(b_{ij})$, the
{\it Frobenius (inner) product} $C=A\bullet B$ is defined as
$\mbox{trace}(\transpose{A}B)=\sum_{i,j\le n} a_{ij}b_{ij}$. 
\end{enumerate}

\subsubsection{Realizations and rigidity} 
\label{s:prelim:rigid}
The definitions below give enough information to define the concept of
rigid graph, but there are several definitions concerning rigidity
concepts. For a more extensive discussion, see Sect.~\ref{s:rigid}.
\begin{enumerate}
\item Given a graph $G=(V,E)$ and a manifold $M\subseteq\mathbb{R}^K$,
  a function $x:G\to M$ is an {\it embedding} of $G$ in $M$ if: (i)
  $x$ maps $V$ to a set of $n$ points in $M$; (ii) $x$ maps $E$ to a
  set of $m$ simple arcs (i.e.~homeomorphic images of $[0,1]$) in $M$;
  (iii) for each $\{u,v\}\in E$, the endpoints of the simple arc
  $x_{uv}$ are $x_u$ and $x_v$. We remark that $x$ can also be seen as
  a vector in $\mathbb{R}^{nK}$ or as an $K\times n$ real
  matrix. \label{embedding}
\item An embedding such that $M=\mathbb{R}^K$ and the simple arcs are
  line segments is called a {\it realization} of the graph in
  $\mathbb{R}^K$. A realization is {\it valid} if it satisfies
  Eq.~\eqref{eq:mdgp}. In practice we ignore the action of $x$ on $E$
  and only denote realizations as functions $x:V\to\mathbb{R}^K$.
\item Two realizations $x,y$ of a graph $G=(V,E)$ are {\it
  congruent} if for every $u,v\in V$ we have
  $\|x_u-x_v\|=\|y_u-y_v\|$. If $x,y$ are not congruent then they are
  {\it incongruent}. If $R$ is a rotation, translation or reflection
  and $Rx=(Rx_1,\ldots,Rx_{n})$, then $Rx$ is congruent to
  $x$ \cite{blumenthal}.
\item A {\it framework} in $\mathbb{R}^K$ is a pair $(G,x)$ where $x$
  is a realization of $G$ in $\mathbb{R}^K$.
\item A {\it displacement} of a framework $(G,x)$ is a continuous
  function $y:[0,1]\to\mathbb{R}^{nK}$ such that: (i) $y(0)=x$; (ii)
  $y(t)$ is a valid realization of $G$ for all $t\in[0,1]$.
\item A {\it flexing} of a framework $(G,x)$ is a displacement $y$ of
  $x$ such that $y(t)$ is incongruent to $x$ for any $t\in(0,1]$.
\item A framework is {\it flexible} if it has a flexing, otherwise it
  is {\it rigid}. \label{rigid} 
\item Let $(G,x)$ be a framework. Consider the linear system $R
  \alpha=0$, where $R$ is the $m\times nK$ matrix each $\{u,v\}$-th
  row of which has exactly $2K$ nonzero entries $x_{ui}-x_{vi}$ and
  $x_{vi}-x_{ui}$ (for $\{u,v\}\in E$ and $i\le K$), and
  $\alpha\in\mathbb{R}^{nK}$ is a vector of indeterminates. The
  framework is {\it infinitesimally rigid} if the only solutions of
  $R\alpha=0$ are translations or rotations \cite{tay-whiteley}, and
  {\it infinitesimally flexible} otherwise. By \cite[Thm.~4.1]{gluck},
  infinitesimal rigidity implies rigidity. \label{infrigid}
\item By \cite[Thm.~2.1]{Hen92}, if a graph has a unique
  infinitesimally rigid framework, then almost all its frameworks are
  rigid. Thus, it makes sense to define a {\it rigid graph} as a graph
  having an infinitesimally rigid framework. The notion of a graph
  being rigid independently of the framework assigned to it is also
  known as {\it generic rigidity} \cite{connelly}.
\end{enumerate}

A few remarks on the concept of embedding and congruence, which are of
paramount importance throughout this survey, are in order. The
definition of an embedding (Item \ref{embedding}) is similar to that
of a {\it topological embedding}. The latter, however, also satisfies
other properties: no graph vertex is embedded in the interior of any
simple arc ($\forall v\in V,\{u,w\}\in E\; (x_v\not\in x_{uw}^\circ)$,
where $S{}^\circ$ is the interior of the set $S$), and no two simple
arcs intersect ($\forall\{u,v\}\not=\{v,z\}\in E\;(x_{uv}^\circ\cap
x_{vz}^\circ=\varnothing)$). The graph embedding problem on a given
manifold, in the topological sense, is the problem of finding a
topological embedding for a graph in the manifold: the constraints are
not given by the distances, but rather by the requirement that no two
edges must be mapped to intersecting simple arcs. Garey and Johnson
list a variant of this problem as the open problem {\sc Graph
Genus} \cite[OPEN3]{gareyjohnson}. The problem was subsequently shown
to be {\bf NP}-complete by Thomassen in 1989 \cite{thomassen}.

The definition of congruence concerns pairs of points: two distinct
pairs of points $\{x_1,x_2\}$ and $\{y_1,y_2\}$ are congruent if the
distance between $x_1$ and $x_2$ is equal to the distance between
$y_1$ and $y_2$. This definition is extended to sets of points $X,Y$
in a natural way: $X$ and $Y$ are congruent if there is a surjective
function $f:X\to Y$ such that each pair $\{x_1,x_2\}\subseteq X$ is
congruent to $\{f(x_1),f(x_2)\}$. Set congruence implies that $f$ is
actually a bijection; moreover, it is an equivalence
relation \cite[Ch.~II \S 12]{blumenthal}.

\subsection{A taxonomy of problems in distance geometry}
\label{s:map}
\begin{figure}[!ht]
\begin{center}
\psfrag{DGP}{$\!\!$\scriptsize DGP}
\psfrag{PSNMR}{\scriptsize PSRD}
\psfrag{MDGP}{\scriptsize MDGP}
\psfrag{DVOP}{\scriptsize DVOP}
\psfrag{DDGP}{\scriptsize DDGP}
\psfrag{kTRILATERATION}{\scriptsize $K$-TRILAT}
\psfrag{kDMDGP}{\scriptsize $\!\!\!\!\!{}^{\mbox{\sf\tiny K}}$DMDGP}
\psfrag{DMDGPk}{\scriptsize $\!\!\!\!$DMDGP${}_K$}
\psfrag{DMDGP}{\scriptsize DMDGP}
\psfrag{DDGPk}{\scriptsize DDGP${}_K$}
\psfrag{iDGP}{\scriptsize {\it i}$\,$DGP}
\psfrag{iDMDGP}{$\!\!\!$\scriptsize {\it i}$\,$DMDGP}
\psfrag{iMDGP}{\scriptsize {\it i}$\,$MDGP}
\psfrag{MCP}{\scriptsize MCP}
\psfrag{EDMCP}{\scriptsize $\!\!$EDMCP}
\psfrag{EDM}{\scriptsize EDM}
\psfrag{PSDMCP}{\scriptsize $\!\!$PSDMCP}
\psfrag{PSD}{\scriptsize PSD}
\psfrag{WSNL}{\scriptsize WSNL}
\psfrag{RIGID}{\scriptsize GRP}
\psfrag{RAP}{\scriptsize IKP}
\psfrag{MDS}{\scriptsize MDS}
\psfrag{molecules}{$\!\!$\scriptsize molecular structure}
\psfrag{interval}{\hspace*{-0.5cm}\scriptsize interval dist.}
\psfrag{discrete}{$\!\!\!$\scriptsize exact distances}
\psfrag{matrices}{$\!\!$\scriptsize matrices}
\psfrag{robotics}{$\!\!$\scriptsize robotics}
\psfrag{statics}{$\!\!$\scriptsize statics}
\psfrag{visualization}{\scriptsize visualization}
\psfrag{sensor networks}{\scriptsize sensor networks}
\includegraphics[width=14cm]{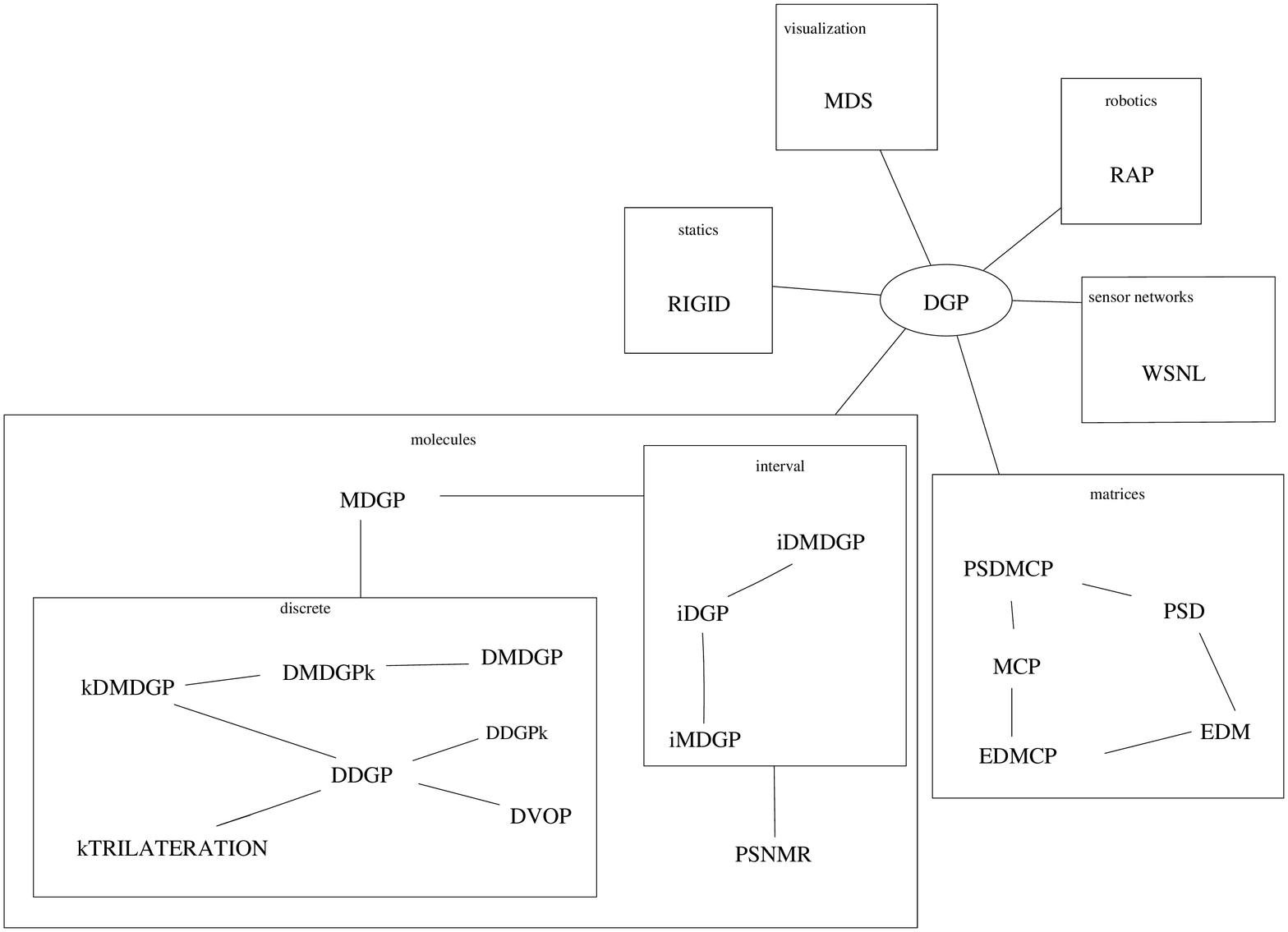}
{\normalsize
\begin{tabular}{l|l}
{\sc Acronym} & {\sc Full Name} \\ \hline
\multicolumn{2}{c}{\it Distance Geometry} \\ \hline
DGP & Distance Geometry Problem \cite{blumenthal} \\ [-0.1em] 
MDGP & Molecular DGP (in 3 dimensions) \cite{CH88} \\ [-0.1em]  
DDGP & Discretizable DGP \cite{dvop} \\ [-0.1em]
DDGP${}_K$ & DDGP in fixed dimension \cite{ddgp} \\ [-0.1em]
${}^{\mbox{\sf\scriptsize K}}$DMDGP & Discretizable MDGP
  (a.k.a.~GDMDGP \cite{powerof2-conf}) \\  [-0.1em]
DMDGP${}_K$ & DMDGP in fixed dimension \cite{bppolybook} \\ [-0.1em]
DMDGP & DMDGP${}_K$ with $K=3$ \cite{dmdgp} \\ [-0.1em]
{\it i}$\,$DGP & interval DGP \cite{CH88} \\ [-0.1em]
{\it i}$\,$MDGP & interval MDGP \cite{morewu2}\\ [-0.1em]
{\it i}$\,$DMDGP & interval DMDGP \cite{bpinterval}\\ \hline
\multicolumn{2}{c}{\it Vertex orders} \\ \hline
DVOP & Discretization Vertex Order Problem \cite{dvop} \\ [-0.1em]
$K$-TRILAT & $K$-Trilateration order problem \cite{eren04} \\ \hline
\multicolumn{2}{c}{\it Applications} \\ \hline
PSRD & Protein Structure from Raw Data \\  [-0.1em]
MDS & Multi-Dimensional Scaling \cite{deleeuw} \\ [-0.1em]
WSNL & Wireless Sensor Network Localization \cite{yemini78} \\ [-0.1em]
IKP & Inverse Kinematic Problem \cite{tolani} \\ \hline
\multicolumn{2}{c}{\it Mathematics} \\ \hline
GRP & Graph Rigidity Problem \cite{yemini}\\ [-0.1em]
MCP & Matrix Completion Problem \cite{mcp} \\ [-0.1em]
EDM & Euclidean Distance Matrix problem \cite{blumenthal} \\ [-0.1em]
EDMCP & Euclidean Distance MCP \cite{laurent97} \\ [-0.1em]
PSD & Positive Semi-Definite determination \cite{laurent00} \\ [-0.1em]
PSDMCP & Positive Semi-Definite MCP \cite{laurent97} 
\end{tabular}
}
\end{center}
\caption{Relation map for problems related to distance geometry.}
\label{f:map}
\end{figure}

Given the broad scope of the presented material (and the considerable
number of acronyms attached to problem variants), we believe that the
reader will appreciate this introductory taxonomy, which defines the
problems we shall discuss in the rest of this paper. Fig.~\ref{f:map}
contains a graphical depiction of the logical/topical existing
relations between problems. Although some of our terminology has
changed from past papers, we are now attempting to standardize the
problem names in a consistent manner. 

We sometimes emphasize problem variants where the dimension $K$ is
``fixed''. This is common in theoretical computer science: it simply
means that $K$ is a given constant which is not part of the problem
input. The reason why this is important is that the worst-case
complexity expression for the corresponding solution algorithms
decreases. For example, in Sect.~\ref{s:dvop} we give an $O(n^{K+3})$
algorithm for a problem parametrized on $K$. This is exponential
time whenever $K$ is part of the input, but it becomes polynomial when
$K$ is a fixed constant.

\begin{enumerate}
\item {\sc Distance Geometry Problem} (DGP) \cite[Ch.~IV \S
  36-42]{blumenthal}, \cite{dmdgpejor}: given an integer $K>0$ and a
  nonnegatively weighted simple undirected graph, find a realization
  in $\mathbb{R}^K$ such that Euclidean distances between pairs of
  points are equal to the edge weigths (formal definition in
  Sect.~\ref{s:intro}). We denote by DGP${}_K$ the subclass of DGP
  instances for a fixed $K$.
\item {\sc Protein Structure from Raw Data} (PSRD): we do not mean
  this as a formal decision problem, but rather as a practical
  problem, i.e.~given all possible raw data concerning a protein, find
  the protein structure in space. Notice that the ``raw data'' might
  contain raw output from the NMR machinery, covalent bonds and
  angles, a subset of torsion angles, information about the secondary
  structure of the protein, information about the potential energy
  function and so on \cite{schlick} (discussed above).
\item {\sc Molecular Distance Geometry Problem} (MDGP) \cite[\S
  1.3]{CH88}, \cite{mdgpsurvey}: same as DGP${}_3$ (discussed in
  Sect.~\ref{s:mdgp}).
\item {\sc Discretizable Distance Geometry Problem} (DDGP)
  \cite{dvop}: subset of DGP instances for which a vertex order is
  given such that: (a) a realization for the first $K$ vertices is also
  given; (b) each vertex $v$ of rank $>\!K$ has $\ge\!K$ adjacent
  predecessors (discussed in Sect.~\ref{s:ddgp}).
\item {\sc Discretizable Distance Geometry Problem} in fixed
  dimension (DDGP${}_K$) \cite{ddgp}: subset of DDGP for which the
  dimension of the embedding space is fixed to a constant value $K$
  (discussed in Sect.~\ref{s:ddgp}). The case $K=3$ was specifically
  discussed in \cite{ddgp}.
\item {\sc Discretization Vertex Order Problem} (DVOP) \cite{dvop}:
  given an integer $K>0$ and a simple undirected graph, find a vertex
  order such that the first $K$ vertices induce a clique and each
  vertex of rank $>\!K$ has $\ge\!K$ adjacent predecessors (discussed
  in Sect.~\ref{s:dvop}).
\item {\sc $K$-Trilateration} order problem ($K$-TRILAT)
  \cite{eren04}: like the DVOP, with ``$K$'' replaced by ``$K+1$''
  (discussed in Sect.~\ref{s:discr}).
\item {\sc Discretizable Molecular Distance Geometry Problem}
  (${}^{\mbox{\sf\scriptsize K}}$DMDGP) \cite{powerof2-conf}: subset
  of DDGP instances for which the $K$ immediate predecessors of $v$
  are adjacent to $v$ (discussed in Sect.~\ref{s:discr}).
\item {\sc Discretizable Molecular Distance Geometry Problem} in fixed
  dimension (DMDGP${}_K$) \cite{bp-poly}: subset of
  ${}^{\mbox{\sf\scriptsize K}}$DMDGP for which the dimension of the
  embedding space is fixed to a constant value $K$ (discussed in
  Sect.~\ref{s:discr}).
\item {\sc Discretizable Molecular Distance Geometry Problem} (DMDGP)
  \cite{dmdgp}: the DMDGP${}_K$ with $K=3$ (discussed in
  Sect.~\ref{s:discr}).
\item {\sc interval Distance Geometry Problem} ({\it i}$\,$DGP)
  \cite{CH88,dmdgpejor}: given an integer $K>0$ and a simple
  undirected graph whose edges are weighted with intervals, find a
  realization in $\mathbb{R}^K$ such that Euclidean distances between
  pairs of points belong to the edge intervals (discussed in
  Sect.~\ref{s:interv}).
\item {\sc interval Molecular Distance Geometry Problem} ({\it
  i}$\,$MDGP) \cite{morewu2,dmdgpejor}: the {\it i}$\,$DGP with $K=3$
  (discussed in Sect.~\ref{s:interv}).
\item {\sc interval Discretizable Molecular Distance Geometry Problem}
  ({\it i}$\,$DMDGP) \cite{iBP-conf}: given: (i) an integer $K>0$;
  (ii) a simple undirected graph whose edges can be partitioned in
  three sets $E_N, E_S, E_I$ such that edges in $E_N$ are weighted
  with nonnegative scalars, edges in $E_S$ are weighted with finite
  sets of nonnegative scalars, and edges in $E_I$ are weighted with
  intervals; (iii) a vertex order such that each vertex $v$ of rank
  $>\!K$ has at least $K$ immediate predecessors which are adjacent to
  $v$ using only edges in $E_N\cup E_S$, find a realization in
  $\mathbb{R}^3$ such that Euclidean distances between pairs of points
  are equal to the edge weights (for edges in $E_N$), or belong to the
  edge set (for edges in $E_S$), or belong to the edge interval (for
  edges in $E_I$) (discussed in Sect.~\ref{s:interv}).
\item {\sc Wireless Sensor Network Localization} problem (WSNL)
  \cite{yemini78,savvides,eren04}: like the DGP, but with a subset $A$
  of vertices (called {\it anchors}) whose position in $\mathbb{R}^K$
  is known {\it a priori} (discussed in Sect.~\ref{s:wsnl}). The
  practically interesting variants have $K$ fixed to $2$ or
  $3$. \label{item:wsnl}
\item {\sc Inverse Kinematic Problem} (IKP) \cite{tolani}: subset of
  WSNL instances such that the graph is a simple path whose endpoints
  are anchors (discussed in Sect.~\ref{s:robot}). \label{item:ikp}
\item {\sc Multi-Dimensional Scaling} problem (MDS) \cite{deleeuw}:
  given a set $X$ of vectors, find a set $Y$ of smaller dimensional
  vectors (with $|X|=|Y|)$ such that the distance between the $i$-th
  and $j$-th vector of $Y$ approximates the distance of the
  corresponding pair of vectors of $X$ (discussed in
  Sect.~\ref{s:mds}).
\item {\sc Graph Rigidity Problem} (GRP) \cite{yemini,laurent97}:
  given a simple undirected graph, find an integer $K'>0$ such that
  the graph is (generically) rigid in $\mathbb{R}^K$ for all $K\ge K'$
  (discussed in Sect.~\ref{s:rigid}).
\item {\sc Matrix Completion Problem} (MCP) \cite{mcp}: given a square
  ``partial matrix'' (i.e.~a matrix with some missing entries) and a
  matrix property $P$, determine whether there exists a completion of
  the partial matrix that satisfies $P$ (discussed in
  Sect.~\ref{s:mathapps}).
\item {\sc Euclidean Distance Matrix} problem (EDM)
  \cite{blumenthal}: determine whether a given matrix is a Euclidean
  distance matrix (discussed in Sect.~\ref{s:mathapps}).
\item {\sc Euclidean Distance Matrix Completion Problem} (EDMCP)
  \cite{laurent97,laurent00,pardalos-edm}: subset of MCP instances
  with $P$ corresponding to ``Euclidean distance matrix for a set of
  points in $\mathbb{R}^K$ for some $K$'' (discussed in
  Sect.~\ref{s:mathapps}).
\item {\sc Positive Semi-Definite} determination (PSD)
  \cite{laurent00}: determine whether a given matrix is positive
  semi-definite (discussed in Sect.~\ref{s:mathapps}).
\item {\sc Positive Semi-Definite Matrix Completion Problem} (PSDMCP)
  \cite{laurent97,laurent00,pardalos-edm}: subset of MCP instances
  with $P$ corresponding to ``positive semi-definite matrix''
  (discussed in Sect.~\ref{s:mathapps}).
\end{enumerate}

\subsection{DGP variants by inclusion}
\label{s:inclusion}
\begin{figure}[!ht]
\begin{center}
\psfrag{iDGP}{{\it i}$\,$DGP}
\psfrag{iMDGP}{{\it i}$\,$MDGP}
\psfrag{iDMDGP}{{\it i}$\,$DMDGP}
\psfrag{DMDGPk}{DMDGP${}_K$}
\psfrag{MDGP}{MDGP}
\psfrag{DGP}{DGP}
\psfrag{DDGP}{DDGP}
\psfrag{DDGP3}{DDGP${}_K$}
\psfrag{GDMDGP}{${}^{\mbox{\sf\scriptsize K}}$DMDGP}
\psfrag{DMDGP}{DMDGP}
\includegraphics[width=5cm]{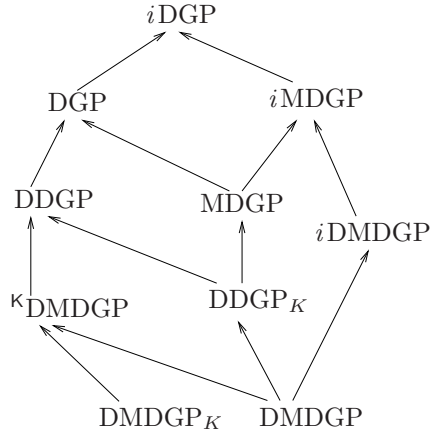}
\end{center}
\caption{Inclusionwise lattice of DGP variants (arrows mean $\subset$).}
\label{f:lattice}
\end{figure}

The research carried out by the authors of this survey focuses mostly
on the subset of problems in the {\it Distance Geometry} category
mentioned in Fig.~\ref{f:map}. These problems, seen as sets of
instances, are related by the inclusionwise lattice shown in
Fig.~\ref{f:lattice}. For reasons relating to our own development of
these ideas, the names of some problems in this paper are different
than those given in previously published papers; the definitions,
however, coincide.

\section{The mathematics of distance geometry}
\label{s:mathapps}
This section will briefly discuss some fundamental mathematical
notions related to DG. As is well known, DG has strong connections to
matrix analysis, semidefinite programming, convex geometry and graph
rigidity \cite{dattorro}. On the other hand, the fact that G\"odel
discussed extensions to differentiable manifolds is perhaps less known
(Sect.~\ref{s:goedel}), as well as perhaps the exterior algebra
formalization (Sect.~\ref{s:exterior}).

Given a set $\mathcal{U}=\{p_0,\ldots,p_K\}$ of
$K+1$ points in $\subseteq\mathbb{R}^K$, the volume of the $K$-simplex
defined by the points in $\mathcal{U}$ is given by the so-called
Cayley-Menger formula \cite{menger28,menger31,blumenthal}:
\begin{equation}
 \Delta_K(\mathcal{U})=\sqrt{\frac{(-1)^{K+1}}{2^K(K!)^2} \CM{\mathcal{U}}},
   \label{eq:deltadef}
\end{equation}
where $\CM{\mathcal{U}}$ is the Cayley-Menger
determinant \cite{menger28,menger31,blumenthal}:
\begin{equation}
\CM{\mathcal{U}}= \left|\begin{array}{ccccc}
    0 & 1 & 1 & \ldots & 1 \\
    1 & 0 & d_{01}^2 & \ldots & d_{0K}^2 \\
    1 & d_{01}^2 & 0 & \ldots & d_{1K}^2 \\
    \vdots & \vdots & \vdots &\ddots & \vdots \\
    1 & d_{0K}^2 & d_{1K}^2 & \ldots & 0
\end{array}\right|, \label{cayley-menger}
\end{equation}
with $d_{uv}=\|p_u-p_v\|$ for all $u,v\in\{0,\ldots,K\}$. The
Cayley-Menger determinant is proportional to the quantity known as the
{\it oriented volume} \cite{CH88} (sometimes also called the {\it
signed volume}), which plays an important role in the theory of
oriented matroids \cite{orientedmatroids}. Opposite signed values of
simplex volumes correspond to the two possible orientations of a
simplex keeping one of its facets fixed (see e.g.~the two positions
for vertex 4 in Fig.~\ref{f:4vtx}, center). In \cite{yang_06}, a
generalization of DG is proposed to solve spatial constraints, using
an extension of the Cayley-Menger determinant.

\subsection{The Euclidean Distance Matrix problem}
Cayley-Menger determinants were used in \cite{blumenthal} to give
necessary and sufficient conditions for the EDM problem,
i.e.~determining whether for a given $n\times n$ matrix $D=(d_{ij})$
there exists an integer $K$ and a set $\{p_1,\ldots,p_n\}$ of points
of $\mathbb{R}^K$ such that $d_{ij}=\|p_i-p_j\|$ for all $i,j\le n$.
Necessary and sufficient conditions for a matrix to be a Euclidean
distance matrix are given in \cite{sippl}. 
\ifsiam \begin{theorem}[Thm.~4 in \cite{sippl}] \else \begin{thm}[Thm.~4 in \cite{sippl}] \fi 
A $n\times n$ distance matrix $D$ is embeddable in $\mathbb{R}^K$ but
not in $\mathbb{R}^{K-1}$ if and only if: (i) there is a principal
$(K+1)\times(K+1)$ submatrix $R$ of $D$ with nonzero Cayley-Menger
determinant; (ii) for $\mu\in\{1,2\}$, every principal $(K+\mu)\times
(K+\mu)$ submatrix of $D$ containing $R$ has zero Cayley-Menger
determinant.\label{thm:edm}
\ifsiam \end{theorem} \else \end{thm} \fi
In other words, the two conditions of this theorem state that there
must be a $K$-simplex $S$ of reference with nonzero volume in
$\mathbb{R}^K$, and all $(K+1)$- and $(K+2)$-simplices containing $S$
as a face must be contained in $\mathbb{R}^K$.

\subsection{Differentiable manifolds}
\label{s:goedel}
Condition (ii) in Thm.~\ref{thm:edm} fails to hold in the cases of
(curved) manifolds. G\"odel showed that, for $K=3$, the condition can
be updated as follows (paper {\it 1933h} in \cite{goedel1}): for any
quadruplet $\mathcal{U}_n$ of point sequences $p_u^n$ (for
$u\in\{0,\ldots,3\}$) converging to a single non-degenerate point
$p_0$, the following holds:
\begin{equation*}
  \lim_{n\to\infty} \frac{\CM{\mathcal{U}_n}}{\sum\limits_{u<v}
  \|p_u^n-p_v^n\|^6} = 0.
\end{equation*}

In a related note, G\"odel also showed that if
$\mathcal{U}=\{p_0,\ldots,p_3\}$ with $\CM{\mathcal{U}}\not=0$, then
the distance matrix over $\mathcal{U}$ can be realized on the surface
of a 2-sphere where the distances between the points are the lengths
of the arcs on the spherical surface (paper {\it 1933b}
in \cite{goedel1}). This observation establishes a relationship
between DG and the Kissing Number Problem \cite{knpdam} and, more in
general, to coding theory \cite{conwaysloane}.

\subsection{Exterior algebras}
\label{s:exterior}
Cayley-Menger determinants are exterior products \cite{auslander}. The
set of all possible exterior products of a vector space forms an {\it
exterior algebra}, which is a special type of Clifford
algebra \cite{chevalley}; specifically, exterior algebras are tensor
algebras modulo the ideal generated by $x^2$. The fact that any square
element of the algebra is zero implies
$0=(x+y)^2=x^2+xy+yx+y^2=xy+yx$, and hence $xy=-yx$.  Accordingly,
exterior algebras are used in the study of alternating multilinear
forms. The paper \cite{dress_93} gives an in-depth view of the
connection between DG and Clifford algebras.

In the setting of distance geometry, we define the product of vectors
$x_1,\ldots,x_n \in\mathbb{R}^K$ (for $n\ge K$) by the corresponding
Cayley-Menger determinant on $\mathcal{U}=\{x_0,\ldots,x_n\}$ where
$x_0$ is the origin. It is clear that, if $x_i=x_j$ for some
$i\not=j$, then the corresponding $n$-simplex is degenerate and
certainly has volume 0 in $\mathbb{R}^K$ (even if $n=K$), hence
$\CM{\mathcal{U}}=0$. Equivalently, if a product $\prod_i x_i$ can be
written as $x_j^2\prod\limits_{i\not=j} x_i$, then it belongs to the
ideal $\langle x^2\rangle$ and is replaced by 0 in the exterior
algebra. This immediately implies that the Cayley-Menger determinant
is an alternating form. 

Abstract relationships between an exterior algebra and its
corresponding vector space are specialized to relationships between
Cayley-Menger determinants and vectors in $\mathbb{R}^K$. Thus, for
example, one can derive a well-known result in linear algebra:
$x_1,\ldots,x_K$ are linearly independent if and only if
$\CM{\mathcal{U}}\not=0$ where $\mathcal{U}=\{x_0,\ldots,x_K\}$ with
$x_0$ being the origin \cite{auslander,chevalley}. A more interesting
example consists in deriving certain invariants expressed in {\it
Pl\"ucker coordinates} \cite{chevalley}: given a basis
$x_1,\ldots,x_K$ of $\mathbb{R}^K$ and a basis $y_1,\ldots,y_h$ of
$\mathbb{R}^h$ where $h\le K$, it can be shown that for any subset $S$
of $\{1,\ldots,K\}$ of size $h$ there exist constants $\alpha_S$ such
that $\sum_{S} \alpha_S \prod\limits_{i\in S} x_i=\prod\limits_{i\le
h} y_i$. In our setting, product vectors correspond to Cayley-Menger
determinants derived from the given points $x_1,\ldots,x_K$ and an
origin $x_0$. It turns out that the ratios of various $\alpha_S$'s are
invariant over different bases $y_1',\ldots,y_h'$ of $\mathbb{R}^h$,
which allows their employment as a convenient coordinate system for
$\mathbb{R}^h$.  Invariants related to the Pl\"ucker coordinates are
exploited in \cite{CH88} to find realizations of chirotopes
(orientations of vector configurations \cite{orientedmatroids}).

\subsection{Bideterminants}
For sets of more than $K+1$ points, the determination of the relative
orientation of each $K$-simplex in function of a $K$-simplex of
reference (see e.g.~Fig.~\ref{fig:cliques}, center and right) is
important. Such relative orientations are given by the {\it
Cayley-Menger bideterminant} of two $K$-simplices
$\mathcal{U}=\{p_0,\ldots,p_{K}\}$ and
$\mathcal{V}=\{q_0,\ldots,q_{K}\}$, with $d_{ij}=\|p_i-q_j\|$:
\begin{equation}
\CM{\mathcal{U},\mathcal{V}}= \left|\begin{array}{ccccc}
    0 & 1 & \ldots & 1 \\
    1 & d_{00}^2 & \ldots & d_{0K}^2 \\
    1 & d_{10}^2 & \ldots & d_{1K}^2 \\
    \vdots & \vdots & \ddots & \vdots \\
    1 & d_{K0}^2 & \ldots & d_{KK}^2
\end{array}\right|. \label{cayley-menger-bi}
\end{equation}
These bideterminants allow, for example, the determination of
stereoisometries in chemistry \cite{orientedmatroids}. 

\subsection{Positive semidefinite and Euclidean distance matrices}
\label{s:schoenberg}
Schoenberg proved in \cite{schoenberg} that there is a one-to-one
relationship between Euclidean distance matrices and positive
semidefinite matrices. Let $D=(d_{ij})$ be an $(n+1)\times (n+1)$ matrix and
$A=(a_{ij})$ be the $(n+1)\times(n+1)$ matrix given by
$a_{ij}=\frac{1}{2}(d_{0i}^2+d_{0j}^2-d_{ij}^2)$.

The bijection given by Thm.~\ref{thm:psd} below can be exploited to
show that solving the PSD and the EDM is essentially the same
thing \cite{sippl2}.
\ifsiam \begin{theorem}[Thm.~1 in \cite{sippl2}] \else \begin{thm}[Thm.~1 in \cite{sippl2}] \fi 
A necessary and sufficient condition for the matrix $D$ to be a
Euclidean distance matrix with respect to a set
$\mathcal{U}=\{p_0,\ldots,p_n\}$ of points in $\mathbb{R}^K$ but not
in $\mathbb{R}^{K-1}$ is that the quadratic form $\transpose{x}Ax$
(where $A$ is given above) is positive semidefinite of rank
$K$. \label{thm:psd}
\ifsiam \end{theorem} \else \end{thm} \fi
Schoenberg's theorem was cast in a very compact and elegant form
in \cite{dattorro2}: 
\begin{equation}
  \mathbb{EDM}
  = \mathbb{S}_h\cap(\mathbb{S}_c^{\bot}-\mathbb{S}_+), \label{eq:dattorro}
\end{equation}
where $\mathbb{EDM}$ is the set of $n\times n$ Euclidean distance
matrices, $\mathbb{S}$ is the set of $n\times n$ symmetric matrices,
$\mathbb{S}_h$ is the projection of $\mathbb{S}$ on the subspace of
matrices having zero diagonal, $\mathbb{S}_c$ is the kernel of the
matrix map $Y\to Y{\bf 1}$ (with ${\bf 1}$ the all-one $n$-vector),
$\mathbb{S}_c^{\bot}$ is the orthogonal complement of $\mathbb{S}_c$,
and $\mathbb{S}_+$ is the set of symmetric positive semidefinite
$n\times n$ matrices. The matrix representation in \eqref{eq:dattorro}
was exploited in the Alternating Projection Algorithm (APA) discussed
in Sect.~\ref{s:apa}.

\subsection{Matrix completion problems}
\label{s:mcp}
Given an appropriate property $P$ applicable to square matrices, the
Matrix Completion Problem (MCP) schema ask whether, given an $n\times
n$ partial matrix $A'$, this can be completed to a matrix $A$ such
that $P(A)$ holds.  MCPs are naturally formulated in terms of graphs:
given a weighted graph $G=(V,E,a')$, with $a':E\to\mathbb{R}$, is
there a complete graph $K$ on $V$ (possibly with loops) with an edge
weight function $a$ such that $a_{uv}=a'_{uv}$ for all $(u,v)\in E$?

MCPs are an interesting class of inverse problems which find
applications in the analysis of data, such as for example the
reconstruction of 3D images from several 2D projections on random
planes in cryo-electron microscopy \cite{singer2}. When $P(A)$ is the
(informal) statement ``$A$ has low rank'', there is an interesting
application is to recommender systems: voters submit rankings for a
few items, and consistent rankings for all items are required. Since
few factors are believed to impact user's preferences, the data matrix
is expected to have low rank \cite{singer1}.

Two celebrated specializations of this problem schema are the
Euclidean Distance MCP (EDMCP) and the Positive Semidefinite MCP
(PSDMCP). These two problems have a strong link by virtue of
Thm.~\ref{thm:psd}, and, in fact, there is a bijection between EDMCP
and PSDMCP instances \cite{laurent97}. MCP variants where $a'_{ij}$ is
an interval and the condition (i) is replaced by $a_{ij}\in a'_{ij}$
also exist (see e.g.~\cite{pardalos-edm}, where a modification of the
EDMCP in this sense is given).

\subsubsection{Positive semidefinite completion}
\label{s:psdmcp}
Laurent \cite{laurent00} remarks that the PSDMCP is an instance of the
Semidefinite Programming (SDP) feasibility problem: given integral
$n\times n$ symmetric matrices $Q_0,\ldots,Q_m$, determine whether
there exist scalars $z_1,\ldots,z_m$ satisfying $Q_0+\sum\limits_{i\le
m} z_iQ_i\succeq 0$. Thus, by Thm.~\ref{thm:psd}, the EDMCP can be
seen as an instance of the SDP feasibility problem too. The complexity
status of this problem is currently unknown, and in particular it is
not even known whether this problem is in {\bf NP}. The same holds for
the PSDMCP, and of hence also for the EDMCP. If one allows
$\varepsilon$-approximate solutions, however, the situation
changes. The following SDP formulation correctly models the PSDMCP:
\begin{equation*}
\left.\begin{array}{rrcl}
\max & \sum_{(i,j)\not\in E} a_{ij} && \\
     & A = (a_{ij}) &\succeq& 0 \\ 
\forall i\in V & a_{ii} &=& a'_{ii} \\
\forall \{i,j\}\in E  & a_{ij} &=& a'_{ij}.
\end{array}
\right\}
\end{equation*}
Accordingly, SDP-based formulations and techniques are common in DG
(see Sect.~\ref{s:sdp}).

Polynomial cases of the PSDMCP are discussed
in \cite{laurent97,laurent00} (and citations therein). These include
chordal graphs, graphs without $K_4$ minors, and graphs without
certain induced subgraphs (e.g.~wheels $W_n$ with $n\ge
5$). Specifically, in \cite{laurent00} it is shown that if a graph $G$
is such that adding $m$ edges makes it chordal, then the PSDMCP is
polynomial on $G$ for fixed $m$. All these results naturally extend to
the EDMCP.

Another interesting question is, aside from actually {\it solving} the
problem, to determine conditions on the given partial matrix to bound
the cardinality of the solution set (specifically, the cases of one or
finitely many solutions are addressed). This question is addressed
in \cite{pardalos-edm}, where explicit bounds on the number of
non-diagonal entries of $A'$ are found in order to ensure uniqueness
or finiteness of the solution set.

\subsubsection{Euclidean distance completion}
\label{s:edmcp}
The EDMCP differs from the DGP in that the dimension $K$ of the
embedding space is not provided as part of the input. An upper bound
to the minimum possible $K$ that is better than the trivial one
($K\le n$) was given in \cite{barvinok} as:
\begin{equation}
  K \le \frac{\sqrt{8|E|+1}-1}{2} \label{eq:barvinok}.
\end{equation}
Because of Thm.~\ref{thm:psd}, the EDMCP inherits many of the
properties of the PSDMCP. We believe that Menger was the first to
explicitly state a case of EDMCP in the
literature: in \cite[p.~121]{menger28} (also
see \cite[p.~738]{menger31}) he refers to the matrices appearing in
Cayley-Menger determinants with one missing entry. These,
incidentally, are also used in the dual Branch-and-Prune (BP)
algorithm (see Sect.~\ref{s:dualbp}).

As mentioned in Sect.~\ref{s:psdmcp}, the EDMCP can be solved in
polynomial time on chordal graphs
$G=(V,E)$ \cite{grone,laurent97}. This is because a graph is chordal
if and only if it has a {\it perfect elimination order}
(PEO) \cite{dirac}, i.e.~a vertex order on $V$ such that, for all
$v\in V$, the set of adjacent successors $N(v)\cap\eta(v)$ is a clique
in $G$. PEOs can be found in $O(|V|+|E|)$ \cite{tarjan2}, and can be
used to construct a sequence of graphs $G=(V,E)=G_0,G_1,\ldots,G_s$
where $G_s$ is a clique on $V$ and $E(G_i)=E(G_{i-1})\cup\{\{u,v\}\}$,
where $u$ is the maximum ranking vertex in the PEO of $G_{i-1}$ such
that there exists $v\in\eta(u)$ with $\{u,v\}\not\in
E(G_{i-1})$. Assigning to $\{u,v\}$ the weight
$d_{uv}=\sqrt{d_{1u}^2+d_{1v}^2}$ guarantees that the weighted
(complete) adjacency matrix of $G_s$ is a distance matrix completion
of the weighted adjacency matrix of $G$, as required
\cite{grone}. This result is introduced in \cite{grone} (for the
PSDMCP rather than the EDMCP) and summarized in \cite{laurent97}.


\section{Molecular Conformation}
\label{s:molecular}
According to the authors' personal interest, this is the largest
section in the present survey. DG is mainly (but not
exclusively \cite{brunak}) used in molecular conformation as a model
of an inverse problem connected to the interpretation of NMR data. We
survey continuous search methods, then focus on discrete search
methods, then discuss the extension to interval distances, and finally
present recent results specific to the NMR application.

\subsection{Test instances} 
\label{s:test}
The methods described in this section have been empirically tested
according to different instance sets and on different computational
testbeds, so a comparison is difficult. In general,
researchers in this area try to provide a ``realistic'' setting; the
most common choices are the following.
\begin{itemize}
\item {\bf Geometrical instances}: instances are generated randomly from a
geometrical model that is also found in nature, such as grids \cite{morewu}.
\item {\bf Random instances}: instances are generated randomly from a
physical model that is close to reality, such as \cite{Lav05,dvnsjogo}.
\item {\bf Dense PDB instances}: real protein conformations are downloaded from
the Protein Data Bank (PDB) \cite{pdb}, and then, for each residue,
all within-residue distances as well as all distances between each
residue and its two neighbours are
generated \cite{morewu2,hoaian,hoaian2};
\item {\bf Sparse PDB instances}: real protein conformations are downloaded from
the Protein Data Bank (PDB) \cite{pdb}, and then all distances within
a given threshold are generated \cite{legrand,dmdgp}.
\end{itemize}
When the target application is the analysis of NMR data, as in the
present case, the best test setting is provided by sparse PDB
instances, as NMR can only measure distances up to a given threshold.
Random instances are only useful when the underlying physical model is
meaningful (as is the case in \cite{Lav05}). Geometrical instance
could be useful in specific cases, e.g.~the analysis of crystals. The
problem with dense PDB instances is that, using the notions given in
Sect.~\ref{s:discr} and the fact that a residue contains more than 3
atoms, it is easy to show that the backbone order on these protein
instances induces a $3$-trilateration order in $\mathbb{R}^3$ (see
Sect.~\ref{s:uniquereal}). Since graphs with such orders can be
realized in polynomial time \cite{eren04}, they do not provide a
particularly hard class of test instances. Moreover, since there are
actually nine backbone atoms in each set of three consecutive
residues, the backbone order is actually a 7-trilateration order. In
other words there is a surplus of distances, and the problem is
overdetermined.

Aside from a few early papers (e.g.~\cite{lln1,lln5,dvnsjogo}) we (the
authors of this survey) always used test sets consisting mostly of
sparse PDB instances. We also occasionally used geometric and (hard)
random instances, but never employed ``easy'' dense PDB instances.

\subsubsection{Test result evaluation}
The test results always yield: a realization $x$ for the given
instance; accuracy measures for $x$, which quantify either how far is
$x$ from being valid, or how far is $x$ from a known optimal solution;
and a CPU time taken by the method to output $x$. Optionally, certain
methods (such as the BP algorithm, see Sect.~\ref{s:bp}) might also
yield a whole set of valid realizations.  Different methods are
usually compared according to their accuracy and speed.

There are three popular accuracy measures. The {\it penalty} is the
evaluation of the function defined in \eqref{eq:mdgpf} for a given
realization $x$. {\it The Largest Distance Error} (LDE) is a scaled,
averaged and square-rooted version of the penalty, given by
$\frac{1}{|E|}\sum_{\{u,v\}\in E} \frac{|\|x_u-x_v\|-d_{uv}|}{d_{uv}}$.
The {\it Root Mean Square Deviation} (RMSD) is a difference measure
for sets of points in Euclidean space having the same center of
mass. Specifically, if $x,y$ are embeddings of $G=(V,E)$, then
$\mbox{RMSD}(x,y)=\min_{T}\|y-Tx\|$, where $T$ varies over all
rotations and translations in $\mathbb{R}^K$.  Accordingly, if $y$ is
the known optimal configuration of a given protein, different
realizations of the same protein yield different RMSD
values. Evidently, RMSD is a meaningful accuracy measure only for test
sets where the optimal conformations are already known (such as PDB
instances).

\subsection{The Molecular Distance Geometry Problem}
\label{s:mdgp}
The MDGP is the same as DGP${}_3$. The name ``molecular'' indicates
that the problem originates from the study of molecular structures.

The relationship between molecules and graphs is probably the deepest
one existing between chemistry and discrete mathematics: a wonderful
account thereof is given in \cite[Ch.~4]{graphhistory}. Molecules were
initially identified by {\it atomic formul{\ae}} (such as H${}_2$O)
which indicate the relative amounts of atoms in each molecule. When
chemists started to realize that some compounds with the same atomic
formula have different physical properties, they sought the answer in
the way the same amounts of atoms were linked to each other through
chemical bonds. Displaying this type of information required more than
an atomic formula, and, accordingly, several ways to represent
molecules using diagrams were independently invented. The one which is
still essentially in use today, consisting in a set of atom symbols
linked by segments, is originally described in \cite{crumbrown}. The
very origin of the word ``graph'' is due to the representation of
molecules \cite{sylvester}.

The function of molecules rests on their chemical composition and 
three-dimensional shape in space (also called {\it structure} or {\it
  conformation}). As mentioned in Sect.~\ref{s:intro}, NMR
experiments can be used to determine a subset of short Euclidean
distances between atoms in a molecule. These, in turn, can be used to
determine its structure, i.e.~the relative positions of atoms in
$\mathbb{R}^3$. The MDGP provides the simplest model for this inverse
problem: $V$ models the set of atoms, $E$ the set of atom pairs for
which a distance is avaiable, and the function $d:E\to\mathbb{R}_+$
assigns distance values to each pair, so that $G=(V,E)$ is the graph
of the molecule. Assuming the input data is correct, the set $X$ of
solutions of the MDGP on $G$ will yield all the structures of the
molecule which are compatible with the observed distances.

In this section we review the existing methods for solving the MDGP
with exact distances on general molecule graphs.

\subsubsection{General-purpose approaches}
\label{s:mdgpgen}
Finding a solution of the set of nonlinear equations \eqref{eq:mdgp}
poses several numerical difficulties. Recent (unpublished) tests
performed by the authors of this survey determined that tiny, randomly
generated weighted graph instances with fewer than 10 vertices could
not be solved using Octave's nonlinear equation solver {\tt
fsolve} \cite{octave}. Spatial Branch-and-Bound (sBB) codes such as
{\sc Couenne} \cite{couenne} could solve instances with
$|V|\in\{2,3,4\}$ but no larger in reasonable CPU times: attaining
feasibility of local iterates with respect to the nonlinear manifold
defined by \eqref{eq:mdgp} is a serious computational challenge. This
motivates the following formulation using Mathematical Programming
(MP):
\begin{equation}
  \min_{x\in\mathbb{R}^K} \sum_{\{u,v\}\in E} (\|x_u-x_v\|^2 - d_{uv}^2)^2.
  \label{eq:mdgpgo}
\end{equation}
The Global Optimization (GO) problem \eqref{eq:mdgpgo} aims to
minimize the squared infeasibility of points in $\mathbb{R}^K$ with
respect to the manifold \eqref{eq:mdgp}. Both terms in the squared
difference are themselves squared in order to decrease floating point
errors ({\tt NaN} occurrences) while evaluating the objective function
of \eqref{eq:mdgpgo} when $\|x_u-x_v\|$ is very close to 0. We remark
that \eqref{eq:mdgpgo} is an unconstrained nonconvex Nonlinear Program
(NLP) whose objective function is a nonnegative polynomial of fourth
degree, with the property that $x\in X$ if and only if the evaluation
of the objective function at $x$ yields 0.

In \cite{lln1}, we tested formulation \eqref{eq:mdgpgo} and some
variants thereof with three GO solvers: a Multi-Level Single Linkage
(MLSL) multi-start method \cite{sobolopt}, a Variable Neighbourhood
Search (VNS) meta-heuristic for nonconvex NLPs \cite{vnssolver}, and
an early implementation of sBB \cite{ooops,liberti,comparison} (the
only solver in the set that guarantees global optimality of the
solution to within a given $\varepsilon>0$ tolerance). We found that
it was possible to solve artificially generated, but realistic protein
instances \cite{Lav05} with up to 30 atoms using the sBB solver,
whereas the two stochastic heuristics could scale up to 50 atoms, with
VNS yielding the best performance.

\subsubsection{Smoothing based methods}
\label{s:smooth}
A {\it smoothing} of a multivariate multimodal function $f(x)$ is a
family of functions $F_\lambda(x)$ such that $F_0(x)=f(x)$ for all
$x\in\mathbb{R}^K$ and $F_\lambda(x)$ has a decreasing number of local
optima as $\lambda$ increases. Eventually $F_\lambda$ becomes convex,
or at least invex \cite{invex}, and its optimum $x^\lambda$ can be
found using a single run of a local NLP solver. A homotopy
continuation algorithm then traces the sequence $x^\lambda$ in reverse
as $\lambda\to 0$, by locally optimizing
$F_{\lambda-\Delta\lambda}(x)$ for a given step $\Delta\lambda$ with
$x^\lambda$ as a starting point, hoping to identify the global optimum
$x^\ast$ of the original function $f(x)$
\cite{smoothing}. A smoothing operator based on the many-dimensional
diffusion equation $\Delta F = \frac{\partial F}{\partial \lambda}$,
where $\Delta$ is the Laplacian $\sum_{i\le n} \partial^2/\partial
x_i^2$, is derived in \cite{smoothing} as the Fourier-Poisson formula
\begin{equation}
  F_\lambda(x)=\frac{1}{\pi^{n/2}\lambda^{n}}
  \int\nolimits_{\mathbb{R}^{n}}f(y)
  e^{-\frac{||y-x||^{2}}{\lambda ^{2}}} dy, \label{gaus} 
\end{equation}
also called {\it Gaussian transform} in \cite{morewu}. The Gaussian
transform with the homotopy method provides a successful methodology
for optimizing the objective function:
\begin{equation}
  f(x) = \sum_{\{u,v\}\in E} (\|x_u-x_v\|^2 - d_{uv}^2)^2, \label{eq:mdgpf}
\end{equation}
where $x\in\mathbb{R}^3$. More information on continuation and
smoothing-based methods applied to the {\it i}\,MDGP can be found in
Sect.~\ref{s:interv}.

\begin{figure}[!ht]
\begin{center}
\includegraphics[width=5cm]{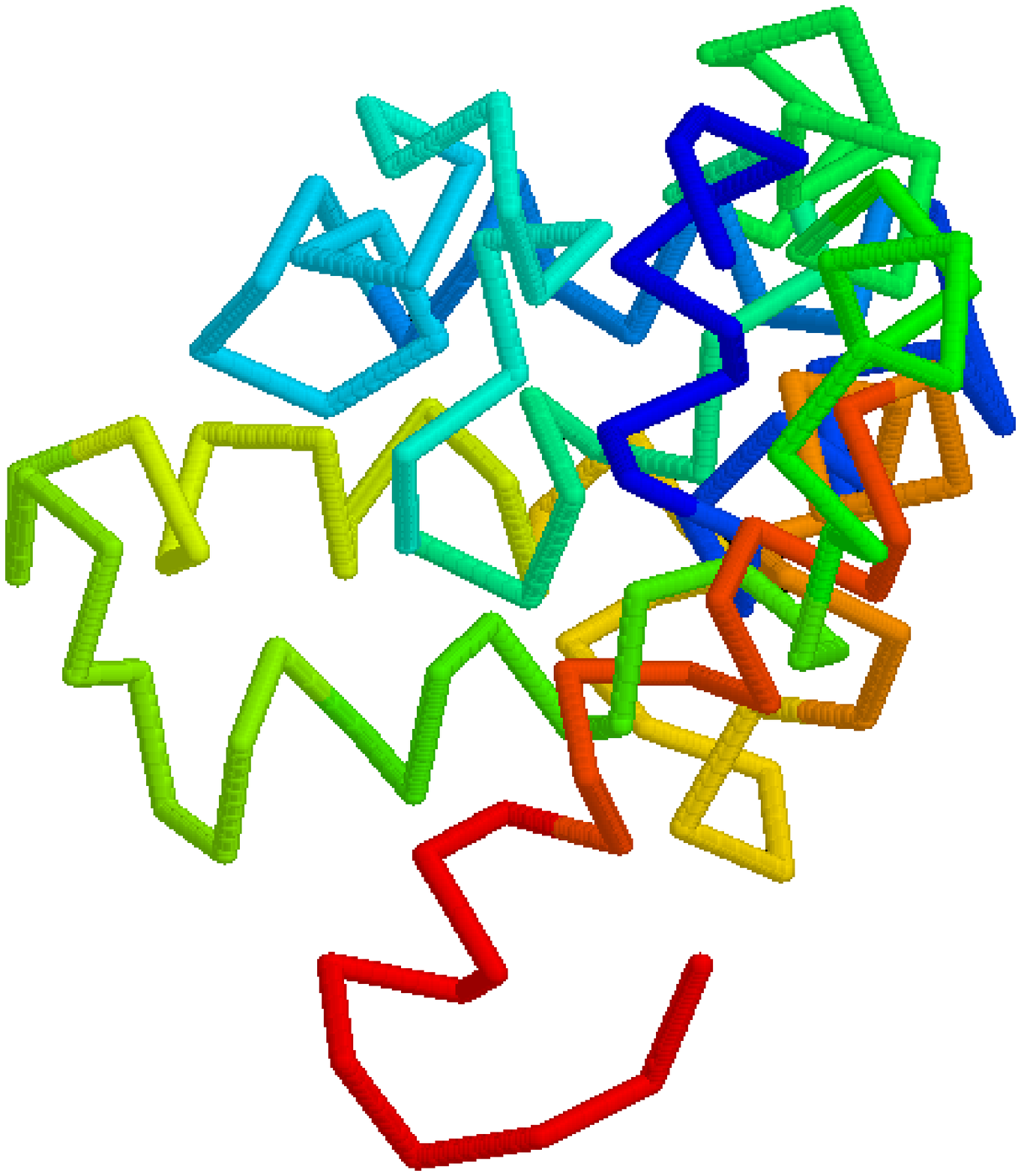}
\hspace*{1cm}
\includegraphics[width=5cm]{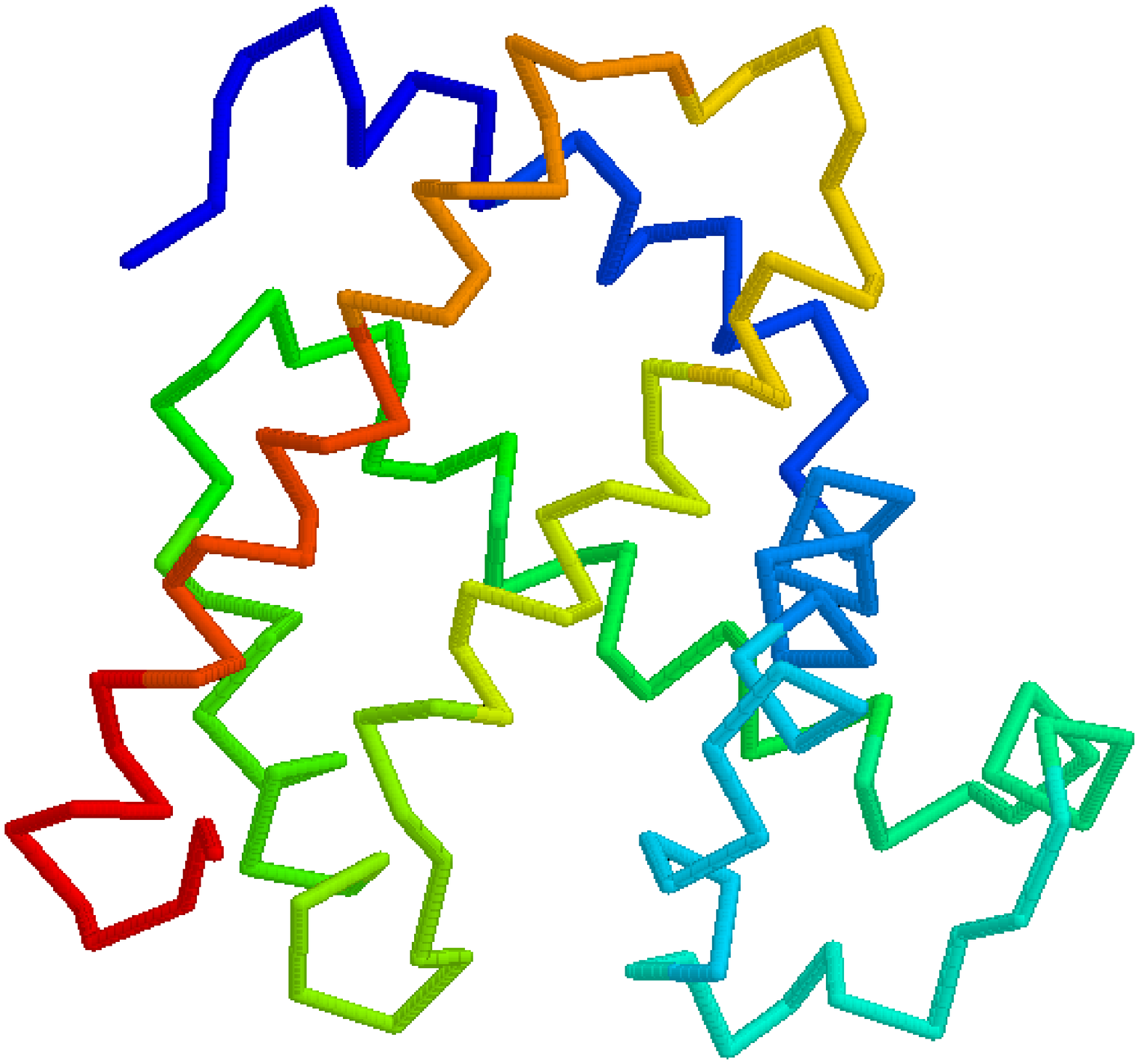}
\end{center}
\caption{Comparison of a wrong molecular conformation for {\tt 1mbn}
found by DGSOL (left) with the correct one found by the BP
Alg.~\ref{alg:bp} (right).}
\label{f:dgsol}
\end{figure}

\label{s:dgsol}
In \cite{morewu}, it is shown that the closed form of the Gaussian
transform applied to \eqref{eq:mdgpf} is:
\begin{equation}
  \langle f\rangle_\lambda=f(x)+10\lambda^2\sum_{\{u,v\}\in E}
  (\|x_u-x_v\|^2-6d_{uv}^2\lambda^2)+ 15\lambda^4|E|.
  \label{eq:mdgpgauss}
\end{equation}
Based on this, a continuation method is proposed and successfully
tested on a set of cubical grids. The implementation of this method,
DGSOL, is one of the few MDGP solution codes that are freely available
(source included):
see \url{http://www.mcs.anl.gov/~more/dgsol/}. DGSOL has several
advantages: it is efficient, effective for small to medium-sized
instances, and, more importantly, can naturally be extended to solve
{\it i}\,MDGP instances (which replace the real edge weights with
intervals). The one disadvantage we found with DGSOL is that it does
not scale well to large-sized instances: although the method is
reasonably fast even on large instances, the solution quality
decreases. On large instances, DGSOL often finds infeasibilities that
denote not just an offset from an optimal solution, but a completely
wrong conformation (see Fig.~\ref{f:dgsol}).

\label{s:dca}
In \cite{hoaian,hoaian2} an exact reformulation of a Gaussian
transform of \eqref{eq:mdgpgo} as a {\it difference of convex} (d.c.)
functions is proposed, and then solved using a method similar to
DGSOL, but where the local NLP solution is carried out by a different
algorithm, called DCA. Although the method does not guarantee global
optimality, there are empirical indications that the DCA works well in
that sense. This method has been tested on three sets of data: the
artificial data from Mor\'{e} and Wu \cite{morewu} (with up to $4096$
atoms), $16$ proteins in the PDB \cite{pdb} (from $146$ up to $4189$
atoms), and the data from Hendrickson \cite{Hen95} (from $63$ up to
$777$ atoms).

\label{s:dvs}
In \cite{dvnsjogo}, VNS and DGSOL were combined into a heuristic
method called {\it Double VNS with Smoothing} (DVS). DVS consists in
running VNS twice: first on a smoothed version $\langle
f\rangle_\lambda$ of the objective function $f(x)$
of \eqref{eq:mdgpgo}, and then on the original function $f(x)$ with
tightened ranges. The rationale behind DVS is that $\langle
f\rangle_\lambda$ is easier to solve, and the homotopy defined by
$\lambda$ should increase the probability that the global optimum
$x^\lambda$ of $\langle f\rangle_\lambda$ is close to the global
optimum $x^\ast$ of $f(x)$. The range tightening that allows VNS to be
more efficient in locating $x^\ast$ is based on a ``Gaussian transform
calculus'' that gives explicit formul{\ae} that relate $\langle
f\rangle_\lambda$ to $f(x)$ whenever $\lambda$ and $d$ change. These
formul{\ae} are then used to identify smaller ranges for $x^\ast$. DVS
is more accurate but slower than DGSOL.

It is worth remarking that both DGSOL and the DCA methods were tested
using (easy) dense PDB instances, whereas the DVS was tested using
geometric and random instances (see Sect.~\ref{s:test}).

\subsubsection{Geometric build-up methods}
\label{s:buildup}
In \cite{dong03}, a combinatorial method called {\it geometric
  build-up} (GB) algorithm is proposed to solve the MDGP on
  sufficiently dense graphs. A subgraph $H$ of $G$, initially chosen
  to only consist of four vertices, is given together with a valid
  realization $\bar{x}$. The algorithm proceeds iteratively by finding
  $x_v$ for each vertex $v\in V(G)\smallsetminus V(H)$. When $x_v$ is
  determined, $v$ and $\delta_H(v)$ are removed from $G$ and added to
  $H$. For this to work, at every iteration two conditions must hold:
\begin{enumerate}
\item $|\delta_H(v)|\ge 4$; \label{buildup1}
\item at least one subgraph $H'$ of $H$, with
  $V(H')=\{u_1,u_2,u_3,u_4\}$ and $|\delta_{H'}(v)|=4$, must be such
  that the realization $\bar{x}$ restricted to $H'$ is
  non-coplanar. \label{buildup2}
\end{enumerate}
These conditions ensure that the position $x_v$ can be determined
using triangulation. More specifically, let
$\bar{x}|_{H'}=\{x_{u_i}\;|\;i\le 4\}\subseteq\mathbb{R}^3$. Then $x_v$ is
a solution of the following system:
\begin{eqnarray*}
||x_v-x_{u_1}|| &=&d_{vu_1}, \\
||x_v-x_{u_2}|| &=&d_{vu_2}, \\
||x_v-x_{u_3}|| &=&d_{vu_3}, \\
||x_v-x_{u_4}|| &=&d_{vu_4}.
\end{eqnarray*}
Squaring both sides of these equations, we have:
\begin{eqnarray*}
||x_v||^{2}-2\transpose{x_v}x_{u_1}+||x_{u_1}||^{2} &=&d_{vu_1}^{2}, \\
||x_v||^{2}-2\transpose{x_v}x_{u_2}+||x_{u_2}||^{2} &=&d_{vu_2}^{2}, \\
||x_v||^{2}-2\transpose{x_v}x_{u_3}+||x_{u_3}||^{2} &=&d_{vu_3}^{2}, \\
||x_v||^{2}-2\transpose{x_v}x_{u_4}+||x_{u_4}||^{2} &=&d_{vu_4}^{2}.
\end{eqnarray*}
By subtracting one of the above equations from the others, one obtains
a linear system that can be used to determine $x_v$.  For example,
subtracting the first equation from the others, we obtain
\begin{equation}
Ax=b,  \label{system}
\end{equation}
where
\[
A=-2\left(
\begin{array}{c}
\transpose{\left( x_{u_1}-x_{u_2}\right)} \\
\transpose{\left( x_{u_1}-x_{u_3}\right)} \\
\transpose{\left( x_{u_1}-x_{u_4}\right)}
\end{array}
\right)
\]
and
\[
b=\left(
\begin{array}{c}
\left(d_{vu_1}^{2}-d_{vu_2}^{2}\right)-
  \left(||x_{u_1}||^{2}-||x_{u_2}||^{2}\right)  \\
\left(d_{vu_1}^{2}-d_{vu_3}^{2}\right)-
  \left(||x_{u_1}||^{2}-||x_{u_3}||^{2}\right)  \\
\left(d_{vu_1}^{2}-d_{vu_4}^{2}\right)-
  \left(||x_{u_1}||^{2}-||x_{u_4}||^{2}\right)
\end{array}%
\right).
\]%
Since $x_{u_1},x_{u_2},x_{u_3},x_{u_4}$ are non-coplanar,
\eqref{system} has a unique solution.

The GB is very sensitive to numerical errors \cite{dong03}. In
\cite{wu07}, Wu and Wu propose an updated GB algorithm
where the accumulated errors can be controlled. Their algorithm was
tested on a set of sparse PDB instances consisting of $10$ proteins
with $404$ up to $4201$ atoms. The results yielded RMSD measures
ranging from $O(10^{-8})$ to $O(10^{-13})$. It is interesting to
remark that if $G$ is a complete graph and $d_{uv}\in\mathbb{Q}_+$ for
all $\{u,v\}\in E$, this approach solves the MDGP in linear time
$O(n)$ \cite{dongwu}. A more complete treatment of MDGP instances
satisfying the $K$-dimensional generalization of conditions
\ref{buildup1}-\ref{buildup2} above is given in \cite{eren04,eren06} in
the framework of the WSNL and $K$-TRILAT problems.

An extension of the GB that is able to deal with sparser graphs (more
precisely, $\delta_H(v)\ge 3$) is given in \cite{protti}; another
extension along the same lines is given in \cite{wuwuyuan}. We remark
that the set of graphs such that $\delta_H(v)\ge 3$ and the
condition \ref{buildup2}.~above hold are precisely the instances of
the DDGP such that $K=3$ (see Sect.~\ref{s:ddgp}): this problem is
discussed extensively in \cite{ddgp}. The main conceptual difference
between these GB extensions and the Branch-and-Prune (BP) algorithm
for the DDGP \cite{ddgp} (see Sect.~\ref{s:discr} below) is that BP
exploits a given order on $V$ (see Sect.~\ref{s:prelim:orders}). Since
the GB extensions do not make use of this order, they are heuristic
algorithms: if $\delta_H(v)<3$ at iteration $v$, then the GB stops,
but there is no guarantee that a different choice of ``next vertex''
might not have carried the GB to termination. A
very recent review on methods based on the GB approach and on the
formulation of other DGPs with inexact distances is given in
\cite{vollerbook}. The BP algorithm (Alg.~\ref{alg:bp}) marks a
striking difference insofar as the knowledge of the order guarantees
the exactness of the algorithm.

\subsubsection{Graph decomposition methods}
\label{s:abbie}
Graph decomposition methods are mixed-combinatorial algorithms based
on graph decomposition: the input graph $G=(V,E)$ is partitioned or
covered by subgraphs $H$, each of which is realized independently (the
local phase). Finally, the realizations of the subgraphs are
``stitched together'' using mathematical programming techniques (the
global phase). The global phase is equivalent to applying MDGP
techniques to the minor $G'$ of $G$ obtained by contracting each
subgraph $H$ to a single vertex. The nice feature of these methods is
that the local phase is amenable to efficient yet exact solutions. For
example, if $H$ is uniquely realizable, then it is likely to be
realizable in polynomial time. More precisely, a graph $H$ is {\it
uniquely realizable} if it has exactly one valid realization in
$\mathbb{R}^K$ modulo rotations and translations, see
Sect.~\ref{s:uniquereal}. A graph $H$ is {\it uniquely localizable} if
it is uniquely realizable and there is no $K'>K$ such that $H$ also
has a valid realization affinely spanning $\mathbb{R}^{K'}$. It was
shown in \cite{ye} that uniquely localizable graphs are realizable in
polynomial time (see Sect.~\ref{s:sdp}). On the other hand, no graph
decomposition algorithm currently makes a claim to overall exactness:
in order to make them practically useful, several heuristic steps must
also be employed.

In ABBIE \cite{Hen95}, both local and global phases are solved using
local NLP solution techniques. Once a realization for all subgraphs
$H$ is known, the coordinates of the vertex set $V_H$ of $H$ can be
expressed relatively to the coordinates of a single vertex in $V_H$;
this corresponds to a starting point for the realization of the minor
$G'$.  ABBIE was the first graph decomposition algorithm for the DGP,
and was able to realize sparse PDB instances with up to 124 amino
acids, a considerable feat in 1995.

In DISCO \cite{Leung}, $V$ is covered by appropriately-sized subgraphs
sharing at least $K$ vertices. The local phase is solved using an SDP
formulation similar to the one given in \cite{biswas2004}. The local
phase is solved using the positions of common vertices: these are
aligned, and the corresponding subgraph is then rotated, reflected and
translated accordingly.

In \cite{mdgpsdp}, $G$ is covered by appropriate subgraphs $H$ which
are determined using a swap-based heuristic from an initial
covering. Both local and global phases are solved using the SDP
formulation in \cite{biswas2004}. A version of this algorithm
targeting the WSNL (see Sect.~\ref{s:wsnl}) was proposed
in \cite{wsnlsdp}: the difference is that, since the positions of some
vertices is known {\it a priori}, the subgraphs $H$ are clusters
formed around these vertices (see Sect.~\ref{s:sdp}).

In \cite{krislocksiam}, the subgraphs include one or more
$(K+1)$-cliques. The local phase is very efficient, as cliques can be
realized in linear time \cite{sippl,dongwu}. The global phase is
solved using an SDP formulation proposed in \cite{wolkowicz} (also see
Sect.~\ref{s:sdp}).

A very recent method called 3D-ASAP \cite{singer3}, designed to be
scalable, distributable and robust with respect to data noise, employs
either a weak form of unique localizability (for exact distances) or
spectral graph partitioning (for noisy distance data) to identify
clusters. The local phase is solved using either local NLP or SDP
based techniques (whose solutions are refined using appropriate
heuristics), whilst the global phase reduces to a 3D synchronization
problem, i.e.~finding rotations in the special orthogonal group
$SO(3,\mathbb{R})$, reflections in $\mathbb{Z}_2$ and translations in
$\mathbb{R}^3$ such that two similar distance spaces have the best
possible alignment in $\mathbb{R}^3$. This is addressed using a 3D
extension of a spectral technique introduced in \cite{singer4}. A
somewhat simpler version of the same algorithm tailored for the case
$K=2$ (with the WSNL as motivating application, see
Sect.~\ref{s:wsnl}) is discussed in \cite{singer5}.

\subsection{Discretizability}
\label{s:discr}
Some DGP instances can be solved using mixed-combinatorial algorithms
such as GB-based (Sect.~\ref{s:buildup}) and graph decomposition based
(Sect.~\ref{s:abbie}) methods. Combinatorial methods offer several
advantages with respect to continuous ones, for example accuracy and
efficiency. In this section, we shall give an in-depth view of
discretizability of the DGP, and discuss at length an exact
combinatorial algorithm for finding {\it all solutions} to those DGP
instances which can be discretized.

We let $X$ be the set of all valid realizations in $\mathbb{R}^K$ of a
given weighted graph $G=(V,E,d)$ modulo rotations and translations
(i.e.~if $x\in X$ then no other valid realization $y$ for which there
exists a rotation or translation operator $T$ with $y=Tx$ is in
$X$). We remark that we allow reflections for technical reasons: much
of the theory of discretizability is based on partial reflections, and
since any reflection is also a partial (improper) reflection,
disallowing reflections would complicate notation later on. In
practice, the DGP system \eqref{eq:mdgp} can be reduced modulo
translations by fixing a vertex $v_1$ to $x_{v_1}=(0,\ldots,0)$ and
modulo rotations by fixing an appropriate set of components out of the
realizations of the other $K-1$ vertices $\{v_2,\ldots,v_K\}$ to
values which are consistent with the distances in the subgraph of $G$
induced by $\{v_i\;|\;1\le i\le K\}$.

Assuming $X\not=\varnothing$, every $x\in X$ is a solution of the
polynomial system:
\begin{equation}
  \forall\{u,v\}\in E \quad \|x_u-x_v\|^2 = d_{uv}^2, \label{eq:mdgp2}
\end{equation}
and as such it has either finite or uncountable cardinality (this
follows from a fundamental result on the structure of semi-algebraic
sets \cite[Thm.~2.2.1]{benedetti}, also see \cite{mishra}). This
feature is strongly related to graph rigidity (see
Sect.~\ref{s:prelim:rigid}, \ref{s:rigid_generic}): specifically,
$|X|$ is finite for a rigid graph, and almost all non-rigid graphs
yield uncountable cardinalities for $X$ whenever $X$ is non-empty. If
we know that $G$ is rigid, then $|X|$ is finite, and {\it a
posteriori}, we only need to look for a finite number of realizations
in $\mathbb{R}^K$: a combinatorial search is better suited than a
continuous one.

\begin{figure}[!ht]
\begin{center}
\psfrag{1}{$1$}
\psfrag{2}{$2$}
\psfrag{3}{$3$}
\psfrag{4}{$4$}
\includegraphics[width=3.5cm]{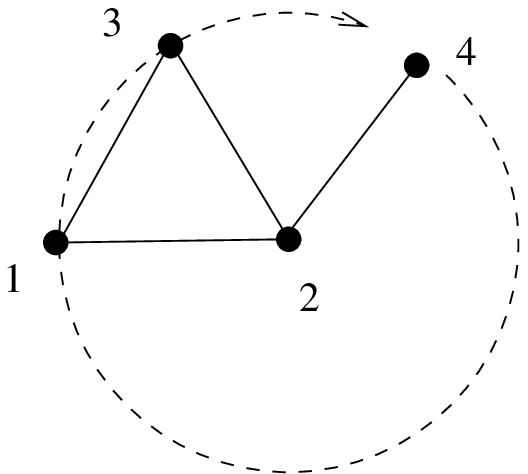}
\hfill
\psfrag{1}{$1$}
\psfrag{2}{$2$}
\psfrag{3}{$3$}
\psfrag{4}{$4$}
\psfrag{4'}{$4'$}
\includegraphics[width=3.5cm]{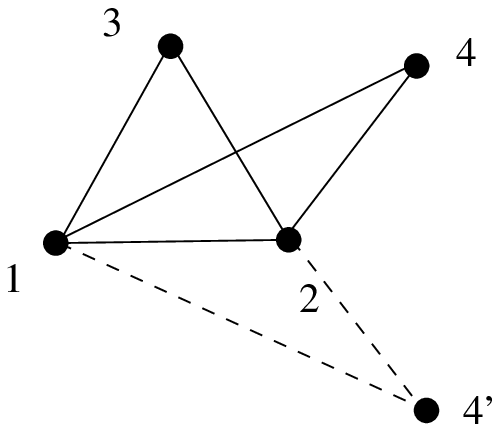}
\hfill
\psfrag{1}{$1$}
\psfrag{2}{$2$}
\psfrag{3}{$3$}
\psfrag{4}{$4$}
\includegraphics[width=3.5cm]{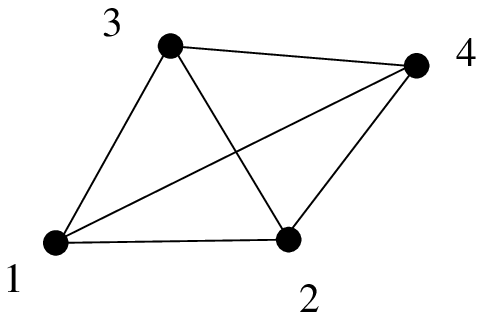}
\end{center}
\caption{A flexible framework (left), a rigid graph
  (center), and a uniquely localizable (rigid) graph (right).}
\label{f:4vtx}
\end{figure}

When $K=2$, it is instructive to inspect a graphical representation of
the situation (Fig.~\ref{f:4vtx}).
The framework for the graph
$(\{1,2,3,4\},\{\{1,2\},\{1,3\},\{2,3\},\{2,4\}\})$ shown in
Fig.~\ref{f:4vtx} (left) is flexible: any of the uncountably many
positions for vertex 4 (shown by the dashed arrow) yield a valid
realization of the graph. If we add the edge $\{1,4\}$ there are
exactly two positions for vertex 4 (Fig.~\ref{f:4vtx}, center), and if
we also add $\{3,4\}$ there is only one possible position
(Fig.~\ref{f:4vtx}, right). Accordingly, if we can only use one
distance $d_{24}$ to realize $x_4$ in Fig.~\ref{f:4vtx} (left) $X$ is
uncountable, but if we can use $K=2$ distances (Fig.~\ref{f:4vtx},
center) or $K+1=3$ distances (Fig.~\ref{f:4vtx}, right) then $|X|$
becomes finite. The GB algorithm \cite{dong03} and the triangulation
method in \cite{eren04} exploit the situation shown in
Fig.~\ref{f:4vtx} (right); the difference between these two methods is
that the latter exploits a vertex order given {\it a priori} which
ensures that a solution could be found for every realizable graph.

The core of the work that the authors of this survey have been
carrying out (with the help of several colleagues) since 2005 is
focused on the situation shown in Fig.~\ref{f:4vtx} (center): we do
not have {\it one} position to realize the next vertex $v$ in the
given order, but (in almost all cases) {\it two}: $x^0_v,x^1_v$, so
that the graph is rigid but not uniquely so. In order to disregard
translations and rotations, we assume a realization $\bar{x}$ of the
first $K$ vertices is given as part of the input. This means that
there will be two possible positions for $x_{K+1}$, four for
$x_{K+2}$, and so on. All in all, $|X|=2^{n-K}$. The situation becomes
more interesting if we consider additional edges in the graph, which
sometimes make one or both of $x_v^0,x_v^1$ infeasible with respect to
Eq.~\eqref{eq:mdgp}. A natural methodology to exploit this situation
is to follow the binary branching process whenever possible, pruning a
branch $x^\ell_v$ ($\ell\in\{0,1\}$) only when there is an additional
edge $\{u,v\}$ whose associated distance $d_{uv}$ is incompatible with
the position $x^\ell_v$. We call this methodology {\it
Branch-and-Prune} (BP).

Our motivation for studying non-uniquely rigid graphs arises from
protein conformation: realizing the protein backbone in $\mathbb{R}^3$
is possibly the most difficult step to realizing the whole protein
(arranging the side chains can be seen as a subproblem
\cite{santana07,ijbra}). As discussed in the rest of this section,
protein backbones conveniently also supply a natural atomic ordering,
which can be exploited in various ways to produce a vertex order that
will guarantee exactness of the BP. The edges necessary to pruning are
supplied by NMR experiments. A definite advantage of the BP is that it
offers a theoretical guarantee of finding {\it all} realizations in
$X$, instead of just {\it one} as most other methods do.

\subsubsection{Rigid geometry hypothesis and molecular graphs}
\label{s:molgraphs}
Discretizability of the search space turns out to be possible only if
the molecule is rigid in physical space, which fails to be the case in
practice. In order to realistically model the flexing of a molecule in
space, it is necessary to consider the bond-stretching and
bond-bending effects, which increase the number of variables of the
problem and also the computational effort to solve it. However, it is
common in molecular conformational calculations to assume that all
bond lengths and bond angles are fixed at their equilibrium values,
which is known as the {\it rigid-geometry
hypothesis} \cite{Gibson_97}.

It follows that for each pair of atomic bonds, say $\{u,v\},\{v,w\}$,
the covalent bond lengths $d_{uv},d_{vw}$ are known, as well as the
angle between them. With this information, it is possible to compute
the remaining distance $d_{uw}$. Every weighted graph $G$ representing
bonds (and their lengths) in a molecule can therefore be trivially
completed with weighted edges $\{u,w\}$ whenever there is a path with
two edges connecting $u$ and $w$. Such a completion, denoted $G^2$, is
called a {\it molecular graph} \cite{jackson_molecule}.  We remark
that all graphs that the BP can realize are molecular, but not vice
versa.

\subsubsection{Development of the Branch-and-Prune algorithm}
\label{s:conception}
To the best of our knowledge, the first discrete search method for the
MDGP that exploits the intersection of three spheres in $\mathbb{R}^3$
was proposed by three of the co-authors of this survey (CL, LL, NM) in
2005 \cite{lln3}, in the framework of a quantum computing algorithm.
Quite independently, the GB algorithm was extended in
2008 \cite{wuwuyuan} to deal with intersections of three rather than
four spheres. Interestingly, as remarked in Sect.~\ref{s:buildup},
another extension to the same case was proposed by a different
research group in the same year \cite{protti}. By contrast, the idea
of a vertex order used to find realizations iteratively was already
present in early works in statics \cite{saviotti_fr,henneberg1886}
(see Sect.~\ref{s:rigid}) and was first properly formalized in
\cite{henneberg1911} (see Sect.~\ref{s:henneberg}).

The crucial idea of combining the intersection of three spheres with a
vertex ordering which would offer a theoretical guarantee of exactness
occurred in june 2005, when two of the co-authors of this survey (CL,
LL) met during an academic visit to Milan. The first version of the BP
algorithm was conceived, implemented and computationally validated
during the summer of 2005: this work, however, only appeared in 2008
\cite{lln5} due to various editorial mishaps. Between 2005 and 2008 we
kept on working at the theory of the DMDGP; we were able to publish an
arXiv technical report in 2006 \cite{lln2}, which was eventually
completed in 2009 and published online in
2011 \cite{dmdgp}. Remarkably, our own early work on BP and an early
version of \cite{wuwuyuan} were both presented at the International
Symposium on Mathematical Programming (ISMP) in Rio de Janeiro already
in 2006.

Along the years we improved and adapted the original BP \cite{lln5} to
further settings. We precisely defined the DGP subclasses on which it
works, and proved it finds {\it all} realizations in $X$ for these
subclasses \cite{lln2,dmdgpsac,dmdgp,ddgp}. We discussed how to
determine a good vertex order automatically \cite{dvop}. We tested and
fine-tuned the BP to proteins \cite{mdgpctw09}. We compared it with
other methods \cite{gecco09}. We tried to decompose the protein
backbone in order to reduce the size of the BP trees \cite{nuccibook}.
We adapted it to work with intervals instead of exact distances
\cite{iccb09,iwcp10,mdgp-togo10,bpinterval}. We engineered it to work
on distances between atoms of given type (this is an important
restriction of NMR
experiments) \cite{csbw09,wco09,mathbalk,ijcb,jogomdgp}. We
generalized it to arbitrary values of $K$ and developed a theory of
symmetries in protein
backbones \cite{powerof2-tr,powerof2-conf,powerof2}. We exploited
these symmetries in order to immediately reconstruct all solutions
from just one \cite{symmBP,symmBPjbcb}. We showed that the BP is
fixed-parameter tractable on protein-like instances and empirically
appears to be polynomial on
proteins \cite{bp-poly,bppolybook}.
We derived a dual 
BP which works in distance rather than realization
space \cite{dBP}. We put all this together so that it would work on
real NMR data \cite{iBP-conf,muchbook}. We started working on
embedding the side chains \cite{ijbra}. We took some first steps
towards applying BP to more general molecular conformation problems
involving energy minimization \cite{gow12}. We provided an
open-source \cite{mdjeep} implementation and tested some parallel
ones \cite{aiccsa10,pco12}. We wrote a number of other surveys
\cite{lln4,mdgpsurvey,dmdgpejor,mda12}, but none as extensive as the present
one. We also edited a book on the subject of distance
geometry and applications \cite{dgpbook}.

\subsubsection{Sphere intersections and probability}
\label{s:sip}
\begin{figure}[!ht]
\begin{center}
\includegraphics[width=5cm]{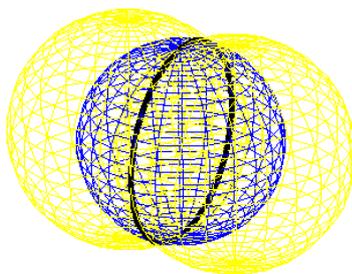}
\end{center}
\caption{When three sphere centers are collinear in 3D, a non-empty
  sphere intersection (the thick circle) has uncountable cardinality.}
\label{f:collinear}
\end{figure}

For a center $c\in\mathbb{R}^K$ and a radius $r\in\mathbb{R}_+$, we
denote by $S^{K-1}(c,r)$ the sphere centered at $c$ with radius $r$ in
$\mathbb{R}^K$. The intersection of $K$ spheres in $\mathbb{R}^K$
might contain zero, one, two or uncountably many points depending on
the position of the centers $x_1,\ldots,x_K$ and the lengths
$d_{1,K+1},\ldots,d_{K,K+1}$ of the radii. Call $P=\bigcap_{i\le K}
S^{K-1}(x_i,d_{i,K+1})$ be the intersection of these $K$ spheres and
$\mathcal{U}^-=\{x_i\;|\;i\le K\}$. If
$\mbox{dim\;aff}(\mathcal{U}^-)<K-1$ then $|P|$ is
uncountable \cite[Lemma 3]{dvop} (see Fig.~\ref{f:collinear}).
Otherwise, if $\mbox{dim\;aff}(\mathcal{U}^-)=K-1$, then
$|P|\in\{0,1,2\}$ \cite[Lemmata 1-2]{dvop}.
We also remark that the condition $\mbox{dim\;aff}(\mathcal{U}^-)<K-1$
corresponds to requiring that $\CM{\mathcal{U}^-}=0$.
See \cite{petitjeanbook} for a detailed treatment of sphere
intersections in molecular modelling.

\begin{figure}[!ht]
\begin{center}
\includegraphics[width=5cm]{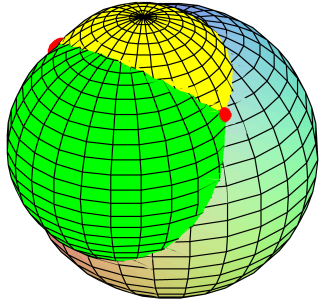}
\end{center}
\caption{General case for the intersection $P$ of three spheres in
  $\mathbb{R}^3$.}
\label{f:sip}
\end{figure}

Now assume $\mbox{dim\;aff}(\mathcal{U}^-)=K-1$, let $x_{K+1}$ be a
given point in $P$ and let
$\mathcal{U}=\mathcal{U}^-\cup\{x_{K+1}\}$. The inequalities
$\Delta_K(\mathcal{U})\ge 0$ (see Eq.~\eqref{eq:deltadef}) are called
{\it simplex inequalities} (or {\it strict} simplex inequalities if
$\Delta_K(\mathcal{U})>0$). We remark that, by definition of the
Cayley-Menger determinant, the simplex inequalities are expressed in
terms of the squared values $d_{uv}$ of the distance function, rather
than the points in $\mathcal{U}$. Accordingly, given a weighted clique
${\bf K}=(U,E,d)$ where $|U|=K+1$, we can also denote the simplex
inequalities as $\Delta_K(U,d)\ge 0$. If the simplex inequalities fail
to hold, then the clique cannot be realized in $\mathbb{R}^K$, and
$P=\varnothing$. If $\Delta_K(U,d)=0$ the simplex has zero volume,
which implies that $|P|=1$ by \cite[Lemma 1]{dvop}. If the strict simplex
inequalities hold, then $|P|=2$ by \cite[Lemma 2]{dvop} (see
Fig.~\ref{f:sip}).
In summary, if $\CM{\mathcal{U}^-}=0$ then $P$ is uncountable,
if $\Delta_K(U,d)=0$ then $|P|=1$, and all other cases lead to
$|P|\in\{0,2\}$. 

Considering the usual probability space on $\mathbb{R}^K$ defined by
the Lebesgue measure, the probability of any sampled point belonging
to any given set having Lebesgue measure zero is equal to zero. Since
both $\{x\in\mathbb{R}^{K^2}\;|\;\CM{\mathcal{U}^-}\}$ and
$\{x\in\mathbb{R}^{K^2}\;|\;\Delta_K(U,d)=0\}$ are (strictly) lower
dimensional manifolds in $\mathbb{R}^{K^2}$, they have Lebesgue
measure zero. Thus the probability of having $|P|=1$ or $P$
uncountable for any given $x\in\mathbb{R}^{K^2}$ is zero. Furthermore,
if we assume $P\not=\varnothing$, then $|P|=2$ with probability 1.  We
extend this notion to hold for any given sentence $\mbox{\sf p}(x)$:
the statement ``$\forall x\in Y\;(\mbox{\sf p}(x)\mbox{ with
probability 1})$'' means that the statement $\mbox{\sf p}(x)$ holds
over a subset of $Y$ that has Lebesgue measure 1. Typically, this
occurs whenever $\mbox{\sf p}$ is a geometrical statement about
Euclidean space that fails to hold for strictly lower dimensional
manifolds. These situations, such as collinearity causing an
uncountable $P$ in Fig.~\ref{f:collinear}, are generally described by
equations. Notice that an event can occur with probability 1
conditionally to another event happening with probability 0. For
example, we shall show in Sect.~\ref{s:symm} that the cardinality of
the solution set of YES instances of the \kDMDGP\; is a power of two
with probability 1, even though a \kDMDGP\; instance has probability 0
of being a YES instance, when sampled uniformly in the set of
all \kDMDGP\; instances.

We remark that our notion of ``statement holding with probability 1''
is different from the genericity assumption which is used in early
works in graph rigidity (see Sect.~\ref{s:rigid} and \cite{connelly}):
a finite set $S$ of real values is {\it generic} if the elements of
$S$ are algebraically independent over $\mathbb{Q}$, i.e.~there exists
no rational polynomial whose set of roots is $S$. This requirement is
sufficient but too stringent for our aims; and besides, since most
computer implementations will only employ (a subset of) rational
numbers, it makes the theory completely inapplicable, as is also
remarked in \cite{Hen95}. The notion we propose might be seen as an
extension to Graver's own definition of genericity, which he
appropriately modified to suit the purpose of combinatorial rigidity:
all minors of the complete rigidity matrix must be nontrivial (see
Sect.~\ref{s:rigid_generic} and \cite{graver}).

\subsubsection{The Discretizable Vertex Ordering Problem}
\label{s:dvop}
The theory of sphere intersections, as described in Sect.~\ref{s:sip},
implies that if there exists a vertex order on $V$ such that each
vertex $v$ such that $\rho(v)>K$ has exactly $K$ adjacent
predecessors, then with probability 1 we have $|X|=2^{n-K}$. If there
are at least $K$ adjacent predecessors, $|X|\le 2^{n-K}$ as either or
both positions $x_v^0,x_v^1$ for $v$ might be infeasible with respect
to some distances. In the rest of the paper, to simplify notation we
identify each vertex $v\in V$ with its (unique) rank $\rho(v)$, 
let $V=\{1,\ldots,n\}$, and write, e.g.~$u-v$ to mean
$\rho(u)-\rho(v)$ or $v>K$ to mean $\rho(v)>K$. 

In this section we discuss the problem of identifying an order with
the properties above. Formally, the DVOP asks to find a vertex order
on $V$ such that $G[\{1,\ldots,K\}]$ is a $K$-clique and such that
$\forall v>K\;(|N(v)\cap\gamma(v)|\ge K)$. We ask that the first $K$
vertices should induce a clique in $G$ because this will allow us to
realize the first $K$ vertices uniquely --- it is a requirement of
discretizable DGPs that a realization should be known for the first
$K$ vertices.

The DVOP is {\bf NP}-complete by trivial reduction from $K$-clique. An
exponential time solution algorithm consists in testing each subset of
$K$ vertices: if one is a clique, then try to build an order by
greedily choosing a next vertex with the largest number of adjacent
predecessors, stopping whenever this is smaller than $K$. This yields
an $O(n^{K+3})$ algorithm. If $K$ is a fixed constant, then of course
this becomes a polynomial algorithm, showing that the DVOP with fixed
$K$ is in {\bf P}. Since DGP applications rarely require a variable
$K$, this is a positive result.

The computational results given in \cite{dvop} show that solving the
DVOP as a pre-processing step sometimes allows the solution of a
sparse PDB instance whose backbone order is not a DVOP order. This may
happen if the distance threshold used to generate sparse PDB instances
is set to values that are lower than usual (e.g.~$5.5${\AA} instead of
$6${\AA}).

\subsubsection{The Discretizable Distance Geometry Problem}
\label{s:ddgp}
The input of the DDGP consists of:
\begin{itemize}
\setlength{\parskip}{-0.05cm}
\item a simple weighted undirected graph $G=(V,E,d)$;
\item an integer $K>0$;
\item an order on $V$ such that:
\begin{itemize}
\setlength{\parskip}{-0.05cm}
\item for each $v>K$, the set $N(v)\cap\gamma(v)$ of adjacent
  predecessors has {\it at least} $K$ elements;
\item for each $v>K$, $N(v)\cap\gamma(v)$ contains a subset $U_v$ of
  {\it exactly} $K$ elements such that:
\begin{itemize}
\setlength{\parskip}{-0.05cm}
\item $G[U_v]$ is a $K$-clique in $G$;
\item strict triangular inequalities $\Delta_{K-1}(U_v,d)>0$ hold (see
Eq.~\eqref{eq:deltadef});
\end{itemize}
\end{itemize}
\item a valid realization $\bar{x}$ of the first $K$ vertices.
\end{itemize} 
The DDGP asks to decide whether $\bar{x}$ can be extended to a valid
realization of $G$ \cite{dvop}. The DDGP with fixed $K$ is denoted by
DDGP${}_K$; the DDGP${}_3$ is discussed in \cite{ddgp}.

We remark that any method that computes $x_v$ in function of its
adjacent predecessors is able to employ a current realization of the
vertices in $U_v$ during the computation of $x_v$. As a consequence,
$\Delta_{K-1}(U_v,d)$ is well defined (during the execution of the
algorithm) even though $G[U_v]$ might fail to be a clique in
$G$. Thus, more DGP instances beside those in the DDGP can be solved
with a DDGP method of this kind. To date, we failed to find a way to
describe such instances aprioristically. The DDGP is {\bf NP}-hard
because it contains the DMDGP (see Sect.~\ref{s:dmdgp} below), and
there is a reduction from {\sc Subset-Sum} \cite{gareyjohnson} to the
DMDGP \cite{dmdgp}.

\subsubsection{The Branch-and-Prune algorithm}
\label{s:bp}
\begin{figure}[!ht]
\begin{center}
\includegraphics[width=14cm]{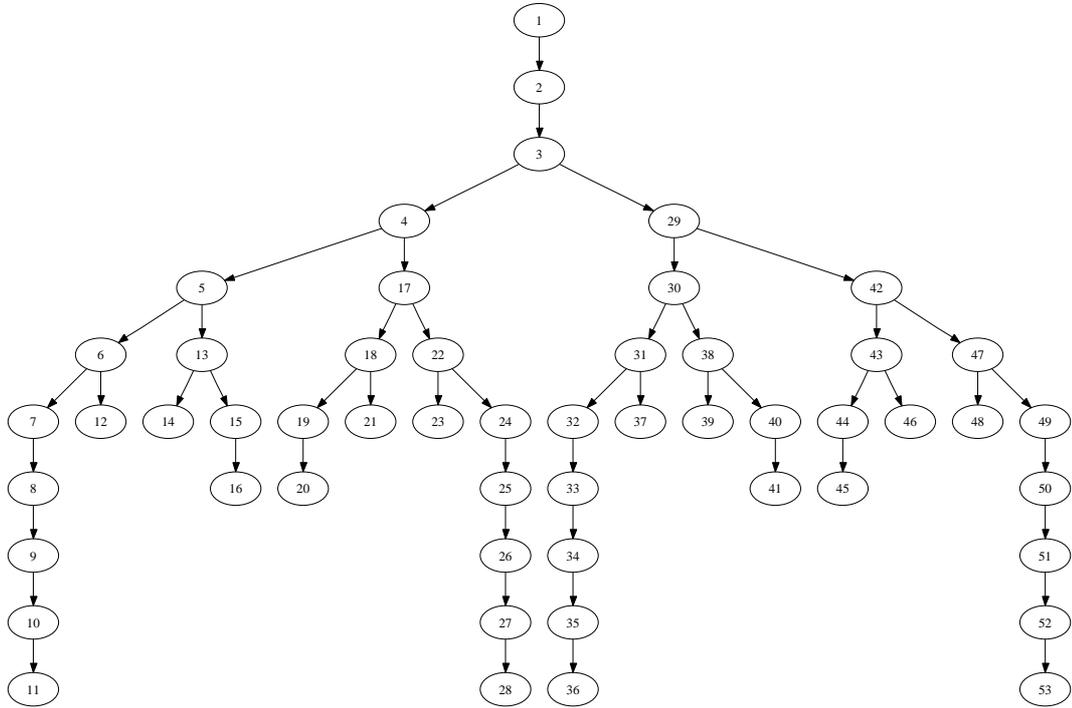}
\end{center}
\caption{An example of BP tree on the random instance {\tt
lavor11\_7} \cite{Lav05}. Pruning edges (see Sect.~\ref{s:pruning})
are as follows: $N(2)=\{9\}$, $N(3)=N(4)=\{8,9,10\}$, $N(5)=\{9,10\}$,
$N(6)=\{10\}$, $N(7)=\{11\}$.}
\label{fig:bptree}
\end{figure}

The recursive step of an algorithm for realizing a vertex $v$ given an
embedding $x'$ for $G[U_v]$, where $U_v$ is as given in
Sect.~\ref{s:ddgp}, is shown in Alg.~\ref{alg:bp}. We recall that
$S^{K-1}(y,r)$ denotes the sphere in $\mathbb{R}^K$ centered at $y$
with radius $r$. By the discretization due to sphere intersections, we
note that $|P|\le 2$.
\begin{algorithm}[!htp]
\begin{algorithmic}[1]
\REQUIRE A vertex~$v\in V\smallsetminus [K]$, an embedding $x'$ for
$G[U_v]$, a set $X$.   
\STATE $P=\bigcap\limits_{u\in N(v)\atop u<v} S^{K-1}(x'_u,d_{uv})$; \label{bp:si}
\FOR{$x_v\in P$}
  \STATE $x=(x',x_v)$
  \IF{$v=n$}
    \STATE $X\leftarrow X\cup\{x\}$
  \ELSE 
    \STATE BP$(v+1, x, X)$ \label{bp:recursion}
  \ENDIF
\ENDFOR
\end{algorithmic}
\caption{{\sc BP}($v$, $\bar{x}$, $X$)} 
\label{alg:bp}
\end{algorithm}
The Branch-and-Prune (BP) algorithm consists in calling
BP$(K+1,\bar{x},\varnothing)$. The BP finds the set $X$ of all valid
realizations of a DDGP instance graph $G=(V,E,d)$ in $\mathbb{R}^K$
modulo rotations and translations \cite{lln5,dmdgp,ddgp}. The
structure of its recursive calls is a binary tree (called the {\it BP
tree}), which contains $2^{n-K}$ nodes in the worst case; this makes
BP a worst-case exponential algorithm. Fig.~\ref{fig:bptree} gives an
example of a BP tree.

Realizations $x\in X$ can also be represented by sequences
$\chi(x)\in\{-1,1\}^n$ such that: (i) $\chi(x)_v=1$ for all $v\le K$;
(ii) for all $v>K$, $\chi(x)_v=-1$ if $ax_v < a_0$ and $\chi(x)_v=1$
if $ax_v\ge a_0$, where $ax=a_0$ is the equation of the hyperplane
through $x(U_v)=\{x_u\;|\; u\in U_v\}$, which is unique with
probability 1. The vector $\chi(x)$ is also known as the {\it
chirality} \cite{CH88} of $x$ (formally, the chirality is defined to
be $\chi(x)_v=0$ if $ax=a_0$, but since this case holds with
probability 0, we disregard it).

The BP (Alg.~\ref{alg:bp}) can be run to termination to find all
possible valid realizations of $G$, or stopped after the first leaf
node at level $n$ is reached, in order to find just one valid
realization of $G$. Compared to most continuous search algorithms we
tested for DGP variants, the performance of the BP algorithm is
impressive from the point of view of both efficiency and reliability,
and, to the best of our knowledge, it is currently the only method
that is able to find all valid realizations of DDGP graphs. The
computational results in \cite{dmdgp}, obtained using sparse PDB
instances as well as hard random instances \cite{Lav05}, show that
graphs with thousands of vertices and edges can be realized on
standard PC hardware from 2007 in fewer than 5 seconds, to an LDE
accuracy of at worst $O(10^{-8})$. Complete sets $X$ of incongruent
realizations were obtained for 25 sparse PDB instances (generation
threshold fixed at $6${\AA}) having sizes ranging from $n=57,m=476$ to
$n=3861,m=35028$. All such sets contain exactly one realization with
RMSD value of at worst $O(10^{-6})$, together with one or more
isomers, all of which have LDE values of at worst $O(10^{-7})$ (and
most often $O(10^{-12})$ or less). The cumulative CPU time taken to
obtain all these solution sets is 5.87s of user CPU time, with one
outlier taking 90\% of the total.

\paragraph{Pruning devices}
\label{s:pruning}
We partition $E$ into the sets $E_D = \{\{u,v\}\in E \;|\; u\in U_v\}$
and $E_P = E\smallsetminus E_D$. We call $E_D$ the {\it discretization
edges} and $E_P$ the {\it pruning edges}. Discretization edges
guarantee that a DGP instance is in the DDGP. Pruning edges are used
to reduce the BP search space by pruning its tree. In practice,
pruning edges might make the set $T$ in Alg.~\ref{alg:bp} have
cardinality 0 or 1 instead of 2, if the distance associated with them
is incompatible with the distances of the discretization edges.

The pruning carried out using pruning edges is called {\it Direct
  Distance Feasibility} (DDF), and is by far the easiest, most
efficient, and most generally useful. Other pruning tests have been
defined. A different pruning technique called {\it Dijkstra Shortest
  Path} (DSP) was considered in \cite[Sect.~4.2]{dmdgp}, based on the
fact that $G$ is a Euclidean network. Specifically, the total weight
of a shortest path from $u$ to $v$ provides an upper bound to the
Euclidean distance between $x_u$ and $x_v$, and can therefore be
employed to prune positions $x_v$ which are too far from $x_u$. The
DSP was found to be effective in some instances but too often very
costly. Other, more effective pruning tests, based on chemical
observations, have been considered in \cite{iBP-conf}.

\subsubsection{Dual Branch-and-Prune}
\label{s:dbp}
There is a close relationship between the DGP${}_K$ and the EDMCP
(see Sect.~\ref{s:edmcp}) with $K$ fixed: each DGP${}_K$ instance $G$
can be transformed in linear time to an EDMCP instance (and vice
versa) by just considering the weighted adjacency matrix of $G$ where
vertex pairs $\{u,v\}\not\in E$ correspond to entries missing from the
matrix. We shall call $\mathscr{M}(G)$ the EDMCP instance
corresponding to $G$ and $\mathscr{G}(A)$ the DGP${}_K$ instance
corresponding to an EDMCP instance $A$.

As remarked in \cite{porta-ikp}, the completion in $\mathbb{R}^3$ of a
distance (sub)matrix $D$ with the following structure:
\begin{equation}
  \left(\begin{array}{ccccc} 
   0 & d_{12} & d_{13} & d_{14} & \mbox{\fbox{$\delta$}} \\
   d_{21} & 0 & d_{23} & d_{24} & d_{25} \\
   d_{31} & d_{32} & 0 & d_{34} & d_{35} \\
   d_{41} & d_{42} & d_{43} & 0 & d_{45} \\
   \mbox{\fbox{$\delta$}} & d_{52} & d_{53} & d_{54} & 0
   \end{array}\right) \label{deltamatrix}
\end{equation}
can be carried out in constant time by solving a quadratic system in
the unknown $\delta$ derived from setting the Cayley-Menger
determinant (Sect.~\ref{s:mathapps}) of the distance space $(X,d)$ to
zero, where $X=\{x_1,\ldots,x_5\}$ and $d$ is given by
Eq.~\eqref{deltamatrix}. This is because the Cayley-Menger determinant
is proportional to the volume of a 4-simplex, which is the (unique, up
to congruences) realization of the weighted 5-clique defined by a full
distance matrix. Since a simplex on 5 points embedded in
$\mathbb{R}^3$ necessarily has 4-volume equal to zero, it suffices to
set the Cayley-Menger determinant of \eqref{deltamatrix} to zero to
obtain a quadratic equation in $\delta$.

We denote the pair $\{u,v\}$ indexing the unknown distance $\delta$ by
$\mbox{\sf e}(D)$, the Cayley-Menger determinant of $D$ by $\CM{D}$,
and the corresponding quadratic equation in $\delta$ by
$\CM{D}(\delta)=0$. If $D$ is a distance matrix, then
$\CM{D}(\delta)=0$ has real solutions; furthermore, in this case it
has two distinct solutions $\delta^1,\delta^2$ with probability 1, as
remarked in Sect.~\ref{s:discr}. These are two valid values for the
missing distance $d_{15}$. This observation extends to general $K$,
where we consider a $(K+1)$-simplex realization of a weighted
near-clique (defined as a clique with a missing edge) on $K+2$
vertices.

\paragraph{BP in distance space}
\label{s:dualbp}
We are given a DDGP instance with a graph $G=(V,E)$ and a partial
embedding $\bar{x}$ for the subgraph $G[[K]]$ of $G$ induced by the
set $[K]$ of the first $K$ vertices. The DDGP order on $V$ guarantees
that the vertex of rank $K+1$ has $K$ adjacent predecessors, hence it
is adjacent to all the vertices of rank $v\in[K]$. Thus, $G[[K+1]]$ is
a full $(K+1)$-clique. Consider now the vertex of rank $K+2$: again,
the DDGP order guarantees that it has at least $K$ adjacent
predecessors. If it has $K+1$, then $G[[K+2]]$ is the full
$(K+2)$-clique. Otherwise $G[[K+2]]$ is a near-clique on $K+2$
vertices with a missing edge $\{u,K+2\}$ for some $u\in[K+1]$. We can
therefore use the Cayley-Menger determinant (see
Eq.~\eqref{deltamatrix} for the special case $K=3$, and
Sect.~\ref{s:mathapps} for the general case) to compute two possible
values for $d_{u,K+2}$. Because the vertex order always guarantees at
least $K$ adjacent predecessors, this procedure can be generalized to
vertices of any rank $v$ in $V\smallsetminus[K]$, and so it defines a
recursive algorithm which:
\begin{itemize}
\item branches whenever a distance can be assigned
two different values;
\item simply continues to the next rank whenever the
subgraph induced by the current $K+2$ vertices is a full clique;
\item prunes all branches whenever the partial distance matrix defined on
the current $K+2$ vertices has no Euclidean completion.
\end{itemize}

\begin{figure}[!ht]
\begin{center}
\includegraphics[width=12cm]{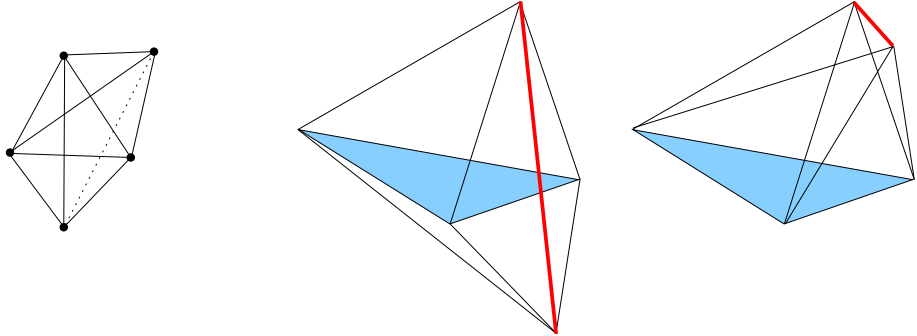}
\end{center}
\caption{On the left, a near clique on 5 vertices with one missing
  edge (dotted line). Center and right, its two possible realizations in
  $\mathbb{R}^3$ (missing distance shown in red).}
\label{fig:cliques}
\end{figure}

In general, this procedure holds for DDGP instances $G$ whenever there
is a vertex order such that each next vertex $v$ is adjacent to $K$
predecessors. This ensures $G$ has a subgraph (containing $v$ and
$K+1$ predecessors) consisting of two $(K+1)$ cliques whose
intersection is a $K$-clique, i.e.~a near-clique with one missing
edge. There are in general two possible realizations in $\mathbb{R}^K$
for such subgraphs, as shown in Fig.~\ref{fig:cliques}.

Alg.~\ref{alg:dbp} presents the dual BP. It takes as input a vertex
$v$ of rank greater than $K+1$, a partial matrix $A$ and a set
$\mathscr{A}$ which will eventually contain all the possible
completions of the partial matrix given as the problem input. For a
given partial matrix $A$, a vertex $v$ of $\mathscr{G}(A)$ and an
integer $\ell\le K$, let $A^{\ell}_{v}$ be the $\ell\times\ell$
symmetric submatrix of $A$ including row and column $v$ that has
fewest missing components. Whenever $A^{K+2}_v$ has no missing
elements, the equation $\CM{A^{K+2}_v,\delta}=0$ is either a tautology
if $A^{K+2}_v$ is a Euclidean distance matrix, or unsatisfiable in
$\mathbb{R}$ otherwise. In the first case, we define it to have
$\delta=d_{uv}$ as a solution, where $u$ is the smallest row/column
index of $A^{K+2}_v$. In the second case, it has no solutions.
\begin{algorithm}[!htp]
\begin{algorithmic}[1]
\REQUIRE A vertex $v\in V\smallsetminus [K+1]$, a partial
matrix $A$, a set $\mathscr{A}$. 
\STATE $P=\{\delta\;|\;\CM{A^{K+2}_{v},\delta}=0\}$ 
  \label{dbp:step0}
\FOR{$\delta\in P$}
  \STATE $\{u,v\}\leftarrow \mbox{\sf e}(A^{K+2}_v)$ \label{dbp:step1}
  \STATE $d_{uv}\leftarrow \delta$ \label{dbp:step2}
  \IF{$A$ is complete}
     \STATE $\mathscr{A}\leftarrow\mathscr{A}\cup\{A\}$
  \ELSE
     \STATE d{\sc BP}($v+1$, $A$, $\mathscr{A}$)
  \ENDIF
\ENDFOR
\end{algorithmic}
\caption{d{\sc BP}($v$, $A$, $\mathscr{A}$)} 
\label{alg:dbp}
\end{algorithm}
\ifsiam\begin{theorem}[\cite{dualbp}]\else\begin{thm}[\cite{dualbp}]\fi
At the end of Alg.~\ref{alg:dbp}, $\mathscr{A}$ contains all
possible completions of the input partial matrix.
\ifsiam\end{theorem}\else\end{thm}\fi

The similarity of Alg.~\ref{alg:bp} and \ref{alg:dbp} is such that it
is very easy to assign dual meanings to the original (otherwise known
as {\it primal}) BP algorithms. This duality stems from the fact that
weighted graphs and partial symmetric matrices are ``dual'' to each
other through the inverse mappings $\mathscr{M}$ and
$\mathscr{G}$. Whereas in the primal BP we decide realizations of the
graph, in the dual BP we decide the completions of partial matrices,
so realizations and distance matrix completions are dual to each
other. The primal BP decides on points $x_v\in\mathbb{R}^K$ to assign
to the next vertex $v$, whereas the dual BP decides on distances
$\delta$ to assign to the next missing distance incident to $v$ and to
a predecessor of $v$; there are at most two choices of $x_v$ as there
are at most two choices for $\delta$; only one choice of $x_v$ is
available whenever $v$ is adjacent to strictly more than $K$
predecessor, and the same happens for $\delta$; finally, no choices
for $x_v$ are available in case the current partial realization cannot
be extended to a full realization of the graph, as well as no choices
for $\delta$ are available in case the current partial matrix cannot
be completed to a Euclidean distance matrix. Thus, point vectors and
distance values are dual to each other. The same vertex order can be
used by both the primal and the dual BP (so the order is self-dual).

There is one clear difference between primal and dual BP: namely, that
the dual BP needs an initial $(K+1)$-clique, whereas the primal BP
only needs an initial $K$-clique. This difference also has a dual
interpretation: a complete Euclidean distance matrix corresponds to
two (rather than one) realizations, one being the reflection of the
other through the hyperplane defined by the first $K$ points (this is
the ``fourth level symmetry'' referred to in \cite[Sect.~2.1]{dmdgp}
for the case $K=3$). We remark that this difference is related to the
reason why the exact SDP-based polynomial method for realizing
uniquely localizable (see Sect.~\ref{s:abbie}) networks proposed
in \cite{ye} needs the presence of at least $K+1$ anchors.

\subsubsection{The Discretizable Molecular Distance Geometry Problem}
\label{s:dmdgp}
The DMDGP is a subset of instances of the DDGP${}_3$; its
generalization to arbitrary $K$ is called \kDMDGP. The difference
between the DMDGP and the DDGP is that $U_v$ is required to be the set
of $K$ {\it immediate} (rather than arbitrary) predecessors of
$v$. So, for example, the discretization edges can also be expressed
as $E_D=\{\{u,v\}\in E\;|\;|u-v|\le K\}$ (see Sect.~\ref{s:pruning}),
and $x(U_v)=\{x_{v-K},\ldots,x_{v-1}\}$. This restriction originates
from the practically interesting case of realizing protein backbones
with NMR data.

\begin{figure}[!ht]
\begin{center}
\psfrag{v}{$v$}
\psfrag{v-1}{$v-1$}
\psfrag{v-2}{$v-2$}
\psfrag{calculated}{\small\it computed}
\includegraphics[width=12cm]{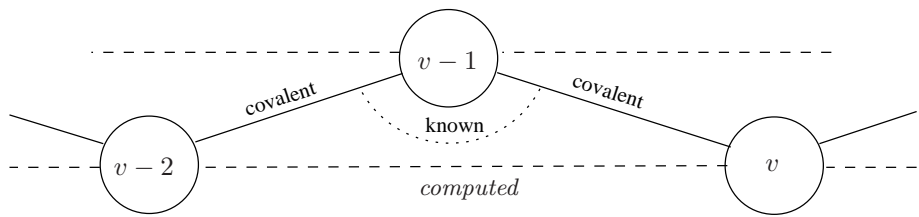}
\end{center}
\caption{Vertex $v$ is adjacent to its two immediate predecessors.}
\label{f:molecular}
\end{figure}

Since such graphs are molecular (see Sect.~\ref{s:molgraphs}), they
have vertex orders guaranteeing that each vertex $v>3$ is adjacent
to {\it two} immediate predecessors, as shown in
Fig.~\ref{f:molecular}.
The distance $d_{v,v-2}$ is computed using the covalent bond lengths
and the angle $(v-2,v-1,v)$, which are known because of the rigid
geometry hypothesis \cite{Gibson_97}. In general, this is only enough
to guarantee discretizability for $K=2$. By exploting further protein
properties, however, we were able to find a vertex order (different
from the natural backbone order) that satisfies the DMDGP definition
(see Sect.~\ref{s:reorders}).

Requiring that all adjacent predecessors of $v$ must be immediate
provides sufficient structure to prove several results about the
symmetry of the solution set $X$ (Sect.~\ref{s:symm}) and about the
fixed-parameter tractabililty of the BP algorithm (Alg.~\ref{alg:bp})
when solving {\kDMDGP}s on protein backbones with NMR data
(Sect.~\ref{s:fpt}). The DMDGP is {\bf NP}-hard by reduction from {\sc
Subset-Sum} \cite{dmdgp}. The result can be generalized to
the \kDMDGP\; \cite{bppolybook}.

\begin{figure}[!ht]
\psfrag{i}{$i-3$}
\psfrag{i+1}{$i-2$}
\psfrag{i+2}{$i-1$}
\psfrag{i+3}{$i$}
\psfrag{ri1}{}
\psfrag{ri2}{}
\psfrag{ri3}{}
\psfrag{dii2}{}
\psfrag{di1i3}{}
\psfrag{ti1}{}
\psfrag{ti2}{}
\psfrag{wi3}{$\phi_i$}
\begin{center}
\includegraphics[width=11cm]{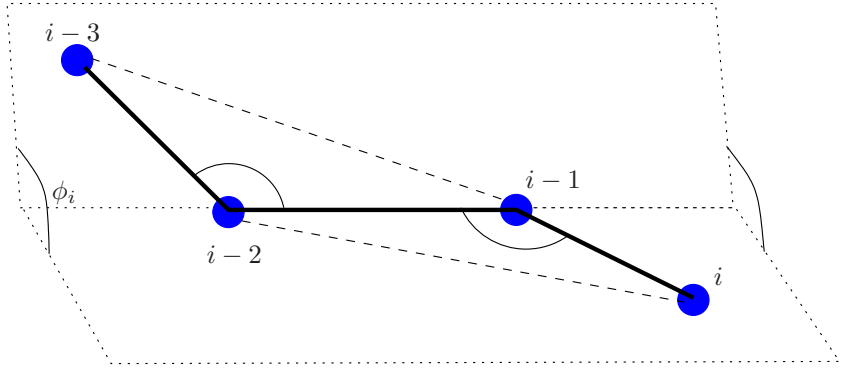}
\end{center}
\caption{The torsion angle $\phi_i$.}
\label{f:tangles}
\end{figure}

\paragraph{Mathematical programming formulation}
\label{s:dmdgpmp}
For completeness, and convenience of mathematical programming versed
readers, we provide here a MP formulation of the DMDGP. We model the
choice between $x_v^0,x_v^1$ by using {\it torsion
angles} \cite{torsionangle}: these are the angles $\phi_v$ defined for
each $v>3$ by the planes passing through $x_{v-3},x_{v-2},x_{v-1}$ and
$x_{v-2},x_{v-1},x_v$ (Fig.~\ref{f:tangles}). More precisely, we
suppose that the cosines $c_v=\cos(\phi_v)$ of such angles are also
part of the input. In fact, the values for
$c:V\smallsetminus\{1,2,3\}\to\mathbb{R}$ can be computed using the
DMDGP structure of the weighted graph in constant time
using \cite[Eq.~(2.15)]{havel}. Conversely, if one is given precise
values for the torsion angle cosines, then every quadruplet
$(x_{v-3},x_{v-2},x_{v-1},x_v)$ must be a rigid framework (for $v>3$).
We let $\alpha:V\smallsetminus\{1,2\}\to\mathbb{R}^3$ be the normal
vector to the plane defined by three consecutive vertices:
\begin{eqnarray*}
  \forall v\ge 3\quad\alpha_v &=& 
    \left|\begin{array}{ccc}
      {\bf i} & {\bf j} & {\bf k} \\
      x_{v-2,1}-x_{v-1,1} & x_{v-2,2}-x_{v-1,2} & x_{v-2,3}-x_{v-1,3} \\
      x_{v,1}-x_{v-1,1} & x_{v,2}-x_{v-1,2} & x_{v,3} - x_{v-1,3}
   \end{array}\right|  \\
  &=& \left(\begin{array}{c}
    (x_{v-2,2}-x_{v-1,2})(x_{v,3}-x_{v-1,3}) - 
      (x_{v-2,3}-x_{v-1,3})(x_{v,2}-x_{v-1,2}) \\
    (x_{v-2,1}-x_{v-1,1})(x_{v,3}-x_{v-1,3}) -
      (x_{v-2,3}-x_{v-1,3})(x_{v,1}-x_{v-1,1}) \\
    (x_{v-2,1}-x_{v-1,1})(x_{v,2}-x_{v-1,2}) -
      (x_{v-2,2}-x_{v-1,2})(x_{v,1}-x_{v-1,1})
      \end{array}\right),
\end{eqnarray*}
so that $\alpha_v$ is expressed a function $\alpha_v(x)$ of $x$ and
represented as a matrix with entries $x_{vk}$. Now, for every $v>3$,
the cosine of the torsion angle $\phi_v$ is proportional to the scalar
product of the normal vectors $\alpha_{v-1}$ and $\alpha_v$:
\begin{equation*}
  \forall v>3\quad \alpha_{v-1}(x)\cdot\alpha_v(x)
  = \|\alpha_{v-1}(x)\|\|\alpha_v(x)\|\cos\phi_v.
\end{equation*}
Thus, the following provides a MP formulation for the DMDGP:
\begin{equation}
  \left. \begin{array}{rrcl}
    \min_x & \sum\limits_{\{u,v\}\in E} (\|x_u-x_v\|^2 - d_{uv}^2)^2  &&\\
    \mbox{s.t.} &   \forall v>3\quad
    \alpha_{v-1}(x)\cdot\alpha_v(x) &=& \|\alpha_{v-1}(x)\|\|\alpha_v(x)\|c_v.
  \end{array} \right\}
  \label{eq:dmdgpmp}
\end{equation}
We remark that generalizations of \eqref{eq:dmdgpmp} to arbitrary
(fixed) $K$ are possible by using Gra{\ss}mann-Pl\"ucker
relations \cite{bokowski} (also see \cite[Ch.~2]{CH88}). 

\subsubsection{Symmetry of the solution set}
\label{s:symm}
When we first experimented with the BP on the DMDGP, we observed that
$|X|$ was always a power of two. An initial conjecture in this
direction was quickly disproved by hand-crafting an instance with 54
solutions derived by the polynomial reduction of the {\sc Subset-Sum}
to the DMDGP used in the {\bf NP}-hardness proof of the
DMDGP \cite{dmdgp}. Notwithstanding, all protein and protein-like
instances we tested yielded $|X|=2^\ell$ for some integer
$\ell$. Years later, we were able to prove that the conjecture holds
on \kDMDGP\; instances with probability 1, and also derived an
infinite (but countable) class of
counterexamples \cite{powerof2-conf}. Aside from explaining our
conjecture arising from empirical evidence, our result is also
important insofar as it provides the core of a theory of partial
reflections for the {\kDMDGP}. References to partial reflections are
occasionally found in the DGP literature \cite{Hen92,ye}, but our
group-theoretical treatment is an extensive addition to the current
body of knowledge.

In this section we give an exposition which is more compact and
hopefully clearer than the one in \cite{powerof2-conf}.  We focus
on \kDMDGP\; and therefore assume that $U_v$ contains the $K$
immediate predecessors of $v$ for each $v>K$. We also assume $G$ is a
YES instance of the \kDMDGP, so that $|P|=2$ with probability 1.

\begin{figure}[!ht]
\begin{center}
\psfrag{nu1}{$\nu_1$}
\psfrag{nu2}{$\nu_2$}
\psfrag{1}{$1$}
\psfrag{2}{$2$}
\psfrag{r}{$5$}
\psfrag{g}{$3$}
\psfrag{b}{$4$}
\psfrag{3}{$\nu_3$}
\psfrag{4}{$\nu_4$}
\psfrag{5}{$\nu_5$}
\psfrag{6}{$\nu_6$}
\psfrag{7}{$\nu_7$}
\psfrag{8}{$\nu_8$}
\psfrag{9}{$\nu_9$}
\psfrag{10}{$\nu_{10}$}
\psfrag{11}{$\nu_{11}$}
\psfrag{12}{$\nu_{12}$}
\psfrag{13}{$\nu_{13}$}
\psfrag{14}{$\nu_{14}$}
\psfrag{15}{$\nu_{15}$}
\psfrag{16}{$\nu_{16}$}
\includegraphics[width=12cm]{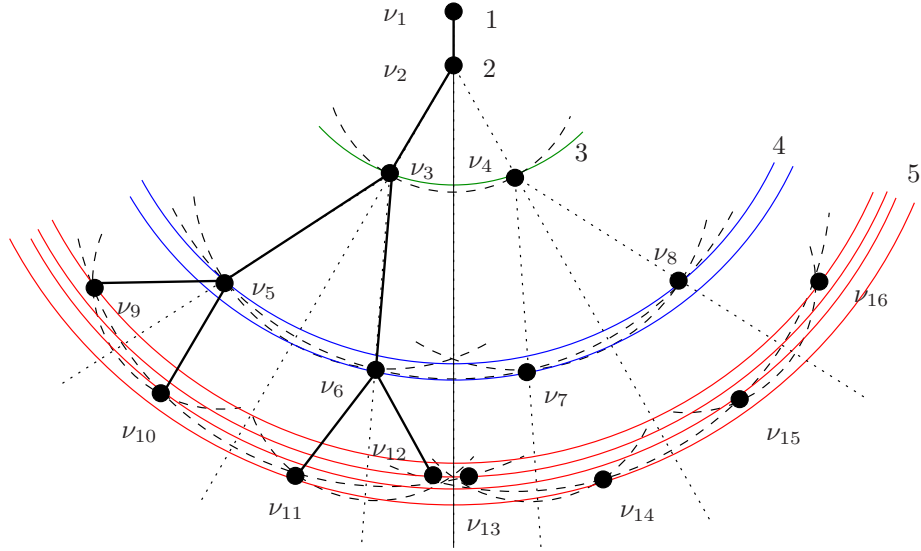}
\end{center}
\caption{A pruning edge $\{1,4\}$ prunes either $\nu_6,\nu_7$ or
  $\nu_5,\nu_8$.}
\label{f:prunedist}
\end{figure}

\paragraph{The discretization group}
Let $G_D=(V,E_D,d)$ be the subgraph of $G$ consisting of the
discretization edges, and $X_D$ be the set of realizations of $G_D$;
since $G_D$ has no pruning edges by definition, the BP search tree for
$G_D$ is a full binary tree and $|X_D|=2^{n-K}$.  The discretization
edges arrange the realizations so that, at level $\ell$, there are
$2^{\ell-K}$ possible positions for the vertex $v$ with rank $\ell$.
We assume that $|P|=2$ (see Alg.~\ref{alg:bp}) at each level $v$ of
the BP tree, an event which, in absence of pruning edges, happens with
probability 1. Let $P=\{x_v^0,x_v^1\}$ be the two possible
realizations of $v$ at a certain recursive call of Alg.~\ref{alg:bp}
at level $v$ of the BP tree; then because $P$ is an intersection of
$K$ spheres, $x_v^1$ is the reflection of $x_v^0$ through the
hyperplane defined by $x(U_v)=\{x_{v-K},\ldots,x_{v-1}\}$. We denote
this reflection operator by $R_x^v$.
\ifsiam
\begin{theorem}[Cor.~4.6 and Thm.~4.9 in \cite{powerof2-conf}]
\else
\begin{thm}[Cor.~4.6 and Thm.~4.9 in \cite{powerof2-conf}]
\fi
\label{pow2lemma}
With probability 1, for all $v>K$ and $u< v-K$ there is a set $H^{uv}$
of $2^{v-u-K}$ real positive values such that for each $x\in X$ we
have $\|x_v-x_u\|\in H^{uv}$. Furthermore, $\forall x'\in X$,
$\|x_v-x_u\|=\|x'_v-x_u\|$ if and only if
$x'_v\in\{x_v,R_x^{u+K}(x_v)\}$.
\ifsiam\end{theorem}\else\end{thm}\fi
We sketch the proof in Fig.~\ref{f:prunedist} for $K=2$; the solid
circles at levels $3,4,5$ mark the locus of feasible realizations for
vertices at rank $3,4,5$ in the \kDMDGP\; order. The dashed circles
represent the spheres $S^x_{uv}$ (see Alg.~\ref{alg:bp}).
Intuitively, two branches from level 1 to level 4 or 5 will have equal
segment lengths but different angles between consecutive segments,
which will cause the end nodes to be at different distances from the
node at level 1. Observe that the number of solid circles at each
level is a power of two where the exponent depends on the level index
$\ell$, and each solid circle contains exactly two realizations (that
are reflections of each other) of the same vertex at rank $\ell$.

We now give a basic result on reflections in $\mathbb{R}^K$. For any
nonzero vector $y\in\mathbb{R}^K$ let $\mathcal{R}(y)$ be the
reflection operator through the hyperplane passing through the origin
and normal to $y$. If $y$ is normal to the hyperplane defined by
$x_{v-K},\ldots,x_{v-1}$, then $\mathcal{R}^y=R_x^v$.
\ifsiam\begin{lemma}[Lemma 4.2 in \cite{bppolybook}]\else\begin{lem}[Lemma 4.2 in \cite{bppolybook}]\fi
Let $x\not=y\in\mathbb{R}^K$ and $z\in\mathbb{R}^K$ such that $z$ is
not in the hyperplanes through the origin and normal to $x,y$. Then
$\mathcal{R}(x)\mathcal{R}(y)z=\mathcal{R}({\mathcal{R}(x)y})\mathcal{R}(x)z$.
\label{conjrefl}
\ifsiam\end{lemma}\else\end{lem}\fi
\begin{figure}[!ht]
\begin{center}
\psfrag{Rx}{$\mathcal{R}(x)$}
\psfrag{Ry}{$\mathcal{R}(y)$}
\psfrag{z}{$z$}
\psfrag{O}{$O$}
\psfrag{x}{$x$}
\psfrag{y}{$y$}
\psfrag{Rxy}{$\mathcal{R}(x)y$}
\psfrag{Ryz}{$\mathcal{R}(y)z$}
\psfrag{Rxz}{$\mathcal{R}(x)z$}
\psfrag{RRxy}{$\mathcal{R}({\mathcal{R}(x)y})$}
\psfrag{RxRyz}{$\mathcal{R}(x)\mathcal{R}(y)z=\mathcal{R}({\mathcal{R}(x)y})\mathcal{R}(x)z$}
\includegraphics[width=7cm]{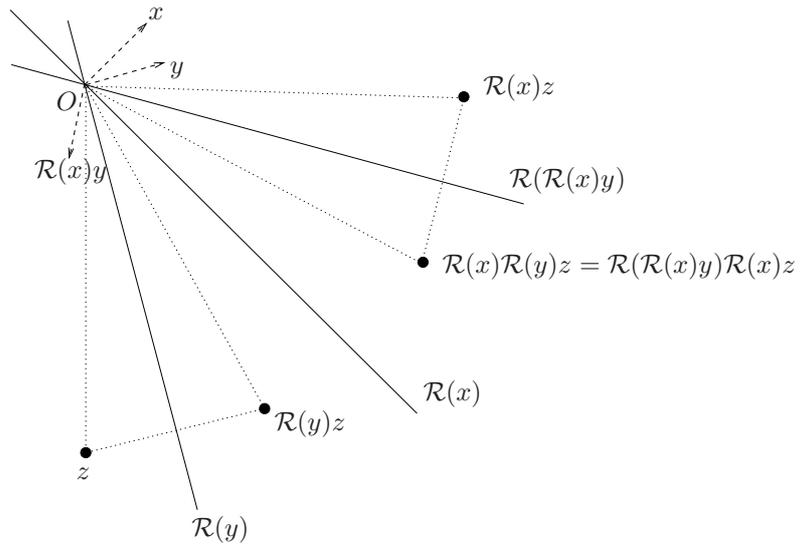}
\caption{Reflecting through $\mathcal{R}(y)$ first and $\mathcal{R}(x)$
  later is equivalent to reflecting through $\mathcal{R}(x)$ first and
  the reflection of $\mathcal{R}(y)$ through $\mathcal{R}(x)$ later.}
\label{f:conjrefl}
\end{center}
\end{figure}
Thm.~\ref{conjrefl} provides a commutativity for reflections acting on
points and hyperplanes. Fig.~\ref{f:conjrefl} illustrates the proof
for $K=2$.

For $v>K$ and $x\in X$ we now define partial reflection operators:
\begin{equation}
  g_v(x)=(x_1,\ldots,x_{v-1},R_x^v(x_v),\ldots,R_x^v(x_n)). \label{prefeq}
\end{equation}
The $g_v$'s map a realization $x$ to its partial reflection with first
branch at $v$. It is easy to show that the $g_v$'s are injective with
probability 1 and idempotent.
\ifsiam\begin{lemma}[Lemma 4.3
in \cite{bppolybook}]\else\begin{lem}[Lemma 4.3 in \cite{bppolybook}]\fi
  For $x\in X$ and $u,v\in V$ such that $u,v>K$,
  $g_ug_v(x)=g_vg_u(x)$.
  \label{lemmcomm}
\ifsiam\end{lemma}\else\end{lem}\fi

We define the {\it discretization group} to be the symmetry group
$\mathcal{G}_D=\langle g_v\;|\;v>K\rangle$ generated by the partial
reflection operators $g_v$.
\ifsiam\begin{corollary}\else\begin{cor}\fi
  With probability 1, $\mathcal{G}_D$ is an Abelian group isomorphic
  to $C_2^{n-K}$ (the Cartesian product consisting of $n-K$ copies of
  the cyclic group of order 2).
\ifsiam\end{corollary}\else\end{cor}\fi
For all $v>K$ let $\gamma_v=(1,\ldots,1,-1_v,\ldots,-1)$ be the vector
consisting of one's in the first $v-1$ components and $-1$ in the last
components. Then the $g_v$ actions are naturally mapped onto the
chirality functions.
\ifsiam\begin{lemma}[Lemma 4.5
in \cite{bppolybook}]\else\begin{lem}[Lemma 4.5
  in \cite{bppolybook}]\fi For all $x\in X$,
  $\chi(g_v(x))=\chi(x)\circ \gamma_v$, where $\circ$ is the Hadamard
  product.
\ifsiam\end{lemma}\else\end{lem}\fi
This follows by definition of $g_v$ and of chirality of a realization.
Since, by Alg.~\ref{alg:bp}, each $x\in X$ has a different chirality,
for all $x,x'\in X$ there is $g\in\mathcal{G}_D$ such that $x'=g(x)$,
i.e.~the action of $\mathcal{G}_D$ on $X$ is transitive. By
Thm.~\ref{pow2lemma}, the distances associated to the discretization
edges are invariant with respect to the discretization group.

\paragraph{The pruning group}
\label{s:pruninggroup}
Consider a pruning edge $\{u,v\}\in E_P$. By Thm.~\ref{pow2lemma},
with probability 1 we have $d_{uv}\in H^{uv}$, otherwise $G$ cannot be
a YES instance (against the initial assumption). Also, again by
Thm.~\ref{pow2lemma}, $d_{uv}=\|x_u-x_v\|\not=\|g_w(x)_u-g_w(x)_v\|$
for all $w\in\{u+K+1,\ldots,v\}$ (e.g.~the distance $\|\nu_1-\nu_9\|$
in Fig.~\ref{f:prunedist} is different from all its reflections
$\|\nu_1-\nu_h\|$, with $h\in\{10,11,12\}$, w.r.t.~$g_4,g_5$).  We
therefore define the {\it pruning group}
\begin{equation*}
  \mathcal{G}_P=\langle g_w\;|\;w>K\land\forall \{u,v\}\in
     E_P\;(w\not\in\{u+K+1,\ldots,v\})\rangle.
\end{equation*}
By definition, $\mathcal{G}_P\leq\mathcal{G}_D$ and the distances
associated with the pruning edges are invariant with respect to
$\mathcal{G}_P$.
\ifsiam\begin{theorem}[Thm.~4.6 in
  \cite{powerof2-conf}]\else\begin{thm}[Thm.~4.6 in
  \cite{powerof2-conf}]\fi  
The action of $\mathcal{G}_P$ on $X$ is transitive with probability
1. \label{transthm}
\ifsiam\end{theorem}\else\end{thm}\fi

\ifsiam\begin{theorem}[Thm.~4.7
  in \cite{bppolybook}]\else\begin{thm}[Thm.~4.7 in \cite{bppolybook}]\fi
With probability 1, $\exists\ell\in\mathbb{N}\;|X|=2^\ell$. \label{pow2thm}
\ifsiam\end{theorem}\else\end{thm}\fi
\begin{proof}
  The argument below holds with probability 1. Since
  $\mathcal{G}_D\cong C_2^{n-K}$, $|\mathcal{G}_D|=2^{n-K}$.  Since
  $\mathcal{G}_P\leq\mathcal{G}_D$, $|\mathcal{G}_P|$ divides the
  order of $|\mathcal{G}_D|$, which implies that there is an integer
  $\ell$ with $|\mathcal{G}_P|=2^\ell$. By Thm.~\ref{transthm}, the
  action of $\mathcal{G}_P$ on $X$ only has one orbit,
  i.e.~$\mathcal{G}_Px=X$ for any $x\in X$. By idempotency, for
  $g,g'\in\mathcal{G}_P$, if $gx=g'x$ then $g=g'$. This implies
  $|\mathcal{G}_Px|=|\mathcal{G}_P|$. Thus, for any $x\in X$,
  $|X|=|\mathcal{G}_Px|=|\mathcal{G}_P|=2^\ell$.
\end{proof}

\paragraph{Practical exploitation of symmetry}
These results naturally find a practical application to speed up the
BP algorithm. The BP proceeds until a first valid realization is
identified. It can be shown that, at that point, a set of generators
for the group $\mathcal{G}_P$ are known. These are used to generate
all other valid realizations of the input graph, up to rotations and
translations \cite{symmBP,symmBPjbcb}. Empirically, this cuts the CPU
time to roughly $2/|X|$ (the factor 2 is due to the fact that the
original BP already takes one reflection symmetry into account,
see \cite[Thm.~2]{dmdgp}).

\subsubsection{Fixed parameter tractability}
\label{s:fpt}
\begin{figure}[!ht]
\begin{center}
\psfrag{2}{1}
\psfrag{4}{2}
\psfrag{8}{4}
\psfrag{16}{8}
\psfrag{32}{16}
\psfrag{64}{32}
\psfrag{v}{$v$}
\psfrag{K+1}{\scriptsize $u\!\!+\!\!K\!\!-\!\!1$}
\psfrag{K+2}{\scriptsize $u\!\!+\!\!K$}
\psfrag{K+3}{\scriptsize $u\!\!+\!\!K\!\!+\!\!1$}
\psfrag{K+4}{\scriptsize $u\!\!+\!\!K\!\!+\!\!2$}
\psfrag{K+5}{\scriptsize $u\!\!+\!\!K\!\!+\!\!3$}
\psfrag{K+6}{\scriptsize $u\!\!+\!\!K\!\!+\!\!4$}
\psfrag{e1}{\scriptsize $0$}
\psfrag{e2}{\scriptsize $1$}
\psfrag{e3}{\scriptsize $2$}
\psfrag{e4}{\scriptsize $3$}
\psfrag{e5}{\scriptsize $4$}
\psfrag{e12}{\scriptsize $0\vee 1$}
\psfrag{e23}{\scriptsize $1\vee 2$}
\psfrag{e34}{\scriptsize $2\vee 3$}
\psfrag{e45}{\scriptsize $3\vee 4$}
\psfrag{e123}{\scriptsize $0\vee 1\vee 2$}
\psfrag{e234}{\scriptsize $1\vee 2\vee 3$}
\psfrag{e345}{\scriptsize $2\!\vee\!3\!\vee\!4$}
\psfrag{e1234}{\scriptsize $0\vee\ldots\vee 3$}
\psfrag{e2345}{\scriptsize $1\vee\ldots\vee 4$}
\psfrag{e12345}{\scriptsize $0\vee\ldots\vee 4$}
\includegraphics[width=15cm]{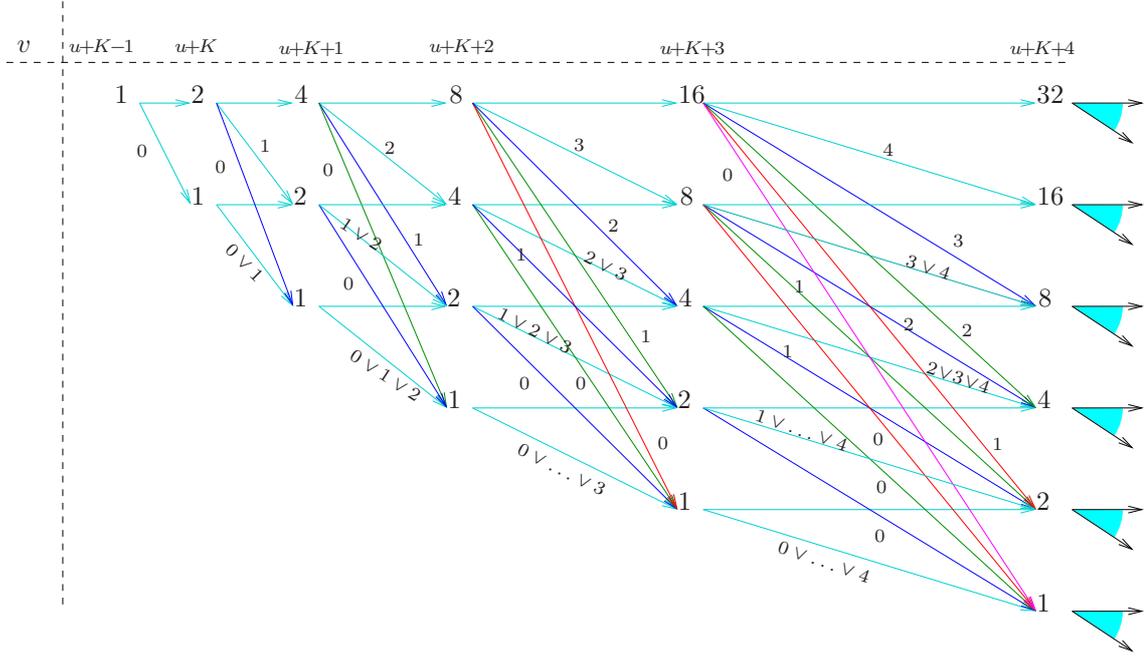}
\end{center}
\caption{Number of valid BP nodes (vertex label) at level $u+K+\ell$
  (column) in function of the pruning edges (path spanning all
  columns).}
\label{f:numsol}
\end{figure}

As the theory of partial reflections, the proof that the BP is
Fixed-Parameter Tractable (FPT) on proteins also stems from empirical
evidence. All the CPU time plots versus instance size for the BP
algorithm on protein backbones look roughly linear, suggesting that
perhaps such instances are a ``polynomial case'' of the DMDGP. The
results that follow provide sufficient conditions for this to be the
case. We were able to verify empirically that PDB proteins conform to
these conditions. These results are a consequence of the theory in
Sect.~\ref{s:symm} insofar as they rely on an exact count of the BP
tree nodes at each level. We formalize this in a DAG
$\mathcal{D}_{uv}$ that represents the number of valid BP search tree
nodes in function of pruning edges between two vertices $u,v\in V$
such that $v>K$ and $u<v-K$ (see Fig.~\ref{f:numsol}).
The first row in Fig.~\ref{f:numsol} shows different values for the
rank of $v$ w.r.t.~$u$; an arc labelled with an integer $i$ implies
the existence of a pruning edge $\{u+i,v\}$ (arcs with
$\vee$-expressions replace parallel arcs with different labels). An
arc is unlabelled if there is no pruning edge $\{w,v\}$ for any
$w\in\{u,\ldots,v-K-1\}$. The vertices of the DAG are arranged
vertically by BP search tree level, and are labelled with the number
of BP nodes at a given level, which is always a power of two by
Thm.~\ref{pow2thm}. A path in this DAG represents the set of pruning
edges between $u$ and $v$, and its incident vertices show the number
of valid nodes at the corresponding levels. For example, following
unlabelled arcs corresponds to no pruning edge between $u$ and $v$ and
leads to a full binary BP search tree with $2^{v-K}$ nodes at level
$v$.

\begin{figure}[!ht]
\begin{center}
\psfrag{2}{\tiny 1}
\psfrag{4}{\tiny 2}
\psfrag{8}{\tiny 4}
\psfrag{16}{\tiny 8}
\psfrag{32}{\tiny 16}
\psfrag{64}{\tiny 32}
\psfrag{v}{}
\psfrag{K+1}{}
\psfrag{K+2}{}
\psfrag{K+3}{}
\psfrag{K+4}{}
\psfrag{K+5}{}
\psfrag{K+6}{}
\psfrag{e1}{}
\psfrag{e2}{}
\psfrag{e3}{}
\psfrag{e4}{}
\psfrag{e5}{}
\psfrag{e12}{}
\psfrag{e23}{}
\psfrag{e34}{}
\psfrag{e45}{}
\psfrag{e123}{}
\psfrag{e234}{}
\psfrag{e345}{}
\psfrag{e1234}{}
\psfrag{e2345}{}
\psfrag{e12345}{}
\includegraphics[width=10cm]{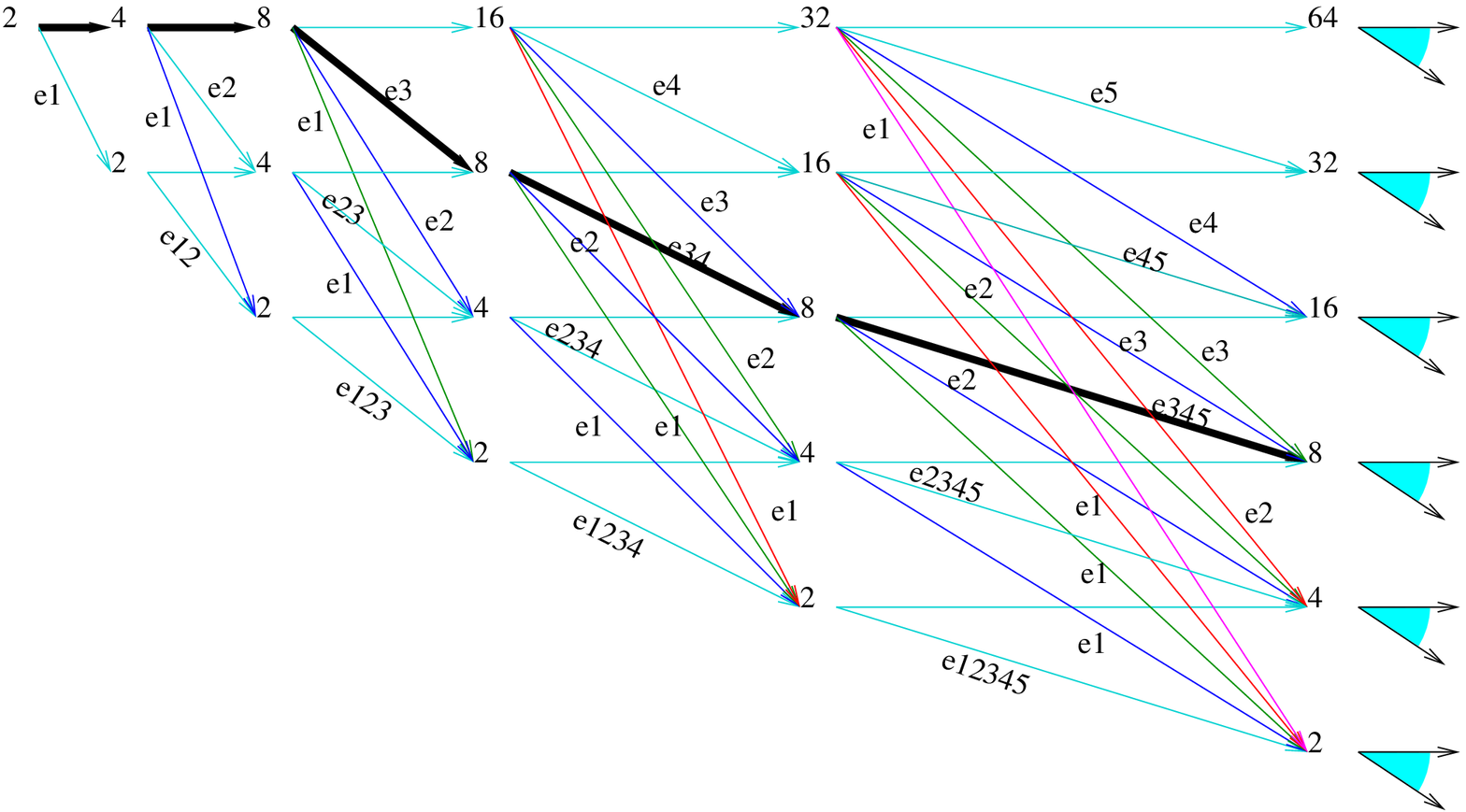} 
\includegraphics[width=10cm]{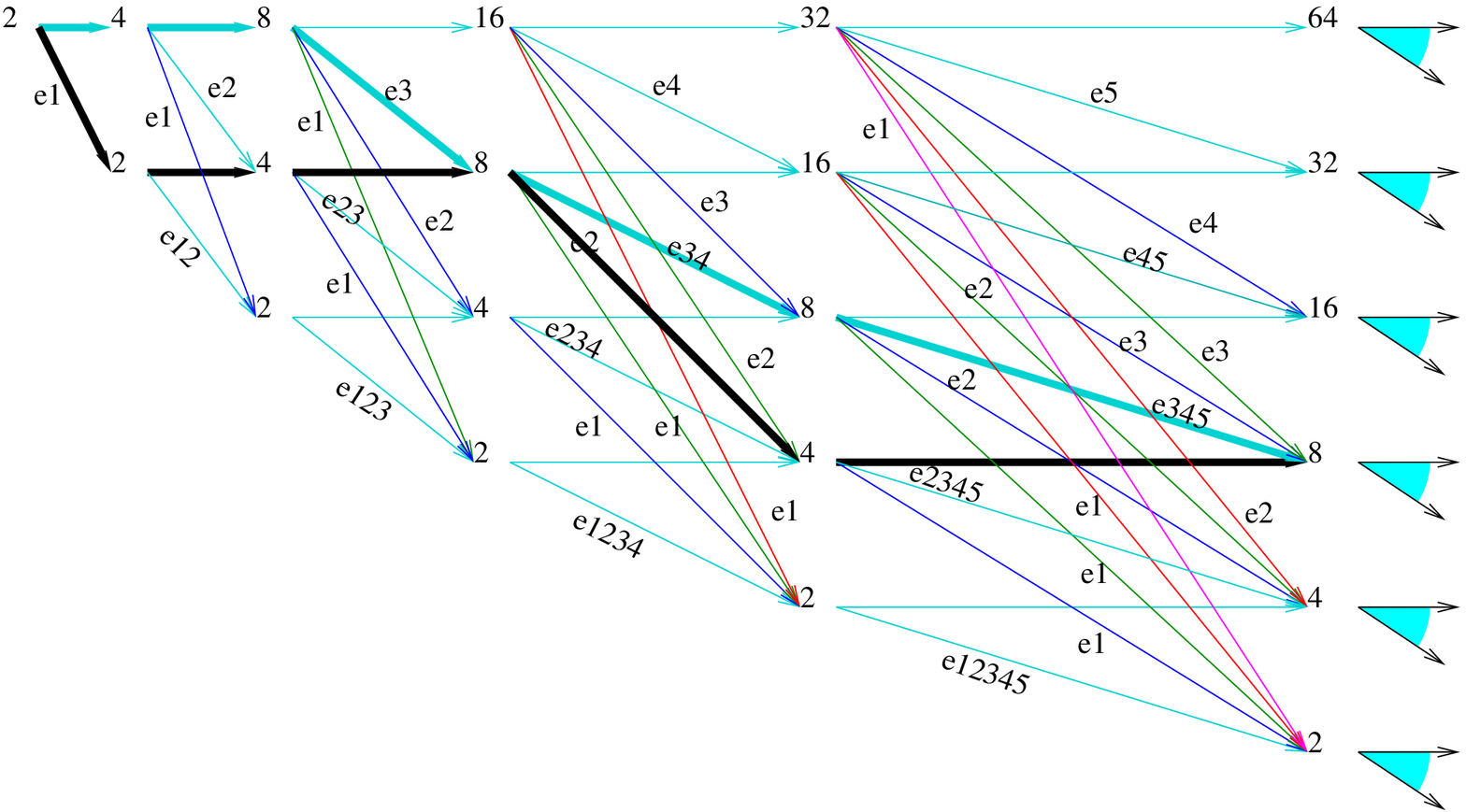}
\end{center}
\caption{A path $\mbox{\sf p}_0$ yielding treewidth $4$ (top) and
  another path below $\mbox{\sf p}_0$ (bottom).}
\label{f:p0}
\end{figure}

\begin{figure}[!ht]
\begin{center}
\includegraphics[width=10cm]{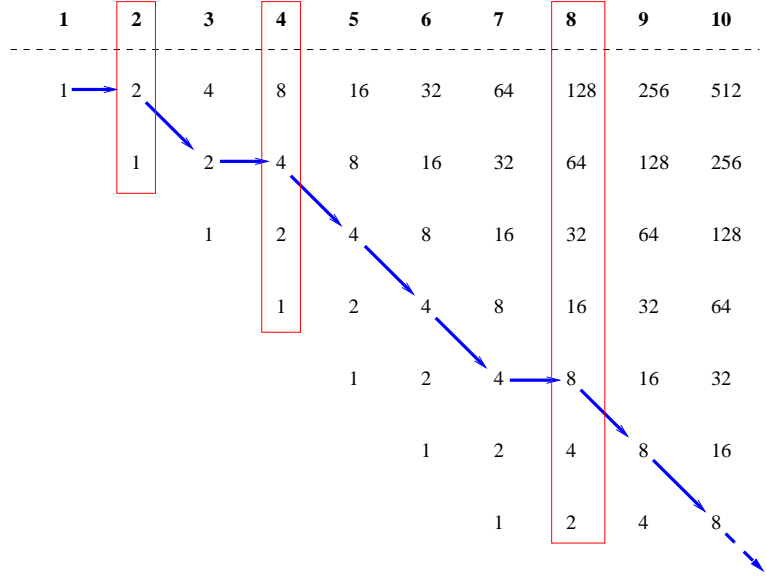}
\end{center}
\caption{A path yielding treewidth $O(n)$.}
\label{f:plog}
\end{figure}

For a given $G_D$, each possible pruning edge set $E_P$ corresponds to
a path spanning all columns in $\mathcal{D}_{1n}$. Instances with
diagonal (Prop.~\ref{bwprop}) or below-diagonal (Prop.~\ref{bwprop2})
$E_P$ paths yield BP trees whose width is bounded by $O(2^{v_0})$
where $v_0$ is small w.r.t.~$n$.
\ifsiam\begin{proposition}[Prop.~5.1 in \cite{bppolybook}]\else\begin{prop}[Prop.~5.1 in \cite{bppolybook}]\fi
\label{bwprop}
If $\exists v_0>K$ s.t.~$\forall v>v_0$ $\exists u<v-K$ with
$\{u,v\}\in E_P$ then the BP search tree width is bounded by
$2^{v_0-K}$.
\ifsiam\end{proposition}\else\end{prop}\fi
This corresponds to a path $\mbox{\sf
  p}_0=(1,2,\ldots,2^{v_0-K},\ldots,2^{v_0-K})$ that follows
unlabelled arcs up to level $v_0$ and then arcs labelled $v_0-K-1$,
$v_0-K-1\vee v_0-K$, and so on, leading to nodes that are all labelled
with $2^{v_0-K}$ (Fig.~\ref{f:p0}, top). 
\ifsiam\begin{proposition}[Prop.~5.2 in \cite{bppolybook}]\else\begin{prop}[Prop.~5.2 in \cite{bppolybook}]\fi
  \label{bwprop2}
  If $\exists v_0>K$ such that every subsequence $s$ of consecutive
  vertices $>\!\!v_0$ with no incident pruning edge is preceded by a
  vertex $v_s$ such that $\exists u_s<v_s\;(v_s-u_s\ge|s|\land
  \{u_s,v_s\}\in E_P)$, then the BP search tree width is bounded by
  $2^{v_0-K}$.
\ifsiam\end{proposition}\else\end{prop}\fi
This situation corresponds to a below-diagonal path (Fig.~\ref{f:p0},
bottom).
In general, for those instances for which the BP search tree width has
a $O(2^{v_0}\log n)$ bound, the BP has a worst-case running time
$O(2^{v_0}L 2^{\log n})=O(Ln)$, where $L$ is the complexity of
computing $T$.  Since $L$ is typically constant in $n$ \cite{dong03},
for such cases the BP runs in time $O(2^{v_0}n)$.
Let $V'=\{v\in V\;|\;\exists \ell\in\mathbb{N}\;(v=2^\ell)\}$. 
\ifsiam\begin{proposition}[Prop.~5.3 in \cite{bppolybook}]\else\begin{prop}[Prop.~5.3 in \cite{bppolybook}]\fi
  If $\exists v_0>K$ s.t.~for all $v\in V\smallsetminus V'$ with
  $v>v_0$ there is $u<v-K$ with $\{u,v\}\in E_P$ then the BP search
  tree width at level $n$ is bounded by $2^{v_0}n$. \label{bwprop3}
\ifsiam\end{proposition}\else\end{prop}\fi
This corresponds to a path roughly along the diagonal apart from
logarithmically many vertices in $V$ (those in $V'$), at which levels
the BP doubles the number of search nodes (Fig.~\ref{f:plog}).
For a pruning edge set $E_P$ as in Prop.~\ref{bwprop3}, or yielding a
path below it, the BP runs in $O(2^{v_0}n^2)$. 

\paragraph{Empirical verification}
On a set of 45 protein instances from the Protein Data Bank (PDB), 40
satisfy Prop.~\ref{bwprop}, and 5 satisfy Prop.~\ref{bwprop2}, all
with $v_0=4$ \cite{bppolybook}. This is consistent with the
computational insight \cite{dmdgp} that BP empirically displays a
polynomial (specifically, linear) complexity on real proteins.

\subsection{Interval data}
\label{s:interv}
In this section we discuss methods that target an MDGP variant, called
{\it i}\,MDGP, which is closer to the real NMR data: edges $\{u,v\}\in
E$ are weighted with real intervals ${\bf d}_{uv}=[d_{uv}^L,d_{uv}^U]$
instead of real values. These intervals occur in practice because, as
all other physical experiments, NMR outputs data with some
uncertainty, which can be modelled using intervals. The {\it i}\,MDGP
therefore consists in finding $x\in\mathbb{R}^K$ that satisfies the
following set of nonlinear inequalities:
\begin{equation}
\forall \{u,v\}\in E \quad d_{uv}^L\le \|x_u-x_v\| \le d_{uv}^U.
\label{eq:imdgp}
\end{equation}
The MP formulation \eqref{eq:mdgpgo} can be adapted to deal with this
situation in a number of ways, such as, e.g.:
\begin{eqnarray}
  \min_x && \sum_{\{u,v\}\in E} (\max (d^L_{uv} - || x_u - x_v||, 0) + 
    \max( || x_u - x_v || - d^U_{uv}, 0)), \label{eq:imdgpgo0} \\
  \min_x && \sum_{\{u,v\}\in E} (\max ((d^L_{uv})^2 - || x_u - x_v||^2, 0) + 
    \max( || x_u - x_v ||^2 - (d^U_{uv})^2, 0), \label{eq:imdgpgo1} \\
  \min_x && \sum_{\{u,v\}\in E} (\max\!{}^2 ((d^L_{uv})^2 - || x_u - x_v||^2, 0) + 
    \max\!{}^2 ( || x_u - x_v ||^2 - (d^U_{uv})^2, 0)). \label{eq:imdgpgo1a} 
\end{eqnarray}
Problem \eqref{eq:imdgpgo1a} is often appropriately modified to avoid
bad scaling (which occurs whenever the observed distances differ in
the order of magnitude):
\begin{equation}
  \min_x \sum_{\{u,v\}\in E} (\max\!{}^2 ( \frac{(d^L_{uv})^2 - || x_u -
    x_v||^2}{(d^L_{uv})^2}, 0) + \max\!{}^2( \frac{|| x_u - x_v ||^2 -
    (d^U_{uv})^2}{(d^U_{uv})^2}, 0)). \label{eq:imdgpgo2}
\end{equation}

\subsubsection{Smoothing-based methods}
\label{s:smoothinterv}
Several smoothing-based methods (e.g.~DGSOL and DCA, see
Sect.~\ref{s:dgsol}) have been trivially adapted to solve
\eqref{eq:imdgpgo1a} and/or \eqref{eq:imdgpgo2}. 

\begin{figure}[!ht]
\begin{center}
\psfrag{F(x,lambda)}{$F(x,\lambda)$}
\psfrag{max(x,0)}{$\max(x,0)$}
\psfrag{x}{$x$}
\includegraphics[width=8cm]{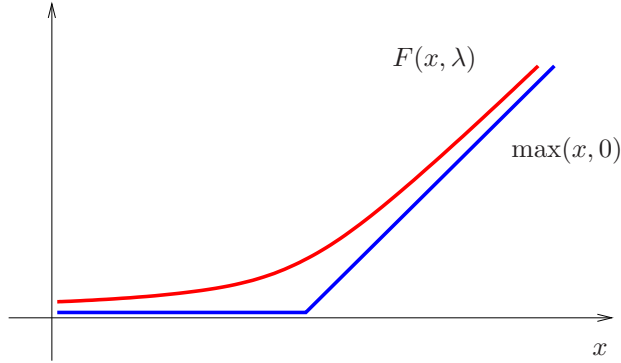}
\end{center}
\caption{The function $\max(x,0)$ and its hyperbolic smoothing
  $F(x,\lambda)$.}
\label{f:hypsmoothing}
\end{figure}

\paragraph{Hyperbolic smoothing}
\label{s:hyperbolic}
The {\it hyperbolic smoothing} described in \cite{souza} is
specifically suited to the shape of each summand in
\eqref{eq:imdgpgo0}, as shown in Fig.~\ref{f:hypsmoothing}.
The actual solution algorithm is very close to the one employed by
DGSOL (see Sect.~\ref{s:dgsol}). Given the fact that the smoothing is
not ``general-purpose'' (as the Gaussian transform is), but is
specific to the problem at hand, the computational results improve. It
should be noted, however, that this approach gives best results for
near cubic grid arrangements.

\subsubsection{The EMBED algorithm}
\label{s:embed}
The EMBED algorithm, proposed by Crippen and Havel \cite{CH88}, first
completes the missing bounds and refines the given bounds using
triangle and tetrangle inequalities. Then, a trial distance matrix
$D'$ is randomly generated, and a solution is sought using a matrix
decomposition method \cite{blumenthal}. Since the distance matrix $D'$
is not necessarily Euclidean \cite{eckart_36}, the solution may not
satisfy \eqref{eq:imdgp}. If this is the case, the final step of the
algorithm is to minimize the distance violations using the previous
solution as the initial guess. More details can be found
in \cite{havel_98,havel_03}.

\subsubsection{Monotonic Basin Hopping}
\label{s:mbh}
\begin{figure}[!ht]
\begin{center}
\psfrag{x1}{$x$}
\psfrag{x2}{$x^1$}
\psfrag{x3}{$x^\ast$}
\psfrag{x4}{$y$}
\psfrag{N1}{$\mathscr{N}(x)$}
\psfrag{N2}{$\mathscr{N}(x^1)$}
\psfrag{N3}{$\mathscr{N}(x^\ast)$}
\psfrag{N4}{$\mathscr{N}(y)$}
\psfrag{f}{$f$}
\includegraphics[width=14cm]{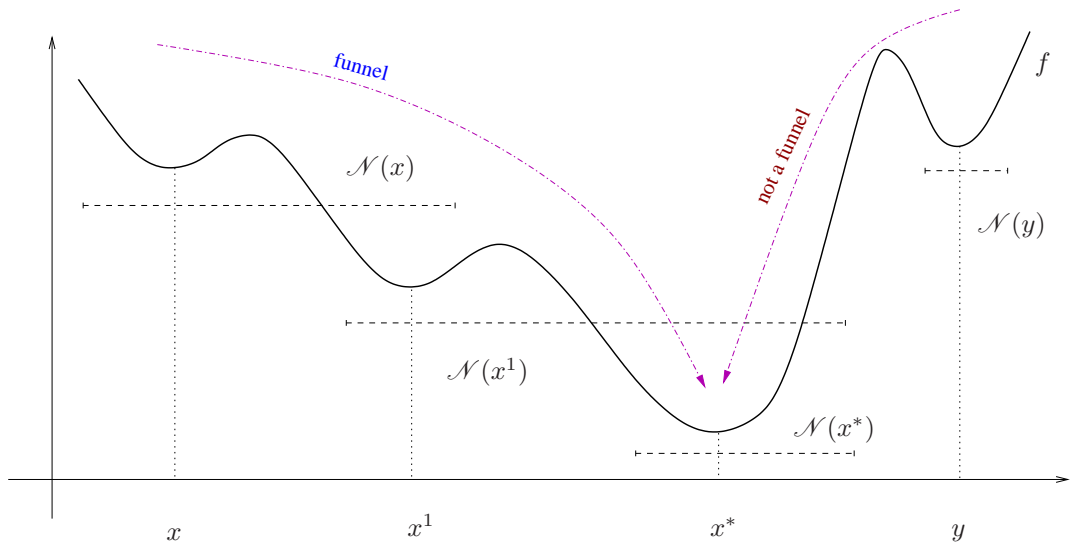} 
\end{center}
\caption{The dashed horizontal lines indicate the extent of the
  neighbourhoods. The set $\mathscr{F}=\{x,x^1,x^\ast\}$ is a funnel,
  because $x\sqsupset x^1\sqsupset x^\ast=\min\mathscr{F}$. The set
  $\{x^\ast,y\}$ is not a funnel, as $y\not\in\mathscr{N}(x^\ast)$.}
\label{f:funnel}
\end{figure}

A Monotonic Basin Hopping (MBH) algorithm for solving
\eqref{eq:imdgpgo1a}-\eqref{eq:imdgpgo2} is employed in \cite{grosso}.
Let $\mathscr{L}$ be the set of local optima of \eqref{eq:mdgpf} and
$\mathscr{N}:\mathbb{R}^3\to\mathscr{P}(\mathbb{R}^3)$ (where
$\mathscr{P}(S)$ denotes the power set of $S$) be some appropriate
neighbourhood structure. A artial order $\sqsupset$ on $\mathscr{L}$
is assumed to exist: $x\sqsupset y$ implies $y\in\mathscr{N}(x)$ and
$f(x)>f(y)$. A {\it funnel} is a subset
$\mathscr{F}\subseteq\mathscr{L}$ such that for each $x\in\mathscr{F}$
there exists a chain $x=x^0\sqsupset x^1\sqsupset \cdots\sqsupset
x^t=\min\mathscr{F}$ (the situation is described in
Fig.~\ref{f:funnel}).
The MBH algorithm is as follows. Starting with a current solution
$x\in\mathscr{F}$, sample a new point $x'\in\mathscr{N}(x)$ and use it
as the starting point for a local NLP solver; repeating this
sufficiently many times will yield the next optimum $x^1$ in the
funnel. This is repeated until improvements are no longer
possible. The MBH is also employed within a population-based
metaheuristic called Population Basin Hopping (PBH), which explores
several funnels in parallel.

\subsubsection{Alternating Projections Algorithm}
\label{s:apa}
\begin{figure}[!ht]
\begin{center}
\includegraphics[width=6cm]{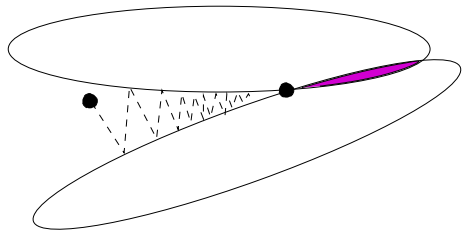}
\end{center}
\caption{The SPM attempts to find a point in the intersection of two
  convex sets.}
\label{f:apa}
\end{figure}

The Alternating Projection Algorithm (APA) \cite{apa} is an
application of the more general Successive Projection Methodology
(SPM) \cite{GrSc00,Ts01} to the {\it i}\,MDGP. The SPM takes a
starting point and projects it alternately on the two convex sets,
attempting to reach a point in their intersection (Fig.~\ref{f:apa}).

In the APA, the starting point is a given pre-distance matrix
$D=(\delta_{uv})$, i.e.~an $n\times n$ symmetric matrix with
non-negative components and zero diagonal. $D$ is generated randomly
so that $d^L_{uv}\le \delta_{uv}\le d^U_{uv}$ for all $\{u,v\}\in E$
and $\delta_{uv}=0$ otherwise. By Schoenberg's Theorem \ref{thm:psd}
and Eq.~\eqref{eq:dattorro}, if we let $P=I-\frac{1}{n}{\bf
1}\transpose{\bf 1}$ and $A=-\frac{1}{2}PDP$, where $I$ is the
$n\times n$ identity matrix and ${\bf 1}$ is the all-one $n$-vector,
$D$ is a Euclidean distance matrix if and only if $A$ is positive
semi-definite. Notice that $P$ is the orthogonal projection operator
on the subspace $M=\{x\in\mathbb{R}^n\;|\;\transpose{x}{\bf 1}=0\}$ of
vectors orthogonal to {\bf 1}, so $D$ is a Euclidean distance matrix
if and only if $D$ is negative semidefinite on $M$ \cite{glunt}. On
the other hand, a necessary condition for any matrix to be\ a
Euclidean distance matrix is that it should have zero diagonal. This
identifies the two convex sets on which the SPM is run: the set
$\mathcal{P}$ of matrices which are negative semidefinite on $M$, and
the set $\mathcal{Z}$ of zero-diagonal matrices. The projection
operator for $\mathcal{P}$ is $Q(D)=PU\Lambda^-UP$, where $U\Lambda U$
is the spectral decomposition of $D$ and $\Lambda^-$ is the
nonpositive part of $\Lambda$, and the projection operator for
$\mathcal{Z}$ is $Q'(D)=D-\mbox{diag}(D)$.

Although the convergence proofs for the SPM assumes an infinite number
of iterations in the worst case, empirical tests suggest that five
iterations of the APA are enough to get satisfactory results. The APA
was tested on the bovine pancreatic trypsin inhibitor protein ({\tt
qlq}), which has 588 atoms including side-chains.

\subsubsection{The GNOMAD iterative method}
\label{s:gnomad}
The GNOMAD algorithm \cite{gromad} (see Alg.~\ref{alg:gnomad}) is a
multi-level iterative method, which tries to arrange groups of atoms
at the highest level, then determines an appropriate order within each
group using the contribution of each atom to the total error, and
finally, at the lowest level, performs a set of atom moves within each
group in the prescribed order. The method exploits several local NLP
searches (in low dimension) at each iteration, as detailed below.
\begin{algorithm}[!ht]
\begin{algorithmic}[1]
\STATE $\{C_1,\ldots,C_\ell\}$ is a vertex cover for $V$ 
\FOR{$i\in \{1,\ldots,\ell\}$}
  \WHILE{termination condition not met}
    \STATE determine an order $<$ on $C_i$
    \FOR{$v\in (C_i,<)$}
      \STATE find search direction $\Delta_v$ for $x_v$ (obtained by
      solving an NLP locally)
      \STATE determine step $s_v$ minimizing constraint infeasibility
             \label{as:inf}
      \STATE $x_v\leftarrow x_v+s_v\Delta_v$
    \ENDFOR
  \ENDWHILE
\ENDFOR
\end{algorithmic}
\caption{GNOMAD}
\label{alg:gnomad}
\end{algorithm}
The constraints exploited in Step \ref{as:inf} are mostly given by van
der Waals distances \cite{schlick}, which are physically inviolable
separation distances between atoms.

\subsubsection{Sthochastic Proximity Embedding heuristic}
\label{s:spe}
\begin{figure}[!ht]
\begin{center}
\psfrag{u}{$u$}
\psfrag{v}{$v$}
\psfrag{d}{$d$}
\psfrag{lambda}{$\lambda$}
\includegraphics[width=8cm]{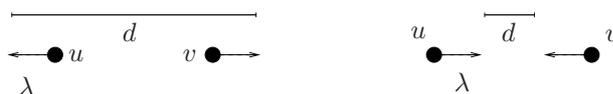}
\end{center}
\caption{Local changes to positions according to discrepancy with
  respect to the corresponding distance.}
\label{f:spe}
\end{figure}

The basic idea of the Stochastic Proximity Embedding (SPE) \cite{spe}
heuristic is as follows. All the atoms are initially placed randomly
into a cube of a given size. Pairs of atoms in $E$ are repeatedly and
randomly selected; for each pair $\{u,v\}$, the algorithm checks
satisfaction of the corresponding constraint in \eqref{eq:imdgp}. If
the constraint is violated, the positions of the two atoms are changed
according to explicit formulae in order to improve the current
embedding (two examples are shown in Fig.~\ref{f:spe}).
\begin{algorithm}[!ht]
\begin{algorithmic}
\WHILE{termination condition not met}
  \STATE Pick $\{u,v\}\in E\;(\|x_u-x_v\|\not\in d_{uv})$
  \STATE Update $\lambda$
  \STATE Let $x_u\leftarrow x_u + \lambda(x_u-x_v)$
  \STATE Let $x_v\leftarrow x_v + \lambda(x_v-x_u)$.
\ENDWHILE
\end{algorithmic}
\caption{SPE Heuristic}
\label{alg:spe}
\end{algorithm}

The SPE heuristic is shown in Alg.~\ref{alg:spe}. SPE offers no
guarantee to obtain a solution satisfying all constraints
in \eqref{eq:imdgp}, however the ``success stories'' reported
in \cite{Izrailev} seem to indicate this as a valid methodology.

\subsection{NMR data}
\label{s:nmr}
Nuclear Magnetic Resonance experiments are performed in order to
estimate distances between some pairs of atoms forming a given
molecule \cite{wuthrich_89}. In solution, the molecule is subjected to
a strong external magnetic field, which induces the alignment of the
spin magnetic moment of the observed nuclei. The analysis of this
process allows the identification of a subset of distances for certain
pairs of atoms, mostly those involving hydrogens, as explained in the
introduction (p.~\pageref{p:nmr}). In proteins, nuclei of carbons and
nitrogens are also sometimes considered.

It is important to remark that some NMR signals may fail to be
precise, because it is not always possible to distinguish between the
atoms of the molecule. We can have this situation, for example, in
proteins containing amino acids such as valines and leucines. In such
a case, the distance {\it restraints} (a term used in proteomics
meaning ``constraints'') involve a ``pseudo-atom'' that is placed
halfway between the two undistinguished atoms \cite{wuthrich_83}.
Once the upper bound for the distance has been chosen when considering
the pseudo-atom, its value is successively increased in order to
obtain an upper bound for the real atoms.

There are also other potential sources of errors that can affect NMR
data. If the molecule is not stable in solution, its conformation may
change during the NMR experiments, and therefore the obtained
information could be inconsistent. Depending on the machine and on the
magnetic field, some noise may spoil the quality of the NMR signals
from which the intervals are derived. Moreover, due to a phenomenon
called ``spin diffusion'', the NMR signals related to two atoms could
also be influenced by neighboring atoms \cite{clore_89}.

Fortunately, for molecules having a known chemical composition, such
as proteins, there are a priori known distances that can be considered
together with the ones obtained through NMR experiments. If two atoms
are chemically bonded, their relative distance is known; this distance
is subject to small variations, but it can still be considered as
fixed in several applications (see the rigid geometry hypothesis,
Sect.~\ref{s:molgraphs}). Moreover, the distance between two atoms
bonded to a common atom can also be estimated, because they generally
form a specific angle that depends upon the kind of involved
atoms. Such distances can therefore be considered precise, and provide
valuable information for the solution of distance geometry problems
(this follows because protein graphs are molecular, see
Sect.~\ref{s:molgraphs}).

As explained in the introduction, on p.~\pageref{p:nmr}, the output of
a Nuclear Magnetic Resonance experiment on a given molecule can be
taken to consist of a set of triplets $(\{a,b\},d,q)$, meaning that
$q$ pairs of atoms of type $a,b$ were observed to have distance
$d$ \cite{berger}. It turns out that NMR data can be further
manipulated so that it yields a list of pairs $\{u,v\}$ of atoms with
a corresponding nonnegative distance $d_{uv}$. Unfortunately this
manipulation is rather error-prone, resulting in interval-type errors,
so that the exact inter-atomic distances $d_{uv}$ are in fact
contained in given intervals $[d_{uv}^L,d_{uv}^U]$ \cite{berger}. For
practical reasons, NMR experiments are most often performed on
hydrogen atoms \cite{berger} (although sometimes carbons and nitrogens
are also considered). Other known molecular information
includes \cite{schlick,donald}: the number and type of atoms in the
molecules, all the covalent bonds with corresponding Euclidean
distances, and all distances between atoms separated by exactly two
covalent bonds.

\subsubsection{Virtual backbones of hydrogens}
\label{s:hydrogens}
In order to address the NMR limitation concerning the lack of data
reliability for inter-atomic distances of non-hydrogen atoms, we
define atomic orders limited to hydrogens, and disregard the natural
backbone order during discretization. Even though we showed that this
approach works on a set of artificially generated instances
\cite{jogomdgp}, we remarked its limitations when we tried to apply it
to real NMR data. These limitations have been addressed by using
re-orders (see Sect.~\ref{s:reorders}).

\subsubsection{Re-orders and interval discretization}
\label{s:reorders}
In \cite{bpinterval} we define an atomic ordering which ensures that
every atom of rank $>3$ is adjacent to its three immediate predecessors
by means of either real-valued distances $d$, or interval distances
$\bar{d}$ that arise from geometrical considerations rather than NMR
experiments.  Specifically, with reference to Fig.~\ref{f:tangles},
the distance $d_{i-3,i}$ belongs to a range determined by the
uncertainty associated with the torsion angle $\phi_i$.

\begin{figure}[!ht]
\begin{center}
\includegraphics[scale=0.30]{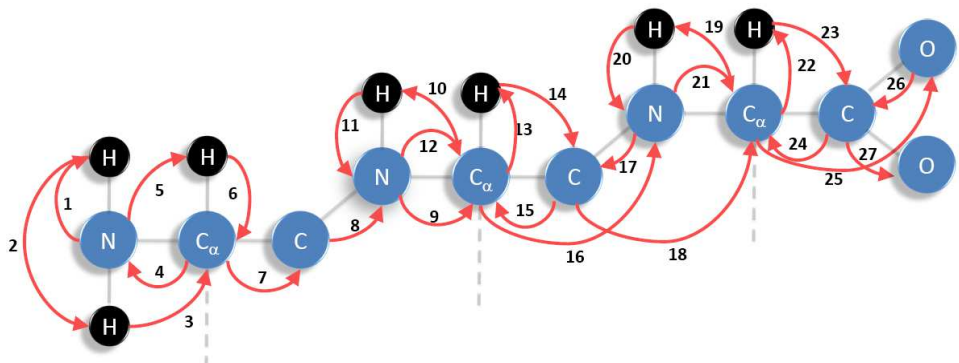}
\end{center}
\caption{The order used for discretizing MDGPs with interval
  data.}
\label{f:ordering}
\end{figure}

We exploited three protein features to this aim: (i) using hydrogen
atoms off the main backbone whenever appropriate, (ii) using the same
atom more than once, (iii) remarking that interval distances $\bar{d}$
can be replaced with finite (small) sets $D$ of real-valued
distances. Considering these properties, we were able to define a new
atomic ordering for which $v$ can be placed in a finite number of
positions in the set $\{0,1,2,2|D|\}$, consistently with the known
positions of the three immediate predecessors of $v$. Feature (i)
allows us to exploit atoms for which NMR data are available. Feature
(ii) allows us to exploit more than just two bond lengths on atoms
with valence $>2$, such as carbons and nitrogens, by defining an order
that includes the atom more than once. Since atoms are repeated in the
order, we call these orders {\it re-orders} \cite{bpinterval}. Feature
(iii) rests on an observation concerning the resolution scope of NMR
experimental techniques \cite{nilges2}. Fig.~\ref{f:ordering} shows a
re-order for a small protein backbone containing 3 amino acids.

Re-orders $(v_1,\ldots,v_p)$ deserve a further remark. We stressed the
importance of strict simplex inequalities in Sect.~\ref{s:sip}, but
requiring that $v_i=v_j$ for some $i\not=j$ introduces a zero distance
$d(v_i,v_j)=0$. If this distance is ever used inappropriately, we
might end up with a triangle with a side of zero length, which might
in turn imply an infinity of possible positions for the next atom. We
recall that, for any $v>K$, strict simplex inequalities
$\Delta_{K-1}(U_v)>0$ in dimension $K-1$ are {\it necessary} to
discretization, as they avoid unwanted affine dependencies (see
e.g.~Fig.~\ref{f:collinear}). By contrast, if
$\Delta_K(U_v\cup\{v\})>0$ hold, then we have a $K$-simplex with
nonzero volume, which has two possible orientations in $\mathbb{R}^K$:
in other words, the two possible positions for $x_v$ are distinct. If
$\Delta_K(U_v\cup\{v\})=0$, however, then there is just one possible
position for $x_v$. Thus, to preserve discretization, zero distances
can never occur between pairs $v_i,v_j$ fewer than $K$ atoms apart,
but they may occur for $|i-j|=K$: in this case we shall have no
branching at level $\max(i,j)$.

Re-orders make it possible to only employ non-NMR distances for
discretization. More precisely, over each set of three adjacent
predecessors, only one is related by an interval distance; this
interval, however, is not due to experimental imprecision in NMR, but
rather to a molecular property of torsion angles. In particular, we
can compute tight lower and upper bounds to these intervals;
consequently, they can be discretized without loss of
precision \cite{bpinterval}. We refer to such intervals as {\it
discretizable}.

\subsubsection{Discrete search with interval distances} 
\label{s:intervals}

\begin{figure}[!ht]
\begin{center}
\psfrag{dL}{$d^L$}
\psfrag{dU}{$d^U$}
\includegraphics[width=12cm]{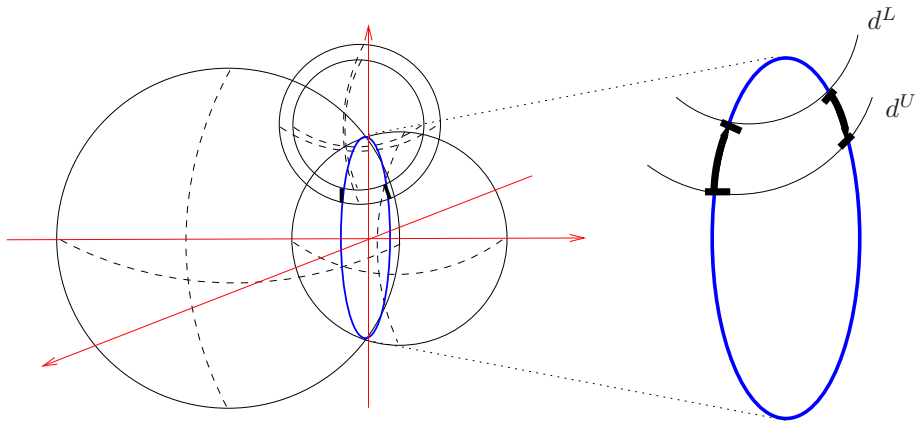}
\end{center}
\caption{The intersection of two spheres with a spherical shell.}
\label{f:approxdiscr}
\end{figure}

The {\it interval} BP ($i$BP) \cite{bpinterval} is an extension of the
BP algorithm which is able to manage interval data. The main idea is
to replace, in the sphere intersections necessary for computing
candidate atomic positions, a sphere by a {\it spherical shell}. Given
a center $c\in\mathbb{R}^K$ and an interval $d=[d^L,d^U]$ the
spherical shell centered at $c$ w.r.t.~$d$ is
$S^{K-1}(c,d^U)\smallsetminus S^{K-1}(c,d^L)$. With $K=3$, the
intersection of two spheres and a spherical shell gives, with
probability one, two disjoint curves in three-dimensional space
(Fig.~\ref{f:approxdiscr}). 
The discretization is still possible if some sample distances are
chosen from the interval associated to the curves \cite{nilges2}.

Similarly to the basic BP algorithm, the two main components of $i$BP
are the branching and the pruning phases. In the branching phase, we
can have 3 different situations, depending on the distance $d(i-3,i)$
(see Fig.~\ref{f:ordering}). If $d(i-3,i)=0$, the current atom $i$
already appeared previously in the order, which means that the only
feasible position for $i$ is the same as $i-3$. If $d(i-3,i)$ is a
precise distance, then 3 spheres are intersected, and only two
positions are found with probability one. Finally, if $d(i-3,i)$ is a
discretizable interval $[d^L_{i-3,i},d^U_{i-3,i}]$, as specified in
Sect.~\ref{s:reorders}, we choose $D$ values from the interval. This
yields a choice of $2D$ candidate atomic solutions for $i$. 

If the discretization order in Fig.~\ref{f:ordering} is employed for
solving NMR instances, (precise) distances derived from the chemical
composition of proteins are used for performing the discretization,
whereas interval distances from NMR experiments are used for pruning
purposes only. The consequent search tree is no longer binary: every
time a discretizable interval is used for branching, the current node
has at most $2D$ subnodes. The advantage is that the generation of the
search tree is not affected by experimental errors caused by the NMR
machinery. 

In order to discretize instances related to entire protein
conformations, it is necessary to identify a discretization order for
all side chains for the 20 amino acids that can be involved in the
protein synthesis. This is a nontrivial task, because side chains have
more complex structures with respect to the part which is common to
each amino acid, and they may contain many atoms. However, side chains
can be of fundamental importance in the identification of protein
conformations, because many distances obtained by NMR experiments may
regard hydrogen atoms contained in side chains. First efforts towards
extending the BP algorithm so that it can calculate the whole
three-dimensional structure of a protein, including its side chains,
can be found in \cite{ijbra}.

\section{Engineering applications}
\label{s:engapps}
In this section, we discuss other well-known applications of distance
geometry: wireless networks, statics, data visualization and robotics.
In wireless networks, mobile sensors can usually estimate their
pairwise distance by measure how much battery they use in order to
communicate. These distances are then used to find the positions of
each sensor (see Sect.~\ref{s:wsnl}). Statics is the field of study of
the equilibrium of rigid structures (mostly man-made, such as
buildings or bridges) under the action of external forces. A
well-known model for such structures is the {\it bar-and-joint
framework}, which is essentially a weighted graph. The main problem is
that of deciding whether a given graph, with a given distance function
on the edges, is rigid or not. An associated problem is that of
deciding whether a given graph models a rigid structure independently
of the distance function (see Sect.~\ref{s:rigid}).

\subsection{Wireless sensor networks}
\label{s:wsnl}
The position of wireless mobile sensors (e.g. sm\-art\-phones,
identification badges and so on) is, by its very definition, local to
the sensor carrier at any given time. Notwithstanding, in order to be
able to properly route communication signals, the network routers must
be aware of the sensor positions, and adapt routes, frequencies, and
network ID data accordingly. The information available to solve this
problem is given by the fact that mobile sensors are always aware of
their neighbouring peers (to within a certain radius $r$ from their
positions, which we shall assume constant), as well as of the amount
of battery charge they use in order to communicate with each other
sensor in their neighbourhood. It turns out that this quantity is
strongly correlated with the Euclidean distance between the
communicating sensors \cite{savvides}. Moreover, certain network
elements, such as routers and wireless repeaters, are fixed, hence
their positions are known (such elements are called {\it anchors} or
{\it beacons}). The problem of determining the sensor positions using
these data was deemed as an important one from the very inception of
wireless networks \cite{weiser,forman}. There are several good reasons
why Global Positioning System (GPS) enabled devices may not be a valid
alternative: they are usually too large, they consume too much power,
and they need a line of sight with the satellites, which may not
always be the case in practice (think for example of localizing
sensors within a building) \cite{savvides}. This problem is formalized
as the WSNL (see Item \ref{item:wsnl} in the list of
Sect.~\ref{s:map}).

In practice, $K\in\{2,3\}$. The 3D case might occur when a single
network is spread over several floors of a building, or whenever a
mobile battlefield network is parachuted over a mountainous
region. Moreover, because the realization represents a practically
existing network, an important question is to determine what amount of
data suffices for the graph to have a {\it unique} realization in
$\mathbb{R}^K$. This marks a striking difference with the application
of DG techniques to molecular conformation, where molecules can exist
in different isomers. 

The earliest connections of WSNL with DG are an SDP formulation
\cite{doherty} for a relaxation of the problem where the Euclidean
distance between two sensors is {\it at most} the corresponding edge
weight, and an in-depth theoretical study of the WSNL from the point
of view of graph rigidity \cite{eren04} (see
Sect.~\ref{s:rigid}). 

\subsubsection{Unique realizability}
\label{s:uniquereal}
In \cite{eren04,eren06}, the WSNL is defined to be {\it solvable} if
the given graph has a unique valid realization, a notion which is also
known as global rigidity. A graph is {\it globally rigid} if it has a
generic realization $x$, and for all other realizations $x'$, $x$ is
congruent to $x'$. For example, if a graph has a $K$-trilateration
order, then it is globally rigid: comparing with DVOP orders, where
each vertex is adjacent to $K$ predecessors, the additional adjacency
makes it possible to identify {\it at most one} position in
$\mathbb{R}^K$ where the next vertex in the order will be placed, if a
position for all predecessors is already known. Any graph possessing a
$K$-trilateration order is called a {\it $K$-trilateration
graph}. Such graphs are globally rigid, and can be realized in
polynomial time by simply remarking that the BP would never branch on
such instances. 

A graph $G=(V,E)$ is {\it redundantly rigid} if $(V,E\smallsetminus
\{e\})$ is rigid for all $e\in E$. It was shown in
\cite{globrigid2,connelly} that $G$ is globally rigid for $K=2$ if and
only if either $G$ is the $2$-clique or $3$-clique, or $G$ is
3-connected and redundantly rigid. Hendrickson had conjectured in
\cite{Hen92} that these conditions would be sufficient for any value
of $K$, but this was disproved by Connelly \cite{connelly2}. He also
proved, in \cite{connelly}, that if a generic framework $(G,x)$ has a
self-stress (see Sect.~\ref{s:infrigid}) $\omega:E\to\mathbb{R}$ such
that the $n\times n$ {\it stress matrix}, with $(u,v)$-th entry
$(-\omega_{uv})$ if $\{u,v\}\in E$, $\sum_{t\in\delta(v)}\omega_{ut}$
if $u=v$, and 0 otherwise, has rank $n-K-1$, then $(G,x)$ is globally
rigid in any dimension $K$. Some graph properties ensuring global
rigidity for $K\in\{2,3\}$ are given in
\cite{easilyul}. A related problem, that of choosing a given subset of
vertices to take the role of anchors, such that the resulting sensor
network is uniquely localizable (see Sect.~\ref{s:abbie}), is
discussed in \cite{fj}. Several results on global rigidity (with
particular attention to the case $K=2$) are surveyed
in \cite{jjsurvey}. In particular, it is shown
in \cite[Thm.~11.3]{jjsurvey} that Henneberg type II steps (replace an
edge $\{u,w\}$ by two edges $\{u,v\}$ and $\{v,w\}$, where $v$ is a
new vertex, then add new edges from $v$ to $K-1$ other vertices
different from $u,w$) are related to global rigidity in a similar way
as Henneberg type I steps (see Sect.~\ref{s:henneberg}) are related to
rigidity: if a globally rigid graph $H$ is derived from a graph $G$
with at least $K+2$ vertices using a Henneberg type II step in
$\mathbb{R}^K$, then $G$ is also globally rigid.

There is an interesting variant of unique localizability which yields
a subclass of DGP instances that can be realized in polynomial
time. Recall that the DGP is strongly {\bf NP}-hard \cite{saxe79} in
general. Moreover, it remains {\bf NP}-hard even when the input is a
unit disk graph (Sect.~\ref{s:udg}) \cite{eren06}, and there exists no
randomized efficient algorithm even when it is known that the input
graph is globally rigid \cite{aspnes}. The problem becomes tractable
under the equivalent assumptions of $K$-unique localizability (a sort
of unique localizability for fixed $K$) \cite{ye} and universal
rigidity \cite{univrigid} (see Sect.~\ref{s:abbie}). Specifically, a
graph is {\it $K$-uniquely localizable} if: (i) it has a unique
realization $x:V\to\mathbb{R}^K$, (ii) it has a unique realization
$y^\ell:V\to\mathbb{R}^\ell$ for all $\ell>K$, and (iii) for all $v\in
V,\ell>K$ we have $y_v^\ell=(x_v,{\bf 0})$, where ${\bf 0}$ is the
zero vector in $\mathbb{R}^{\ell-K}$. Anchors play a crucial role in
ensuring that the graph should be globally rigid in $\mathbb{R}^K$:
the subgraph induced by the anchors should yield a generic globally
rigid framework in $\mathbb{R}^K$, thus the set of anchors must have
at least $K+1$ elements. Under these assumptions, an exact polynomial
algorithm (exploiting the SDP formulation and its dual) for realizing
$K$-uniquely localizable graphs was described in \cite{ye}.

\subsubsection{Semidefinite Programming}
\label{s:sdp}
Most of the recent methods addressing the WNSL make use of SDP
techniques. This is understandable in view of the relationship between
DG and SDP via Thm.~\ref{thm:psd}, and because PSD completion is
actually a special case of the general SDP feasibility problem (see
Sect.~\ref{s:psdmcp}). We also mention that most SDP methods can
target DGP problem variants where the edge weight $d$ maps into
bounded intervals, not only reals, and are therefore suitable for
applications where distance measurements are not precise.

We believe \cite{psdcpapprox} is the first reference in the literature
that proposes an SDP-based method for solving MCPs (specifically, the
PSDMCP). In \cite{wolkowicz}, the same approach is adapted to a
slightly different EDMCP formulation. Instead of a partial matrix, an
$n\times n$ {\it pre-distance matrix} $A$ is given, i.e.~a matrix with
zero diagonal and nonnegative off-diagonal elements. We look for an
$n\times n$ Euclidean distance matrix $D$ that minimizes $\|H\circ
(A-D)\|_F$, where $H$ is a given matrix of weights, $\circ$ is the
Hadamard product, and $\|\cdot\|_F$ is the Frobenius norm
($\|Q\|_F=\sqrt{\sum_{i,j\le n} q_{ij}^2}$). An optional linear
constraint can be used to fix some of the values of $D$. A
reformulation of the constraint ``$D$ is a Euclidean distance matrix''
to $X\succeq 0$, is derived by means of the statement that $D$ is a
Euclidean distance matrix if and only if $D$ is negative semidefinite
on the orthogonal complement of the all-one vector \cite{gower,apa}
(see Sect.~\ref{s:apa}). In turn, this is related to
Thm.~\ref{thm:psd}.

In \cite{boydlmi,doherty}, interestingly, the connection with SDP is
{\it not} given by Thm.~\ref{thm:psd}, but rather because the WSNL
variants mentioned in the paper make use of convex norm constraints
which are reformulated using Linear Matrix Inequalities (LMI). For
example, if there is a direct communication link between two nodes
$u,v\in V$, then $\|x_u-x_v\|\le r$, where $r$ is a scalar threshold
given by the maximum communication range, the inequality can be
reformulated to the following LMI: 
\begin{equation*}
  \left(\begin{array}{cc} rI_2 & x_u-x_v \\ \transpose{(x_u-x_v)} & r
  \end{array}\right) \succeq 0,
\end{equation*}
where $I_K$ is the $K\times K$ identity matrix (with $K=2$).

Biswas and Ye proposed in \cite{biswas2004} an SDP formulation of the
WSNL problem which then gave rise to a series of papers
\cite{biswasacm,biswas2006,biswas2006ieee,biswasphd,wsnlsdp,mdgpsdp} focusing
on algorithmic exploitations of their formulation. In the spirit of
\cite{refmathprog}, this can be derived from the ``classic'' WSNL
feasibility formulation below by means of a sequence of basic
reformulations: 
\begin{eqnarray*}
  \forall \{u,v\}\in E \quad (\|x_u-x_v\|_2 &=& d_{uv}) \\
  \forall u\in A, v\not\in A \quad (\{u,v\}\in E\to \|a_u-x_v\| &=& d_{uv}), 
\end{eqnarray*}
where $A\subseteq V$ is the set of anchors whose positions
$\{a_u\;|\;u\in A\}\subseteq\mathbb{R}^K$ are known {\it a priori}.
Let $X$ be the $K\times n$ decision variable matrix whose $v$-th
column is $x_v$. The authors remark that:
\begin{itemize}
\item for all $u<v\in V$,
  $\|x_u-x_v\|^2=\transpose{e_{uv}}\transpose{X}Xe_{uv}$,
  where $e_{uv}=1$ at component $u$, $-1$ at component $v$, and 0 elsewhere;
\item for all $u\in A,v\in V$,
  $\|a_u-x_v\|^2=\transpose{(a_u;e_v)}\transpose{[I_K;X]}[I_K;X](a_u;e_v)$,
  where $(a_u;e_v)$ is the column $(K+n)$vector consisting of $a_u$ on top
  of $e_v$, with $e_V=1$ at component $v$ and 0 elsewhere, and
  $[I_K;X]$ is the $K\times (K+n)$ matrix consisting of $I_K$ followed
  by $X$;
\item $\transpose{[I_K;X]}[I_K;X]=\left(\begin{array}{cc} I_K & X
  \\ \transpose{X} & \transpose{X}X \end{array}\right)$, a
  $(K+n)\times (K+n)$ matrix denoted by $Z$;
\item the scalar products of decision variable vectors in
  $\transpose{X}X$ (rows of $\transpose{X}$ by columns of $X$) can be
  linearized, replacing each $x_ux_v$ by $y_{uv}$, which results in
  substituting $\transpose{X}X$ by an $n\times n$ matrix $Y=(y_{uv})$
  such that $Y=\transpose{X}X$.
\end{itemize}
This yields the following formulation of the WSNL:
\begin{eqnarray*}
  \forall \{u,v\}\in E \quad \transpose{e_{uv}} Y e_{uv} &=& d^2_{uv} \\
  \forall u\in A, v\not\in A \quad (\{u,v\}\in E\to 
    \transpose{(a_u;e_v)}Z(a_u;e_v) &=& d^2_{uv})\\
  Y=\transpose{X}X.
\end{eqnarray*}
The SDP relaxation of the constraint $Y=\transpose{X}X$, which is
equivalent to requiring that $Y$ has rank $K$, consists in replacing
it with $Y-\transpose{X}X\succeq 0$, which is equivalent to $Z\succeq
0$. The whole SDP can be written in function of the indeterminate
matrix $Z$ as follows, using Matlab-like notation to indicate
submatrices:
\begin{eqnarray}
  Z_{1:K,1:K} &=& I_K \label{yesdp1} \\
  \forall u,v\in V\smallsetminus A\quad (\{u,v\}\in E\to ({\bf 0};
  e_{uv})\transpose{({\bf 0};e_{uv})} \bullet Z &=& d_{uv}^2
  )  \label{yesdp2} \\
  \forall u\in A, v\in V\smallsetminus A\quad (\{u,v\}\in E\to (a_u;
  e_{v})\transpose{(a_u;e_{v})} \bullet Z &=& d_{uv}^2
  )  \label{yesdp3}  \\
  Z &\succeq& 0, \label{yesdp4}
\end{eqnarray}
where $\bullet$ is the Frobenius product. This formulation was exploited
algorithmically in a number of ways. As mentioned in
Sect.~\ref{s:uniquereal} and \ref{s:abbie}, solving the SDP
formulation \eqref{yesdp1}-\eqref{yesdp4} yields a polynomial-time
algorithm for the DGP on uniquely localizable graphs (see
Sect.~\ref{s:abbie}). The proof uses the dual SDP formulation
to \eqref{yesdp1}-\eqref{yesdp4} in order to show that the interior
point method for SDP yields an exact solution \cite[Cor.~1]{ye} and
the fact that the SDP solution on uniquely localizable graphs has rank
$K$ \cite[Thm.~2]{ye}. Another interesting research direction
employing \eqref{yesdp1}-\eqref{yesdp4} is the edge-based SDP (ESDP)
relaxation \cite{edgebased}: this consists in relaxing \eqref{yesdp4}
to only hold on principal submatrices of $Z$ indexed by $A$. To
address the fact that SDP and ESDP formulations are very sensitive to
noisy data, a robust version of the ESDP relaxation was discussed
in \cite{tseng2} (see Sect.~\ref{s:abbie}).

Among the methods based on formulation \eqref{yesdp1}-\eqref{yesdp4},
\cite{wsnlsdp,mdgpsdp} are particularly interesting. They address the
limited scaling capabilities of SDP solution techniques by identifying
vertex clusters where embedding is easier, and then match those
embeddings in space using a modified SDP formulation. The vertex
clusters cover $V$ in such a way that neighbouring clusters share some
vertices (these are used to ``stitch together'' the embeddings
restricted to each cluster). The clustering technique is based on
permuting columns of the distance matrix $(d_{ij})$ so as to try to
pool the nonzeros along the main diagonal. The partial embeddings for
each cluster are computed by first solving an SDP relaxation of the
quadratic system (\ref{eq:imdgp}) restricted to edges in the cluster,
and then applying a local NLP optimization algorithm that uses the
optimal SDP solution as a starting point. When the distances have
errors, there may not exist any valid embedding satisfying all the
distance constraints. In this case, it is likely that the SDP approach
(which relaxes these constraints anyhow) will end up yielding an
embedding $x'$ which is valid in a higher dimensional space
$\mathbb{R}^{K'}$ where $K'>K$. In such cases, $x'$ is projected onto
an embedding $x$ in $\mathbb{R}^K$. Such projected embeddings usually
exhibit clusters of close vertices (none of which satisfies the
corresponding distance constraints), due to correct distances in the
higher dimensional space being ``squeezed'' to their orthogonal
projection into the lower dimensional space. In order to counter this
type of behaviour, a regularization objective $\max\sum_{i,j\in V}
||x_i-x_j||^2$ is added to the feasibility SDP.

In \cite{krislocksiam,krislock}, Krislock and Wolkowicz also exploit
the SDP formulations of \cite{wolkowicz} together with vertex
clustering techniques in order to improve the scaling abilities of SDP
solution methods (also see Sect.~\ref{s:abbie}). Their facial
reduction algorithm identifies cliques in the input graph $G$ and
iteratively expands them using a $K$-trilateration order (see
Sect.~\ref{s:discr}). Rather than ``stitching together'' pieces, as
in \cite{mdgpsdp}, the theory of facial reduction methods works by
considering the SDP relaxation of the whole problem and showing how it
can be simplified in presence of one or more cliques (be they
intersecting or disjoint). The computational results
of \cite{krislocksiam} show that the facial reduction algorithm scales
extremely well (graphs up to 100,000 vertices were embedded in
$\mathbb{R}^2$). A comparison with the BP algorithm (see
Sect.~\ref{s:bp}) appears in \cite[Table 6]{dmdgp}. The BP algorithm
is less accurate (the most common LDE values are $O(10^{-12})$ for BP
and $O(10^{-13})$ for facial reduction) but faster (BP scores between
1\% and 10\% of the time taken by facial reduction).

\subsubsection{Second-order cone programming}
A second-order cone programming (SOCP) relaxation of the WSNL was
discussed in \cite{tseng3}. The NLP formulation \eqref{eq:mdgpgo} is
first reformulated as follows:
\begin{equation}
  \left.\begin{array}{rrcl}
    \min & \sum\limits_{\{u,v\}\in E} z_{uv} && \\
    \forall \{u,v\}\in E & x_u-x_v &=& w_{uv} \\
    \forall \{u,v\}\in E & y_{uv}-z_{uv} &=& d_{uv}^2 \\
    \forall \{u,v\}\in E & \|w_{uv}\|^2 &=& y_{uv} \\
                         & u &\ge & 0.
  \end{array}\right\}\label{eq:presocp}
\end{equation}
Next, the constraint $\|w_{uv}\|^2=y_{uv}$ is relaxed to
$\|w_{uv}\|^2\le y_{uv}$. The SOCP relaxation is weaker than the SDP
one (\eqref{yesdp1}-\eqref{yesdp4}), but scales much better (4000
vs.~500 vertices). It was abandoned by Tseng in favour of the ESDP
\cite{tseng2}, which is stronger than the SOCP relaxation but scales
similarly.

\subsubsection{Unit disk graphs}
\label{s:udg}
Unit disk graphs are intersection graphs of equal circles in the
plane, i.e.~vertices are the circle centers, and there is an edge
between two vertices $u,v$ if their Euclidean distance is at most
twice the radius. Unit disk graphs provide a good model for broadcast
networks, with each center representing a mobile transmitter/receiver,
and the radius representing the range. In \cite{udg}, it is shown that
several standard {\bf NP}-complete graph problems are just as
difficult on unit disk graphs as on general graphs, but that the
maximum clique problem is polynomial on unit disk graphs (the problem
is reduced to finding a maximum independent set in a bipartite
graph). In \cite{breu}, it is shown that even recognizing whether a
graph is a unit disk graph is {\bf NP}-hard. A slightly different
version of the problem, consisting in determining whether a given
weighted graph can be realized in $\mathbb{R}^2$ as a unit disk graph
of given radius, is also {\bf NP}-hard \cite{eren06}. From the point
of view of DG, it is interesting to remark that the DGP, restricted to
sufficiently dense unit disk graphs and provided a partial realization
is known for a subset of at least $K+1$ vertices, can be solved in
polynomial time \cite{ye}. If the graph is sparse, however, the DGP is
still {\bf NP}-hard \cite{aspnes}.

The study of unit disk graphs also arises when packing equal spheres
in a subset of Euclidean space \cite{conwaysloane}: the contact graph
of the sphere configuration are unit disk graphs.

\subsection{Statics}
\label{s:rigid}
Statics is the study of forces acting on physical systems in static
equilibrium. This means that the barycenter of the system undergoes no
linear acceleration (we actually assume the barycenter to have zero
velocity), and that the system does not rotate. Geometrically, with
respect to a frame of reference, the system undergoes no translations
and no rotations. The physical systems we are concerned with are
bar-and-joint structures, i.e.~three-dimensional embodiments of graph
frameworks $(G,x)$ where $G$ is a simple weighted undirected graph and
$x$ is a valid realization thereof: joints are vertices, bars are
edges, and bar lengths are edge weights. The placement of the
structure in physical space provides a valid realization of the
underlying graph. Because we suppose the structures to be stiff, they
cannot undergo reflections, either. In short, the equivalence class of
a rigid graph frameworks modulo congruences is a good representation
of a structure in static equilibrium. Naturally, the supporting
bar-and-joint structures of man-made constructions such as houses,
buildings, skyscrapers, bridges and so on must always be in static
equilibrium, for otherwise the construction would collapse.

Statics was a field of study ever since humans wanted to have rooves
over their heads. The main question is the estimation of reaction
forces that man-made structures have to provide in order to remain in
static equilibrium under the action of external forces. In 1725,
Varignon published a textbook \cite{varignon} which implemented ideas
he had sketched in 1687 about the application of systems of forces to
different points of static structures. By the mid-1800s, there was
both an algebraic and a graphical method for testing rigidity of
structures. Because of the absence of computing machinery, the latter
(called {\it graphical statics}) was preferred to the former
\cite{cremona1872,saviotti1888,henneberg1911}. Cremona proposed a
graphical axiomatization of arithmetic operations in
\cite{cremona1874}, whose purpose was probably that of giving an
implied equivalence between two methods. The algebraic method
attracted some attention notwithstanding its numerical difficulties:
Maxwell proposed a simplified version \cite{maxwell1864} in 1864.

\subsubsection{Infinitesimal rigidity}
\label{s:infrigid}
Since statics is mainly concerned with the physical three-dimensional
world, we fix $K=3$ for the rest of this section.  Consider a function
$F:V\to\mathbb{R}^3$ that assigns a force vector $F_v\in\mathbb{R}^3$
to each point $x_v\in\mathbb{R}^3$ of a framework $(G,x)$. If the
framework is to be stationary, the total force and torque acting on it
must be null to prevent translations (assuming a zero initial velocity
of the barycenter) and rotations. This can be written
algebraically \cite{roth,tay-whiteley} as:
\begin{eqnarray}
  \sum_{v\in V} F_v&=&0 \label{eq:equilforce1} \\
  \forall i<j\le K \quad \sum_{v\in V}(F_{vi}x_{vj}-F_{vj}x_{vi}) &=&0.
  \label{eq:equilforce2}
\end{eqnarray}
A force $F$ satisfying
Eq.~\eqref{eq:equilforce1}-\eqref{eq:equilforce2} is called an {\it
  equilibrium force} (or {\it equilibrium load}). Applied to
bar-and-joint structures, equilibrium forces tend to compress or
extend the bars without moving the joints in space. Since bars are
assumed to be stiff (or equivalently, the graph edge weights are given
constants), the corresponding reaction forces at the endpoint of each
bar should be equal in magnitude and opposite in sign. We can define
these reaction forces by means of an edge weighting
$\omega:E\to\mathbb{R}$ representing the amount of force in each bar
per unit length ($\omega$ is negative for bar tensions and positive
for bar compressions). Stiffness of the structure translates
algebraically to a balance of equilibrium force and reaction:
\begin{equation}
  \forall u\in V\quad F_u+\sum_{v\in N(u)} \omega_{uv}(x_u-x_v)=0.
  \label{eq:resolution}
\end{equation}
A vector $\omega\in\mathbb{R}^{m}$ satisfying
Eq.~\eqref{eq:resolution} is called a {\it resolution}, or {\it
  resolving stress}, of the equilibrium force $F$ \cite{roth}. If
$F=0$, then $\omega$ is a {\it self-stress}.

For the following, we introduce (squared) edge functions and
displacements.  The {\it edge function} of a framework $(G,x)$ is a
function $\phi:\mathbb{R}^{nK}\to\mathbb{R}^m$ given by
$\phi(x)=(\|x_u-x_v\|\;|\;\{u,v\}\in E)$. We denote the {\it squared}
edge function $(\|x_u-x_v\|^2\;|\;\{u,v\}\in E)$ by $\phi^2$.  The
{\it edge displacement} of a framework $(G,x)$, with respect to a
displacement $y$, is a continuous function $\mu:[0,1]\to\mathbb{R}^m$
given by $\mu(t)=(\|y_u(t)-y_v(t)\|\;|\;\{u,v\}\in E)$. We denote the
squared edge displacement $(\|y_u(t)-y_v(t)\|^2\;|\;\{u,v\}\in E)$ by
$\mu^2$.

Eq.~\eqref{eq:resolution} can also be written as
\begin{equation}
  \frac{1}{2}\transpose{(\mbox{\sf d}\phi^2)}\omega = -F, \label{eq:res2}
\end{equation}
where $\mbox{\sf d}\phi^2$ is the matrix whose $\{u,v\}$-th row
encodes the derivatives of the $\{u,v\}$-th component of the squared
edge function $\phi^2(x)$ with respect to each component $x_{vi}$ of
$x$. Observe that the $\{u,v\}$-th row of this matrix only has the six
nonzero components $2(x_{ui}-x_{vi})$ and $2(x_{vi}-x_{ui})$ for
$i\in\{1,2,3\}$ (see \cite[p.~13]{roth}). If we now consider
Eq.~\eqref{eq:res2} applied to a displacement $y(t)$ of $x$,
differentiate it with respect to $t$ and evaluate it at $t=0$, we
obtain the linear system $\omega A=0$ where $A=\frac{1}{2}\mbox{\sf
d}\phi^2$, i.e.~the homogeneous version of Eq.~\ref{eq:res2}. 

Consider now a squared edge displacement $\mu^2(t)$ with respect to a
flexing $y$ of the framework $(G,x)$. By definition of flexing, we
have $\mu^2(t)=(d_{uv}^2\;|\;\{u,v\}\in E)$ for all $t\in
[0,1]$. Differentiating with respect to $t$, we obtain the scalar
product relation $2(y_u(t)-y_v(t))\cdot(\frac{\mbox{\sf
d}y_u(t)}{\mbox{\sf d}t}-\frac{\mbox{\sf d}y_v(t)}{\mbox{\sf d}t})=0$
(because the edge weights $d_{uv}$ are constant with respect to $t$)
for all $\{u,v\}\in E$. Evaluating the derivative at $t=0$ yields
\begin{equation}
  \forall\{u,v\}\in E\quad (x_u-x_v)\cdot(\alpha_u-\alpha_v)=0,
  \label{eq:vel}
\end{equation}
where $\alpha:V\to\mathbb{R}^3$ is a map that assigns initial
velocities $\alpha_v=\frac{\mbox{\sf d}x_u}{\mbox{\sf d}t}|_{0}$ to
each $v\in V$. We remark that the system \eqref{eq:vel} can be written
as $A\alpha=0$ \cite[Thm.~3.9]{gluck}. We therefore have the dual
relationship $\omega A=0=A\alpha$ between $\alpha$ and $\omega$.

By definition, $(G,x)$ is infinitesimally rigid if $\alpha$ only
encodes rotations and translations. The above discussion should give
an intuition as to why this is equivalent to stating that every
equilibrium force has a resolution (see 
\cite{gluck,roth,tay-whiteley} for a full description).
Indeed, infinitesimal rigidity was defined in this dual way by
Whiteley \cite{whiteleyI} (who called it {\it static rigidity}). The
matrix $A$ above is called the {\it rigidity matrix} of the framework
$(G,x)$. Notice that, when a valid realization $x$ is known for $G$,
then even those distances for $\{u,v\}\not\in E$ can be computed for
$G$: when the rows of $A$ are indexed by all unordered pairs $\{u,v\}$
we call $A$ the {\it complete rigidity matrix} of $(G,x)$.

Infinitesimal rigidity is a stricter notion than rigidity: all
infinitesimally rigid frameworks are also rigid
\cite[Thm.~4.1]{gluck}. Counterexamples to the converse of this
statements, i.e.~rigid frameworks which are infinitesimally flexible,
usually turn out to have some kind of degeneracy: a flat triangle, for
example, is rigid but infinitesimally flexible
\cite[Ex.~4.2]{roth}. In general, infinitesimally rigid frameworks in
$\mathbb{R}^K$ (for some integer $K>0$) might fail to be
infinitesimally rigid in higher-dimensional spaces \cite{servatius}.

\subsubsection{Graph rigidity}
\label{s:rigid_generic}
An important practical question to be asked about rigidity is whether
certain graphs give rise to infinitesimally rigid frameworks just
because of their graph topology, independently of their edge
weights. Bar-and-joint frameworks derived from such graphs are
extremely useful in architecture and construction engineering. An
important concept in answering this question is that of genericity: a
realization is {\it generic} if all its vertex coordinates are
algebraically independent over $\mathbb{Q}$. Because the algebraic
numbers have Lebesgue measure zero in the real numbers, this means
that the set of non-generic realizations have Lebesgue measure 0 in
the set of all realizations.

Rigidity and infinitesimal rigidity are defined as properties of
frameworks, rather than of graphs. It turns out, however, that if a
graph possesses a single generic rigid framework, then all its generic
frameworks are rigid \cite[Cor.~2]{asimow1}. This also holds for
infinitesimal rigidity \cite{asimow2}. Moreover, rigidity and
infinitesimal rigidity are the same notion over the set of all generic
frameworks \cite[Sect.~3]{asimow2}. By genericity, this implies that
in almost all cases it makes sense to speak of a ``rigid graph''
(rather than a rigid framework). The {\sc Graph Rigidity Problem}
asks, given a simple undirected graph $G$, whether it is generically
rigid. Notice that the input, in this case, does not involve edge
weights. For example, any graph is almost always flexible for large
enough values of $K$ unless it is a clique \cite[Cor.~4]{asimow1}.

We remark as an aside that, although genericity is required for laying
the theoretical foundations of graph rigidity (see the proof of
\cite[Thm.~6.1]{gluck}), in practice it is too strong. For an edge
weighting to be algebraically independent over $\mathbb{Q}$, at most
one edge weight can be rational (or even algebraic). Since computers
are usually programmed to only represent rational (or at best
algebraic) numbers, no generic realization can be treated exactly in
any practical algorithmic implementation. The conceptual requirement
that genericity is really meant to convey is that an infinitesimally
rigid generic realization will stay rigid even though the edge
weighting is perturbed slightly \cite{servatius}. The definition given
in \cite{graver} is more explicit in this sense: a realization is
generic if all the nontrivial minors of the complete rigidity matrix
have nonzero value. Specifically, notice that the polynomials induced
by each minor are algebraic relations between the values of the
components of each vector in the realization. Naturally, asking for
full algebraic independence with respect to any polynomial in
$\mathbb{Q}$ guarantees Graver's definition, but in fact, as Graver
points out \cite{graverbook}, it is sufficient to enforce algebraic
independence with respect to the system of polynomials induced by the
nontrivial minors of the rigidity matrix (also see
Sect.~\ref{s:sip}). 

Generic graph rigidity can also be described using the graphical
matroid $M(G)$ on $G$: a set of edges is independent if it does not
contain simple cycles. The closure of an edge subset $F\subseteq E$
contains $F$ and all edges which form simple cycles with edges of
$F$. We call the edge set $F$ {\it rigid} if its closure is the clique
on the vertices incident on $F$. A graphical matroid $M(G)$ is an {\it
abstract rigidity matroid} if it satisfies two requirements: (i) if
two edge sets are incident to fewer than $K$ common vertices, the
closure of their union should be the union of their closures; and (ii)
if two edge sets are incident to at least $K$ common vertices, their
union should be a rigid edge set \cite{servatius}. Condition (i)
loosely says that if the two edge sets are not ``connected enough'',
then their union should give rise to flexible frameworks in
$\mathbb{R}^K$, as the common vertices can be used as a ``hinge'' in
$\mathbb{R}^K$ around which the two edge sets can rotate. Condition
(ii) says that when no such hinges can be found, the union of the two
edge sets gives rise to rigid graphs. If the only resolution to the
zero equilibrium force is the zero vector, then the complete rigidity
matrix has maximum rank (i.e.~it has the maximum possible rank over
all embeddings in $\mathbb{R}^{nK}$), and its rows naturally induce a
matroid on the complete set of edges $\{\{u,v\}\;|\;u\not=v\in V\}$,
called the {\it rigidity matroid} of the framework $(G,x)$. It was
shown in \cite{graver} that if $x$ is generic, then the rigidity
matroid is abstract.

\subsubsection{Some classes of rigid graphs}
\label{s:henneberg}
Euler conjectured in 1766 that all graphs given by the edge incidence
of any triangulated polyhedral surface are rigid in
$\mathbb{R}^3$. This conjecture was proven true for special cases but
eventually disproved in general. Cauchy proved in 1813 that the
conjecture holds for strictly convex polyhedra \cite{cauchyrigid},
Alexandrov proved in 1950 that it holds for convex
polyhedra \cite{alexandrov}, and Gluck proved in 1975 that it also
almost always holds for any triangulation of a topological
sphere \cite{gluck}. The general conjecture was finally disproved by
Connelly in 1977 \cite{connelly-countereg} using a skew octahedron.

This does not mean to say that there are no purely topological
characterizations of rigid graphs. In 1911, Henneberg described two
local procedures (or ``steps'') to construct new, larger rigid graphs
from given rigid graphs \cite{henneberg1911} (if a given graph can be
``deconstructed'' by using the same procedures backwards, then the
graph is rigid). The Henneberg type I step is as follows: start with a
$K$-clique and add new vertices adjacent to at least $K$ existing
vertices. This defines a vertex order known as Henneberg type I order
(see Sect.~\ref{s:prelim:orders}). The Henneberg type II step is
somewhat more involved, and we refer the interested reader to the
extensive account of Henneberg and Henneberg-like procedures which can
be found in \cite{tay-whiteley}. Here follows a philological note on
Henneberg type I orders: although they are always referred
to \cite{henneberg1911}, they were actually first defined in a
previous book by Henneberg \cite[p.~267]{henneberg1886}. But in fact,
a picture with a Henneberg type I order in $\mathbb{R}^2$ appeared one
year earlier, in 1885, in \cite[Fig.~30, Pl.~XV]{saviotti_fr}.

Limited to $\mathbb{R}^2$, a characterization of all rigid graphs $G$
in $\mathbb{R}^2$ was described by Laman in 1970 \cite{laman}:
$|E|=2|V|-3$ and for every subgraph $(V',E')$ of $G$, $|E'|\le
2|V'|-3$. Equivalent but more easily verifiable conditions were
proposed in \cite{lovasz-yemini,recski,tay}. Unluckily, such
conditions do not hold for $\mathbb{R}^3$. For $K>2$, no such complete
characterization is known as yet; an account of the current
conjectures can be found in \cite{whiteley_survey,jackson_molecule},
and a heuristic method was introduced in \cite{sitharam}.

\subsection{Other applications}
\label{s:otherapps}
DG is not limited to these applications, however. For example, an
application to the synchronization of clocks from the measure of time
offsets between pairs of clocks is discussed in \cite{singer4}. This,
incidentally, is the only engineering application of the DGP${}_1$ we
are aware of. The solution method involves maximizing a quadratic form
subject to normalization constraints; this is relaxed to the
maximization of the same quadratic form over a sphere, which is solved
by the normalized eigenvector corresponding to the largest
eigenvalue. Another application is the localization and control of
fleets of autonomous underwater vehicles (AUVs) \cite{bahr}. This is
essentially a time-dependent DGP, as the delays in sound measurements
provide an estimate of AUV-to-AUV distance and an indication of how it
varies in time. We remark that GPS cannot be used under water, so AUVs
must resurface in order to determine their positions precisely. A
third application to the quantitative analysis of music and rhythm is
discussed in \cite{demaine}.

In the following section, we briefly discuss two other important
engineering applications of DG: data visualization by means of
multidimensional scaling, and robotics, specifically inverse kinematic
calculations. In the former, we aim to find a projection in the plane
or the space which renders the graph visually as close as possible to
the higher-dimensional picture (see Sect.~\ref{s:mds}). In the latter,
the main issue is to study how a robotic arm (or system of robotic
arms) moves in space in order to perform certain tasks. Known
distances include those from a joint to its neighbouring joints. The
main problem is that of assigning coordinate values to the position
vector of the farthest joint (see Sect.~\ref{s:robot}).

\subsubsection{Data visualization}
\label{s:mds}

Multidimensional Scaling (MDS) \cite{borg_10,everitt_97} is a
visualization tool in data analysis for representing measurements of
dissimilarity among pairs of objects as distances between points in a
low-dimensional space in such a way that the given dissimilarities are
well-approximated by the distances in that space. The choice of
dimension is arbitrary, but the most frequently used dimensions are 2
and 3. MDS methods differ mainly according to the distance model, but
the most usual model is the Euclidean one (in order to represent
correlation measurements, a spherical model can also be used). Other
distances, such as the $\ell_1$ norm (also called Manhattan distance)
are used \cite{arabie_91,zilinskas_09}. The output of MDS provides
graphical displays that allow decision makers to discover hidden
structures in complex data sets.

MDS techniques have been used primarily in psychology. According to
\cite{leeuw_82}, the first important contributions to the theory of
MDS are probably \cite{stumpf_83,stumpf_90}, but they did not lead to
practical methods. The contributions to the MDS methods are due to
Thurstonian approach, summarized in chapter 11 of \cite
{torgerson_58}, although the real computational breakthrough was due
to Shepard \cite{shepard_62a,shepard_62b,shepard_66}. The next
important step was given by Kruskal \cite {kruskal_64a,kruskal_64b},
who puts Shepard's ideas on a formal way in terms of optimization of a
least squares function. Two important contributions after
Shepard-Kruskal works are \cite{carroll_70} and \cite{takane_77}.

Measurements of dissimilarity among $n$ objects can be represented by
a dissimilarity matrix $D=(d_{ij})$. The goal of MDS is to construct a
set of points $x_i\in\mathbb{R}^K$ (for $i\le n$ and $K$ low,
typically $K\in\{2,3\}$) corresponding to those $n$ objects such that
pairwise distances approximate pairwise object dissimilarities (also
see the APA method in Sect.~\ref{s:apa}). MDS is complementary to
Principal Component Analysis (PCA) \cite{gower_66}, in the following
sense. Given a set $X$ of $n$ points in $\mathbb{R}^H$ (with $H$
``high''), PCA finds a $K$-dimensional subspace of $\mathbb{R}^H$
(with $K$ ``small'') on which to project $X$ in such a way that the
variance of the projection is maximum (essentially, PCA attempts to
avoid projections where two distant points are projected very
close). PCA might lose some distance information in the projection,
but the remaining information is not distorted. MDS identifies a
$K$-dimensional subspace $\tau$ of $\mathbb{R}^H$ which minimizes the
discrepancy between the original dissimilarity matrix $D$ of the
points in $X$ and the dissimilarity matrix $D'$ obtained by the
projection on $\tau$ of the points in $X$ \cite{domany}. In other
words, MDS attempts to represents all distance information in the
projection, even if this might mean that the information is distorted.


\subsubsection{Robotics}
\label{s:robot}
Kinematics is the branch of mechanics concerning the geometric
analysis of motion. The kinematic analysis of rigid bodies connected
by flexible joints has many similarities with the geometric analysis
of molecules, when the force effects are ignored.

The fundamental DG problem in robotics is known as the Inverse
Kinematic Problem (IKP --- see Item \ref{item:ikp} in the list of
Sect.~\ref{s:map}). Geometric constructive methods can be applied to
solve the IKP \cite{Fudos_97}, but algebraic techniques are more
suitable to handle more general instances. Reviews of these techniques
in the context of robotics and molecular conformation can be found,
for example, in \cite{Nielsen_99,Emiris_99,rojas_10}. There are three
main classes of methods in this category: those that use algebraic
geometry, those based on continuation techniques, and those based on
interval analysis.

In general, the solution of the IKP leads to a system of polynomial
equations. The methods based on algebraic geometry reduce the
polynomial system to a univariate polynomial, where the roots of this
polynomial yield all solutions of the original
system \cite{Manocha_94,Canny_00}.  Continuation methods, originally
developed in \cite{Roth_63}, start with an initial system, whose
solutions are known, and transform it into the system of interest,
whose solutions are sought. In \cite{Tsai_85}, using continuation
methods, it was shown that the inverse kinematics of the general 6R
manipulator (an arm system with six rotatable bonds with fixed lengths
and angles \cite{Hunt_90}) has 16 solutions; more information can be
found in \cite{Wampler_90}.  

A type of interval method applied to IKP is related to the interval
version of the Newton method \cite{Rao_98}, and others are based on
the iterative division of the distance space of the
problem \cite{dualbp}. An interesting method in the latter
class \cite{Thomas_04} essentially consists in solving a EDMCP whose
entries are intervals (see Sect.~\ref{s:mcp} and \ref{s:edmcp}). When
the distance matrix is complete, the realization of the selected
points can be carried out in polynomial time (see
e.g.~\cite{sippl,dongwu}). In order to determine the values for the
unknown distances, in \cite{Porta_05}, a range is initially assigned
to the unknowns and their bounds are reduced using a branch-and-prune
technique, which iteratively eliminates from the distance space entire
regions which cannot contain any solution. This elimination is
accomplished by applying conditions derived from the theory of
distance geometry. This branch-and-prune technique is different from
the BP algorithm discussed in Sect.~\ref{s:discr} and \ref{s:nmr}, as
the search space is continuous in the former and discrete in the
latter. Another branch-and-prune scheme for searching continuous space
is described in \cite{Zhang_05}. This is applied to molecular
conformational calculations related to computer-assisted drug design.

\section{Conclusion}
\label{s:concl}
Euclidean distance geometry is an extensive field with major
biological, statistical and engineering applications. The foundation
of its theory was laid around a century ago by mathematicians such as
Cayley, Menger, Schoenberg, Blumenthal and G\"odel. Recent extensions,
targeting the inverse problem of determining a distance space given a
partial distance function, contribute further mathematical as well as
applied interest to the field. Because of the breadth and maturity of
this field, our survey makes no claim to completeness; furthermore, we
admit to a personal bias towards applications to molecular
conformation. We strove, however, to give the reader a sufficiently
informative account of the most useful, interesting, and beautiful
results of Euclidean distance geometry.

\section*{Acknowledgments}
We are grateful to Jon Lee, Audrey Lee-St.~John, Therese Malliavin,
Beno\^{\i}t Masson, Michael Nilges and Maxim Sviridenko for
co-authoring some of the papers we wrote on different facets of this
topic. We equally grateful to Leandro Martinez for useful
discussions. We also wish to thank Chiara Bellasio for providing
inspiring dishes, a pleasant atmosphere and lots of patience and
support during many working sessions in Paris. This work was partially
supported by the Brazilian research agencies FAPESP, CNPq, CAPES, and
by the French research agency ANR.


\bibliographystyle{siam}
\bibliography{pcc}

\end{document}